\documentclass[preprint2]{aastex61}
\usepackage{amsmath}
\PassOptionsToPackage{breaklinks=true}{hyperref}
\def\chandra    {\emph{Chandra}}

\def\vla        {\emph{VLA}}

\def\lofar       {\emph{LOFAR}}
\def\gmrt {\emph{GMRT}}

\def\lax{\lesssim}
\def\gax{\gtrsim}

\received{April 30, 2019}
\revised{June 12, 2019}
\accepted{June 12, 2019}
\submitjournal{ApJ}

\shorttitle{Expanding the sample of radio minihalos}
\shortauthors{Giacintucci et al.}

\begin{document}


\title{Expanding the sample of radio minihalos in galaxy clusters}

\correspondingauthor{Simona Giacintucci}
\email{simona.giacintucci@nrl.navy.mil}

\author{Simona Giacintucci}
\affiliation{Naval Research Laboratory, 
4555 Overlook Avenue SW, Code 7213, 
Washington, DC 20375, USA}

\author{Maxim Markevitch}
\affiliation{NASA/Goddard Space Flight Center,
Greenbelt, MD 20771, USA}

\author{Rossella Cassano}
\affiliation{INAF - Istituto di Radioastronomia,
via Gobetti 101, I-40129 Bologna, Italy}

\author{Tiziana Venturi}
\affiliation{INAF - Istituto di Radioastronomia,
via Gobetti 101, I-40129 Bologna, Italy}

\author{Tracy E. Clarke}
\affiliation{Naval Research Laboratory,
4555 Overlook Avenue SW, Code 7213,
Washington, DC 20375, USA}

\author{Ruta Kale}
\affiliation{National Centre for Radio Astrophysics, 
Tata Institute of Fundamental Research, Pune University, 
Pune 411007, India}

\author{Virginia Cuciti}
\affiliation{INAF - Istituto di Radioastronomia,
via Gobetti 101, I-40129 Bologna, Italy}

\begin{abstract}
Radio minihalos are diffuse synchrotron sources of unknown origin
found in the cool cores of some galaxy clusters. We use {\em GMRT}\/ 
and {\em VLA}\/ data to expand the sample of minihalos by reporting 
three new minihalo detections (A\,2667, A\,907 and PSZ1\,G139.61+24.20)
and confirming minihalos in five clusters (MACS\,J0159.8--0849, 
MACS\,J0329.6-0211, RXC\,J2129.6+0005, AS\,780 and A\,3444). With these 
new detections and confirmations, the sample now stands at 23, the 
largest sample to date. For consistency, we also reanalyze archival 
{\em VLA}\/ 1.4 GHz observations of 7 known minihalos. We revisit possible 
correlations between the non-thermal emission and the thermal properties 
of their cluster hosts. Consistently with our earlier findings from a 
smaller sample, we find no strong relation between the minihalo radio 
luminosity and the total cluster mass. Instead, we find a strong positive 
correlation between the minihalo radio power and X-ray bolometric luminosity 
of the cool core ($r<70$ kpc). This supplements our earlier result that 
most if not all cool cores in massive clusters contain a minihalo. Comparison 
of radio and {\em Chandra}\/ X-ray images indicates that the minihalo emission
is typically confined by concentric sloshing cold fronts in the cores
of most of our clusters, supporting the hypothesis that minihalos
arise from electron reacceleration by turbulence caused by core gas
sloshing. Taken together, our findings suggest that the origin of
minihalos should be closely related to the properties of thermal
plasma in cluster cool cores.
\end{abstract}

\keywords{catalogs --- galaxies: clusters: general --- surveys --- X-rays:
galaxies: clusters --- radio continuum: galaxies: clusters}

\section{Introduction}
\label{sec:intro}

Galaxy clusters are filled with tenuous, hot X-ray emitting plasma, which is
their dominant baryonic component. Over the cluster lifetime, it is heated
by merger-generated shocks and turbulence, and cools via X-ray radiation,
which occurs especially fast in the dense cluster cores. This plasma is
permeated by tangled magnetic fields and relativistic particles, which
contribute only a small fraction of the total energy density, but can
strongly affect the plasma behavior as well as provide an interesting window
into the cluster physics. The ultrarelativistic electrons in the
intracluster medium (ICM) cool rapidly via synchrotron and inverse Compton
radiation and have to be continuously energized. Thus, their synchrotron
radio emission provides insight into the particle acceleration processes in
clusters. Disregarding the radio galaxies often found in clusters, the diffuse
radio emission that originates in the ICM comes as three broad phenomena, each
apparently representing different acceleration mechanisms and energy sources
(see, e.g., van Weeren et al.\ 2019 for a recent review). 
Peripheral radio relics are believed to be the result of acceleration at 
ICM shocks, while giant diffuse radio halos, typically found in disturbed 
clusters  and extending over the whole cluster, are probably caused by electron 
reacceleration by merger-induced turbulence (e.g., Brunetti \& Jones 2014).

The third phenomenon, radio minihalos, are confined to cluster cool cores
(e.g., Giacintucci et al.\ 2017, hereafter G17). Minihalos are not directly
powered by radio jets from the active galactic nucleus (AGN) harbored in the
central cluster galaxy, nor they originate from diffusion, or other
transport mechanisms, of relativistic electrons from the AGN. The
radio-emitting electrons are instead generated {\em in situ}\/ either as a
by-product of hadronic collisions of cosmic-ray protons with thermal protons
in the ICM, or reacceleration of seed relativistic electrons
by turbulence in the cool core (e.g., Gitti et al.\ 2002, Pfrommer \&
Ensslin 2004, Fujita et al.\ 2007, Keshet \& Loeb 2010, Fujita \& Ohira
2013, Zandanel et al.\ 2014, ZuHone et al.\ 2013, 2015; for a review see
Brunetti \& Jones 2014).

The distinction among these types of faint radio sources is not always clear
in the radio data because of the finite instrument resolution and interferometric
image quality, as well as projection effects and the presence of unrelated
sources such as radio galaxies, but also because there is a continuum of
source morphologies. A physically-motivated definition of a minihalo that we
adopt is given in G17: it is an extended source in the cluster center that
(1) does not consist of radio lobes or tails nor shows any morphological
connection (such as jets) to the central cluster AGN, and (2) has
a radius between 50 kpc and $0.2R_{500}$.%
\footnote{$R_{500}$ is the radius within which the cluster mean total
  density is 500 times the critical density at the cluster redshift.}
Smaller sources can plausibly be explained by diffusion from the central
AGN (e.g., \S6.4 in Giacintucci et al. 2014a) and $0.2R_{500}$ is 
the typical boundary between the cluster core, where non-gravitational 
processes (cooling, AGN and stellar feedback) become important, 
and the outer, simpler cluster region where density, temperature 
and pressure profiles of the ICM are self-similar (e.g., McDonald et 
al. 2017 and references therein).

Using a complete cluster sample, G17 showed that minihalos are rather
exclusive to cool cores and that {\em most}\/ massive ($M_{\rm 500} > 6\times10^{14}$ $M_{\odot}$)
clusters with cool cores possess minihalos. The cool cores often exhibit sloshing-driven X-ray
cold fronts (e.g., Markevitch \& Vikhlinin 2007). In sloshing cores with a
minihalo, the radio emission is generally found to be co-spatial with the
gas sloshing region (e.g., Mazzotta \& Giacintucci 2008, Hlavacek-Larrondo
et al.\ 2013, Giacintucci et al.\ 2014a,b, Gendron-Marsolais et al.\ 2017, this
paper), suggesting a causal link between sloshing and
minihalos -- e.g., sloshing motions may amplify the magnetic fields (Keshet
et al.\ 2010) and generate turbulence in the cool core, which in turn
reaccelerates the relativistic electrons in minihalos (ZuHone et al.\ 2013).
Gas turbulent motions, that are sufficiently strong for this scenario, have
been measured by the {\em Hitomi} X-ray observatory in the cool core of the
Perseus cluster (Hitomi Collaboration et al.\ 2016), which is host to a
radio minihalo (Sijbring 1993, Burns et al.\ 1992, Gendron-Marsolais et al.\ 
2017).

Even though most minihalos have active radio galaxies at their centers, it
is unclear whether there is a relation between these AGNs and the
surrounding diffuse radio emission. The lack of a morphological connection
to the AGN (i.e., jets or lobes) in high-resolution radio images of
minihalos, when available, suggests that the AGN does not directly replenish
the minihalo with high-energy electrons. These AGNs could still be a source
of electrons for the minihalos by providing seed particles (e.g., from aged,
disrupted radio lobes) that are re-energized by turbulence (Cassano et al.\ 
2008, ZuHone et al.\ 2013). Central AGNs are also a strong candidate source
of the heating required to compensate for the otherwise catastrophic X-ray radiative
cooling of the gas at the cluster centers (see McNamara \& Nulsen 2007 for 
a review). The dissipation of the AGN-driven turbulence is one of the possible
mechanisms by which energy of the AGN outbursts is transferred to the gas
(Zhuravleva et al.\ 2014). Other proposed mechanisms include heating by
cosmic rays (e.g., Guo \& Oh 2008, Pfrommer 2013, Yang \& Reynolds 2016,
Ruszkowski et al.\ 2017) as well as gas sloshing, which can provide a net
heat inflow by bringing the outer, higher-entropy gas into the cool core
(ZuHone et al.\ 2010). If the radio minihalos are generated by turbulence
and/or hadronic cosmic ray interactions, then a connection between the
non-thermal radio emission and the non-gravitational processes in cool cores
(cooling and heating) may exist at a fundamental level (e.g., Fujita \&
Ohira 2013, Bravi et al.\ 2016, Jacob \& Pfrommer 2017). This makes the
minihalos a very interesting physical probe.

From an observational point of view, the detection and study of minihalos is
complicated by a combination of their extent and low surface brightness and
the presence of the often much brighter central radio galaxy. Sensitive,
{\em low}-resolution radio observations with a good sampling of the
interferometric $uv$\/ plane, particularly at short antenna spacings
(ideally, a full-synthesis single-dish imaging), are crucial to properly
image the extended emission. At the same time, high-resolution images are
needed to identify the emission associated with the central galaxy (as well
as other possible radio galaxies in, or projected onto, the cluster core)
and rule out the possibility of the large-scale diffuse emission being part
of the radio galaxy. For most confirmed minihalos, radio images at different
angular resolutions are available (e.g., Govoni et al.\ 2009, Giacintucci et
al.\ 2014a,b). However, the existing images for a number of minihalo
candidates do not sufficiently detail the morphology of the central galaxy
and thus it is not possible to establish whether the AGN is connected to the
surrounding extended emission.

In this paper, we present {\em Giant Metrewave Radio Telescope}\/ ({\em
  GMRT}) images from our high-resolution follow-up of previously known
candidate minihalos in MACS\,J0159.8-0849, MACS\,J0329.6-0211 and
RXC\,J2129.6+0005 (Giacintucci et al.\ 2014a, henceforth G14a; Kale et al.\ 
2015; hereafter K15). We present new {\em GMRT}\/ and {\em Very Large Array}
({\em VLA}) images of the minihalos in AS\,780 and A\,3444 (Venturi et al.\ 
2007, K15). Using {\em GMRT}\/ data, we also report on the detection of new
minihalos in the cool cores of A\,2667, A\,907 and PSZ1\,G139.61+24.20. The
latter is a newly discovered {\em Planck} cluster, where a recent {\em
  LOFAR}%
\footnote{Low Frequency Array, van Haarlem et al.\ (2013).}
observation at 144 MHz has revealed faint ultra-steep spectrum radio
emission, extending beyond the minihalo region (Savini et al.\ 2018). For
consistency with our minihalo studies (G14a, G17, this paper), we also
re-analyze archival, multi-configuration {\em VLA}\/ observations at 1.4 GHz
of 7 clusters, in which minihalos were reported in the literature --- A\,1835,
Ophiuchus, A\,2029, RBS\,797, RX\,J1347.5--1145, MS\,1455.0+2232 and
2A\,0335+096 (Govoni et al.\ 2009, Gitti et al.\ 2006, 2007, Doria et al.\
2012, Venturi et al.\ 2008, Mazzotta \& Giacintucci 2008, Sarazin et al.\
1995). Radio results presented here have been used for the G17 statistical
sample study.

Using our larger sample of confirmed minihalos, we investigate correlations
between the radio emission and the properties of the cluster hosts, derived
from {\em Chandra}\/ X-ray data. Table~\ref{tab:sources} lists the minihalos
used in our statistical analysis. We do not include clusters with a central
extended radio source whose classification as a minihalo is still uncertain
or unconfirmed (G14a, G17, van Weeren et al.\ 2019).

The paper is organized as follows: the radio observations are described in
\S2; our new radio images are presented in \S 3 and \S 4; the analysis of
the {\em Chandra}\/ archival X-ray data is described in \S5, and a brief
discussion of our results is given in \S6. Details of our radio analysis of
the previously known minihalos is presented in the Appendix.

We adopt a $\Lambda$CDM cosmology with H$_0$=70 km s$^{-1}$ Mpc$^{-1}$,
$\Omega_m=0.3$ and $\Omega_{\Lambda}=0.7$. All errors are quoted at the
$68\%$ confidence level. For the radio spectral index $\alpha$, we 
adopt $S_{\nu} \propto \nu^{-\alpha}$, where $S_{\nu}$ 
is the flux density at the frequency $\nu$.


\startlongtable
\begin{deluxetable*}{lrccc}
\tabletypesize{\footnotesize}  
\tablecaption{List of minihalo clusters
\label{tab:sources}}
\tablehead{
\colhead{Cluster name} & \colhead{RA$_{\rm J2000}$} & \colhead{DEC$_{\rm J2000}$} & \colhead{$z$} & \colhead{scale}  \\
\colhead{} & \colhead{(h, m, s)} & \colhead{($^{\circ}, ^{\prime}, ^{\prime \prime}$)}  &  \colhead{} &  \colhead{(kpc$/^{\prime \prime}$)}   \\
}
\startdata
\multicolumn{5}{c}{Clusters analyzed in this paper} \\
\noalign{\smallskip} 
MACS\,J0159.8-0849      & 01 59 48.0 & $-$08 49 00 & 0.405 & 5.413 \\
MACS\,J0329.6-0211      & 03 29 40.8 & $-$02 11 54 & 0.450 & 5.759 \\ 
RXC\,J2129.6+0005       & 21 29 37.9 &   +00 05 39 & 0.235 & 3.734 \\ 
AS\,780                 & 14 59 29.3 & $-18$ 11 13 & 0.236 & 3.746 \\ 
A\,3444                 & 10 23 54.8 & $-$27 17 09 & 0.254 & 3.958  \\
A\,907                  & 09 58 22.2 & $-$11 03 35 & 0.153 & 2.653  \\
A\,2667                 & 23 51 40.7 & $-$26 05 01 & 0.230 & 3.674  \\
PSZ1\,G139.61+24.20\tablenotemark{a} & 06 22 13.9 &   +74 41 39 & 0.267 & 4.102  \\
A1835                   & 14 01 02.3 &  +02 52 48  & 0.253 & 3.944  \\
Ophiuchus               & 17 12 25.9 & $-$23 22 33 & 0.028 & 0.562  \\
A\,2029                 & 15 10 55.0 & +05 43 12   & 0.077 & 1.451 \\
RBS\,797                & 09 47 12.9 & +76 23 13    & 0.354 & 4.894 \\
RX\,J1347.5--1145       & 13 47 30.6 & $-$11 45 10 & 0.451 & 5.771 \\
MS\,1455.0+2232         & 14 57 15.1 & +22 20 34   & 0.258 & 4.001 \\
2A\,0335+096            & 03 38 35.3 & +09 57 55   & 0.036 & 0.722 \\
\hline
\noalign{\smallskip}
\multicolumn{5}{c}{Clusters with literature radio information} \\
\noalign{\smallskip}
A\,478              & 04 13 20.7 & +10 28 35   & 0.088 & 1.646 \\ 
ZwCl\,3146          & 10 23 39.6 & +04 11 10   & 0.290 & 4.350 \\ 
RX\,J1532.9+3021    & 15 32 54.4 & +30 21 11   & \phantom{0}0.362\tablenotemark{b} & 5.048 \\ 
A\,2204             & 16 32 45.7 & +05 34 43   &  0.152  & 2.643 \\ 
Perseus             & 03 19 47.2 & +41 30 47   & 0.018  & 0.366 \\
RXC\,J1504.1--0248  & 15 04 07.5 & $-$02 48 16 & 0.215  & 3.494 \\
RX\,J1720.1+2638    & 17 20 08.9 & +26 38 06   & 0.164   & 2.814  \\
Phoenix\tablenotemark{c}   & 23 44 42.2 & $-$42 43 08 & 0.597   & 6.670 \\
\enddata 
\tablecomments{   
Column 1: cluster name. Columns 2--4: cluster coordinates and redshift from the NASA/IPAC Extragalactic Database (NED).  
Column 5: angular to linear scale conversion. The literature radio information is from G14a, except for Perseus (Sijbring 1993), 
RXC\,J1504.1--0248 (Giacintucci et al. 2011), RX\,J1720.1+2638 (Giacintucci et al.2014b) and Phoenix (van Weeren et al. 2014).}
\tablenotetext{a}{ PSZ1\,G139.61+24.20: Coordinates and redshift are from the {\em Planck} SZ cluster catalog (Planck Collaboration et al. 2014).}
\tablenotetext{b}{ Crawford et al. (1999).}
\tablenotetext{c}{ SPT-CL J2344-4243: coordinates and redshift are from McDonald et al. (2015).}
\end{deluxetable*}


\section{Radio observations}

In this section, we describe the new {\em GMRT} observations at 1.4 GHz and
1.3 GHz of the clusters MACS\,J0159.8-0849, MACS\,J0329.6-0211,
RXC\,J2129.6+0005 and AS\,780.  For RXC\,J2129.6+0005 and AS\,780, we
complement these observations with higher-frequency data retrieved from the
{\em VLA} archive. We also re-analyze archive observations of AS\,780 ({\em
  GMRT}, 610 MHz), A\,3444 ({\em VLA}, 1.4 GHz), A\,907 ({\em GMRT}, 610
MHz) and A\,2667 ({\em GMRT}, 610 MHz and 1.15 GHz). Finally, we present new
{\em GMRT} data at 1.28 GHz and 610 MHz of the {\em Planck} cluster
PSZ1\,G139.61+24.20. Details on all these observations are provided in Table
2.

In Appendix A, we present our re-analysis of archival {\em VLA} observations
at 1.4 GHz of A\,1835, Ophiuchus, A\,2029, RBS\,797, RX\,J1347.5--1145,
MS\,1455.0+2232 and 2A\,0335+096 (Table 8), known to possess a central
minihalo at their center.  New images and individual notes on these clusters
are presented.

All data were reduced using the NRAO\footnote{National Radio Astronomy
  Observatory.} Astronomical Image Processing System (AIPS) package. A
description of the {\em GMRT} and {\em VLA} data reduction is given in
\S \ref{sec:obs} and \S \ref{sec:obs2}, respectively.

For all clusters, the flux densities of individual radio galaxies and
minihalos are summarized in Tables 3 (Table 9 for the newly analyzed
clusters with previously known minihalos). The position, size and flux
density of the unresolved radio galaxies were determined by fitting the
sources with a Gaussian model (JMFIT). For extended radio galaxies, we
measured the total flux density by integration within the $+3\sigma$ surface
brightness contour. As in G14a, the flux density of minihalos was obtained
by measuring the total flux in circular regions of a radius that
progressively increased from the radius encompassing the $+3\sigma$
isocontour until the integrated flux density reached saturation. The
uncertainty on the minihalo flux density was estimated as described in G14a.
For all minihalos, size ($R_{\rm MH}$, measured as in G14a) and luminosity at 1.4 GHz are
summarized in Table 4, which includes also 8 minihalos with literature
information. The table also reports the 1.4 GHz luminosity of the central 
radio galaxy. 

\subsection{GMRT data reduction}\label{sec:obs}

The {\em GMRT} data were collected in spectral-line observing mode, using
the software backend for all clusters (with the exception of A\,907 and
A\,2667), which provides a total observing bandwidth of 33.3 MHz, subdivided
in 256 channels.  The observations of A\,907 at 610 MHz and A\,2667 at 1.15
GHz were taken with the old hardware correlator, using a 256-channel band of
16 MHz width at 1.15 GHz, and two 16 MHz-wide upper-side and lower-side
bands (USB and LSB) at 610 MHz, each having 128 channels.

We accurately inspected the data and found that all observations were in
part impacted by radio frequency interference (RFI). After an initial step
of flux and bandpass calibration using standard calibration sources (3C48,
3C147 and 3C286), we used RFLAG to excise RFI-affected visibilities,
followed by manual flagging to remove residual bad data.  The flux and
bandpass calibration was then recomputed and applied to the data. The flux
densities of the primary calibrators were set in SETJY using the Perley \&
Butler (2013) coefficients. Phase calibrators, observed several times during
the observation, were used to calibrate the data in phase. A number of
phase-only self-calibration cycles and imaging were applied to the target
visibilities to correct residual phase errors. To compensate for the
non-complanarity of the array, we used wide-field imaging in each step of
the self-calibration process and to produce the final images, decomposing
the primary beam area into $\sim$60-80 smaller facets at 610 MHz and
$\sim$30-50 facets at frequencies near 1 GHz. Finally, to improve the
dynamic range of the final images, we used PEELR to ``peel'' problematic
nearby bright sources, whose side lobes affected the area of the target. The
final images were corrected for the \gmrt\ primary beam response using
PBCOR\footnote{www.ncra.tifr.res.in:8081/{\textasciitilde}ngk/primarybeam/beam.html}.
Residual amplitude errors are estimated to be within $5\%$ at 1 GHz and
$8\%$ at 610 MHz (Chandra et al.\ 2004). 

Table 2 provides the restoring beams and root mean square (rms) noise levels
($1\sigma$) of the final images, obtained using a ``robustness'' parameter
(Briggs 1995) ROBUST=0 in IMAGR. For each cluster, we made sets of images with 
different angular resolution and weighting schemes, ranging from pure uniform 
weighting (ROBUST$=-5$) to natural weighting (ROBUST$=+5$), to enhance the 
extended emission. The higher-resolution images were used to identify the 
radio source associated with the BCG and other possible radio galaxies in 
(or projected onto) the cluster core region. We also produced 
images by cutting the innermost region of the $uv$ plane and using the 
remaining long baselines ($ \gax 1-2$ k$\lambda$) to better evaluate 
the contribution of the discrete sources. Low-resolution images were 
used to map the diffuse radio emission.


\startlongtable
\begin{deluxetable*}{lcccccccc}
\tabletypesize{\scriptsize}  
\tablecaption{Details of the radio observations
\label{tab:obs}}
\tablehead{
\colhead{Cluster} & \colhead{Array}    & \colhead{Project} & \colhead{Frequency} & \colhead{Bandwidth} & \colhead{Date} & \colhead{Time}  & \colhead{FWHM, PA}  &   \colhead{rms} \\ 
 \colhead{}       &  \colhead{}        &  \colhead{}       &  \colhead{(GHz)}   &      \colhead{(MHz)} &  \colhead{}   &  \colhead{(min)}&  \colhead{($^{\prime \prime} \times^{\prime \prime}$, $^{\circ}$)\phantom{00}} & \colhead{($\mu$Jy beam$^{-1}$)}  \\
}
\startdata
MACS\,J0159.8$-$0849 & GMRT     & $28_{-}077$ & 1.39  &  33       & 2015 Aug 8     & 240   & $2.4\times2.1$, 18       & 35  \\ 
MACS\,J0329.6$-$0211 & GMRT     & $28_{-}077$ & 1.28  &  33       & 2015 Aug 7     & 123   & $2.7\times2.1$, 61       & 23  \\ 
RXC\,J2129.6+0005    & GMRT     & $28_{-}077$ & 1.30  &  33       & 2015 Jul 4     & 120   & $2.6\times2.1$, 82       & 40  \\ 
                     & VLA$-$B  & AH788       & 4.86  & 50        & 2002 Jul 07    & 60    & $1.3\times1.2$, 10       & 22  \\ 
                     & VLA$-$A  & AE177       & 8.46  & 50        & 1998 Apr 12    & 18    & $0.25\times0.22$, $-14$  & 22  \\
AS\,780              & GMRT     & 07CRA01     & 0.61  & 32        & 2005 Jan 7     & 50    & $6.0\times3.9$, 24       & 65  \\ 
                     & GMRT     & $30_{-}065$ & 1.30  & 33        & 2016 Oct 1     & 76    & $3.2\times1.8$, 33       & 83 \\ 
                     & VLA$-$A  & AR517       & 8.46  & 50        & 2003 Jul 24    & 3     & $0.4\times0.3$, $-1$     & 30 \\
A\,3444              & VLA$-$BnA & AC696      & 1.44  & 50        & 2003 Oct 02    & 200   & $8.3\times4.0$, 3        &  35 \\
                     & VLA$-$DnC & AC696      & 1.44  & 50        & 2004 May 30    & 260   & $42.5\times30.2$, 58     & 50 \\
A\,907               & GMRT     & 15JHC01     & 0.61  &  32       & 2008 Dec 04    & 150   & $6.0\times4.5$, $-55$    & 45  \\ 
A\,2667              & GMRT     & 10CRA01     & 0.61  &  16\tablenotemark{a}   & 2006 Aug 27    & 80    & $7.3\times4.3$, 2        & 50  \\
                     & GMRT     & 13HBA01     & 1.15  &  16       & 2007 Nov 19,20 & 560   & $3.6\times2.3$, 11       & 35  \\ 
                     PSZ1\,G139.61+24.20  & GMRT     & $28_{-}077$ & 0.61  &  33       & 2014 Oct 25    & 286   & $7.6\times5.2$, 27       & 35  \\
                     & GMRT     & $27_{-}025$ & 0.61  &  33       & 2015 Sept 4    & 280   & $6.3\times3.7$, 44       & 70  \\ 
                     & GMRT     & $30_{-}065$ & 1.28  &  33       & 2016 Aug 20    & 310   & $3.5\times2.0$, 53       & 15  \\
\enddata
\tablecomments{Column 1: cluster name. Column 2: radio telescope (for the {\em
  VLA}, the array configuration is also reported).  
Column 3: project identifier. Columns 4--7: observation frequency, bandwidth,
date and total time on source. Column 8: full-width half maximum (FWHM) and
position angle of the array, obtained for
ROBUST=0 in IMAGR. Column 9: image rms level ($1\sigma$). 
}
\tablenotetext{a}{ The cluster was observed using both the USB and LSB, however, due to residual RFI and other errors affecting the LSB data, 
we chose to use only the USB data set.}
\end{deluxetable*}


\subsection{VLA data reduction}\label{sec:obs2}

The {\em VLA} data were calibrated and reduced following standard procedure.
For each cluster, the observations in different array configurations were
processed and imaged separately. Several loops of imaging and
self-calibration were applied to each data set to reduce the effects of
residual phase errors in the images.  Tables 2 and 8 summarize radio beams
and $1\sigma$ noise levels of the final images, made using a ROBUST=0. 
When possible, data from different dates and configurations
were combined together in the $uv$ or image plane. As for the {\em GMRT} data,
we used set of images with different angular resolution, weighting schemes and
$uv$ range to identify discrete radio galaxies and map the diffuse minihalos. 

The flux density scale was set using standard calibration sources (3C48,
3C147 and 3C286) and the Perley \& Butler (2013) coefficients. Residual
amplitude errors are within $5\%$.


\startlongtable
\begin{deluxetable*}{lccccc}
\tabletypesize{\footnotesize}  
\tablecaption{Radio flux density and spectral index
\label{tab:index}}
\tablehead{
  \colhead{Cluster name} &     \colhead{Source} & \colhead{$\nu$} &  \colhead{S$_{\nu}$} &   \colhead{$\alpha$} &  \colhead{Reference}     \\
    \colhead{} &    \colhead{} &    \colhead{(GHz)} &    \colhead{(mJy)} &         \colhead{} &    \colhead{} \\
}
\startdata
MACS\,J0159.8--0849 & S1       & 1.39  & $34.0\pm1.7$  & $-0.57\pm0.04$  & 1 \\ 
                    &          & 1.40  & $35.0\pm1.8$  &                 & 2\\ 
                    &          & 8.46  & $94.8\pm4.8$  &                 & 2 \\ 
                    & MH       & 1.39  &  $\sim 2$     & \nodata         & 1 \\ 
                    &          & 1.40  & $2.4\pm0.2$   &                 & 2 \\
&&&&& \\ 
MACS\,J0329.6--0211 & S1       & 1.30  & $3.9\pm0.2$   &  \nodata &   1 \\ 
                    &          & 1.40  & $3.8\pm0.2$   &           & 2 \\
                    & MH       & 1.30  & $4.5\pm0.4$   &  \nodata  & 1   \\ 
                    &          & 1.40  & $3.8\pm0.4$   &           & 2 \\
&&&&& \\                                                                               
RXC\,J2129.6+0005   & S1       & 0.24  & $87\pm5$      &  $0.77\pm0.02$   & 3 \\
                    &          & 0.61  & $48.9\pm2.5$  &                  & 3 \\
                    &          & 1.30  & $25.3\pm1.3$  &                  & 1  \\ 
                    &          & 1.40  & $24\pm1.2$    &                  & 3 \\
                    &          & 4.86  & $8.6\pm0.3$   &                  & 1  \\ 
                    &          & 8.46  & $5.6\pm0.2$   &                  & 1 \\ 
                    & MH       & 0.24  & $21.0\pm1.6$  & $1.2\pm0.1$      & 3 \\
                    &          & 0.61  & $8.0\pm0.7$   &                  & 3 \\
                    &          & 1.30  & $2.5\pm0.2$   &                  & 1 \\ 
                    &          & 1.40  & $2.4\pm0.2$   &                  & 3 \\
&&&&& \\
AS\,780             & S1       & 0.61  & $102\pm8$    & $-0.17\pm0.04$    &  1       \\
                    &          & 1.30  & $101\pm5$    &                   &  1                     \\
                    &          & 8.46  & $160\pm8$    &                   &  1 \\ 
                    & MH       & 0.61  & $34\pm2$     & $<1.49$           &  1 \\ 
                    &          & 1.30  & $>11$        &                   &  1 \\
&&&&& \\
A\,3444             & S1       & 1.44  & $1.41\pm0.07$ & \nodata          & 1 \\
                    & MH       & 1.44  & $12.1\pm0.9$  & \nodata          & 1 \\
&&&& &\\
A\,907              & S1       &  0.61  &  $117\pm9$     & \nodata  & 1 \\  
                    & S2       &  0.61  &   $42\pm3$      & \nodata & 1 \\ 
                    & MH       &  0.61  &   $34.9\pm5.6$  & \nodata & 1  \\
&&&&& \\
A\,2667             & S1       &  0.61  & $20.1\pm1.6$  &  $0.5\pm0.1$ & 1   \\ 
                    &          &  1.15  & $14.8\pm0.7$  &             & 1  \\ 
                    & S2       &  0.61  & $3.7\pm0.3$   & $0.7\pm0.1$ & 1  \\    
                    &          &  1.15  & $2.4\pm0.1$   &              & 1 \\ 
                    & MH       &  0.61  & $15.3\pm1.2$  &  $1.0\pm0.2$ & 1  \\   
                    &          &  1.15  & $8.3\pm0.7$   &              & 1 \\ 
&&&&&\\
PSZ1\,G139.61+24.20 & S1       &  0.61  & $1.9\pm0.2$    & $1.05\pm0.16$ & 1 \\ 
                    &          &  1.28  & $0.87\pm0.05$ &               &  1 \\
                    & S2       &  0.61  & $0.25\pm0.02$ & $0.14\pm0.02$  & 1 \\
                    &          &  1.28  & $0.18\pm0.01$ &               &  1    \\ 
                    & MH       &  0.14  & $12\pm1.8$    &  $1.33\pm0.08$  & 4 \\
                    &          &  0.61  & $2.3\pm0.8$   &                 & 1     \\ 
                    &          &  1.28  & $0.65\pm0.08$ &                 & 1     \\ 
\enddata
\tablecomments{Column 1: cluster name. Column 2: radio source. Column 3: frequency. 
Column 4: flux density. Column 5: integrated spectral index, computed between the lowest 
and highest frequencies reported in the table. Column 6: reference for the radio flux. 
(1) this work. (2) G14a. (3) K15. (4) Savini et al.\ (2018).}

\end{deluxetable*}



\startlongtable
\begin{deluxetable*}{lcccc}
\tablecaption{Properties of minihalos
\label{tab:halos}}
\tablehead{
  \colhead{Cluster name} &     \colhead{$\rm R_{\rm \, MH}$} & \colhead{$P_{\rm \,MH, \,1.4 \, GHz}$} &  \colhead{$P_{\rm \,BCG, \,1.4 \, GHz}$} &   \colhead{Reference}     \\
    \colhead{} &    \colhead{(kpc)  } &    \colhead{($10^{24}$ W Hz$^{-1}$)} &    \colhead{($10^{24}$ W Hz$^{-1}$)} &         \colhead{}  \\
}
\startdata
\multicolumn{5}{c}{Clusters analyzed in this paper} \\
\noalign{\smallskip}
MACS\,J0159.8--0849 &   90  & $1.40\pm0.14$\phantom{0}   & $20.3\pm0.1$    &  2   \\  
MACS\,J0329.6--0211 &   70  & $2.84\pm0.30$\phantom{0}   &  $2.8\pm0.1$    &  2   \\  
PSZ1\,G139.61+24.20 &   50  & $0.13\pm0.02$\tablenotemark{a}  &  $<0.007$\tablenotemark{a}     &  1     \\ 
A\,907              &   65  & $0.9\pm0.2$\tablenotemark{b}              &  $4.1\pm0.4$\tablenotemark{b}  &  1   \\  
AS\,780             &   50  & $>1.7$\tablenotemark{c}                   &  $17\pm1$\tablenotemark{c}     &   1     \\
A\,3444             &  120  & $2.4\pm0.2$\phantom{0}     & $0.28\pm0.01$ &   1    \\  
RX\,J2129.6+0005    &   80  & $0.40\pm0.03$\phantom{0}   &  $3.1\pm0.2$ &   6    \\ 
A\,2667             &   70  & $1.1\pm0.1$\tablenotemark{d}              & $2.1\pm0.1$\tablenotemark{d}     &    1   \\  
A\,1835             &  240  & $1.19\pm0.25$\phantom{0}   & $6.3\pm0.3$     &   1   \\  
Ophiuchus           &  250  & $0.11\pm0.02$\phantom{0}   & $0.064\pm0.003$ &    1  \\   
A\,2029             &  270  & $0.28\pm0.04$\phantom{0}   & $7.4\pm0.4$     &    1  \\  
RBS\,797            &  120  & $2.20\pm0.24$\phantom{0}   & $7.6\pm0.4$     &     1  \\   
RX\,J1347.5--1145   &  320  & $26.7\pm2.1$\phantom{0}    & $22.7\pm1.1$    &    1    \\ 
MS\,1455.0+2232     &  120  & $1.75\pm0.23$\phantom{0}   & $0.96\pm0.05$   &    1  \\   
2A\,0335+096        &   70  & $0.06\pm0.01$\phantom{0}   & $0.058\pm0.003$ &    1   \\  
\hline\noalign{\smallskip}
\multicolumn{5}{c}{Clusters with literature radio information} \\
\noalign{\smallskip}
A\,478              &  160  & $0.32\pm0.06$   & $0.60\pm0.03$ &   2  \\   
ZwCl\,3146          &   90  & $1.4\pm0.3$     & $0.8\pm0.2$   &   2  \\   
RX\,J1532.9+3021    &  100  & $3.35\pm0.17$   & $7.0\pm0.4$   &   2  \\   
A\,2204             &   50  & $0.54\pm0.05$   & $3.7\pm0.2$   &   2   \\  
Perseus             &  130  & $2.18\pm0.11$   & $13.4\pm0.1$\tablenotemark{e}  &   3   \\  
RXC\,J1504.1--0248  &  140  & $2.70\pm0.14$   & $5.7\pm0.3$   &    4  \\   
RX\,J1720.1+2638    &  140  & $5.33\pm0.32$   & $0.50\pm0.02$ &   5  \\   
Phoenix             &   230 & $9.6\pm3.1$\tablenotemark{f}   & $73.9\pm7.4$\tablenotemark{g}  &    7    \\ 
\enddata
\tablecomments{Column 1: cluster name. Column 2: average radius of the minihalo (estimated as in G14a). 
Column 3: radio luminosity of the minihalo at 1.4 GHz. Column 4: radio luminosity of the BCG at 1.4 GHz. 
Column 5: references. (1) this work. (2) G14a. (3) Sijbring (1993). (4) Giacintucci et al.\ (2011). (5) Giacintucci et al.\ (2014b). 
(6) K15. (7) van Weeren et al.\ (2014).} 
\tablenotetext{a}{ Extrapolated from 1.28 GHz (\S\ref{sec:pg139}).}
\tablenotetext{b}{ Extrapolated from 610 MHz (\S\ref{sec:a907}).}
\tablenotetext{c}{ Extrapolated from 1.30 GHz (\S\ref{sec:as780}).}
\tablenotetext{d}{ Extrapolated from 1.15 GHz (\S\ref{sec:a2667}).}
\tablenotetext{e}{ From NVSS.}
\tablenotetext{f}{ Extrapolated from $S_{\rm \, MH, \,610 \, MHz}= 17\pm5$ mJy assuming $\alpha=1.0\div1.3$.}
\tablenotetext{g}{ Extrapolated from $S_{\rm \,BCG, \,610 \, MHz}= 87.9\pm9.0$ mJy assuming $\alpha=0.6\div0.8$.}
\end{deluxetable*}


\section{Minihalo confirmation}

In G14a, we suggested the presence of a minihalo in the clusters
MACS\,J0329.6--0211 and MACS\,J0159.8--0849, based on {\em VLA} 1.4 GHz
images at $5^{\prime\prime}$ resolution.  In K15, {\em GMRT} images at 235
MHz and 610 MHz revealed a central diffuse radio source in
RXC\,J2129.6+0005, which was classified as a minihalo. A possible minihalo
was reported at the center of AS\,780 and A\,3444 by Venturi et al.\ (2007)
using {\em GMRT} observations at 610 MHz. However, due to the lack of images
of all these clusters with a finer angular resolution, it was not possible
to exclude the possibility that the observed extended emission is sustained
directly by the central AGN through large-scale jets, rather than being an
actual minihalo. Here, we present new high-resolution {\em GMRT} and {\em
  VLA} images that show no evidence of such jets (or other radio features on
the minihalo scale directly connected to the AGN), thus confirming a
minihalo in all these clusters.

%
%
%
\begin{figure}
\centering
\includegraphics[width=7.5cm]{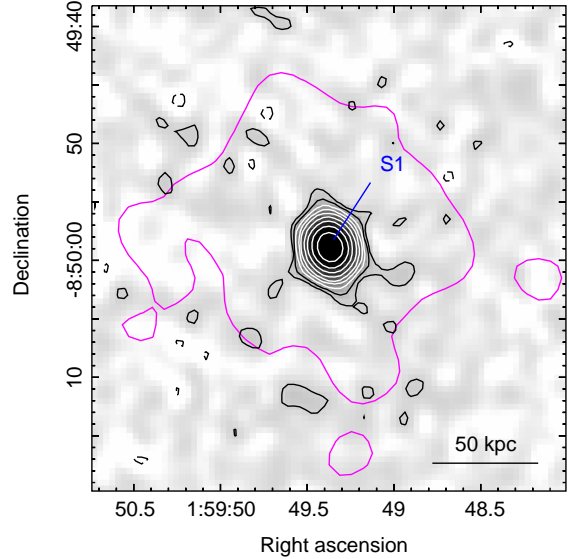}
\smallskip
\caption{MACS\,J0159.8--0849. \gmrt\ image at 1.39 GHz (greyscale and black and white
  contours). The unresolved source S1 is associated with the BCG. The restoring beam 
is $2^{\prime\prime}.4\times2^{\prime\prime}.1$,
  in p.a.  $18^{\circ}$ and rms noise level is $1\sigma=35$ $\mu$Jy
  beam$^{-1}$.  Contours are 0.07, 0.14 (black), 0.28, 0.56, 1, 2, 4, 8, 16,
  and 32 (white) mJy beam$^{-1}$.  Contours at $-0.07$ mJy beam$^{-1}$ are
  shown as dashed. For a comparison, the first contour from the {\em VLA}
  1.4 GHz image of G14a (their Figure ~5) is reported in magenta. The
  contour level is $0.045$ mJy beam$^{-1}$, for a $5^{\prime\prime}$
  circular beam.}
\label{fig:0159}
\end{figure}
%
%
%

%
%
%
\begin{figure*}
\centering
\includegraphics[width=8cm]{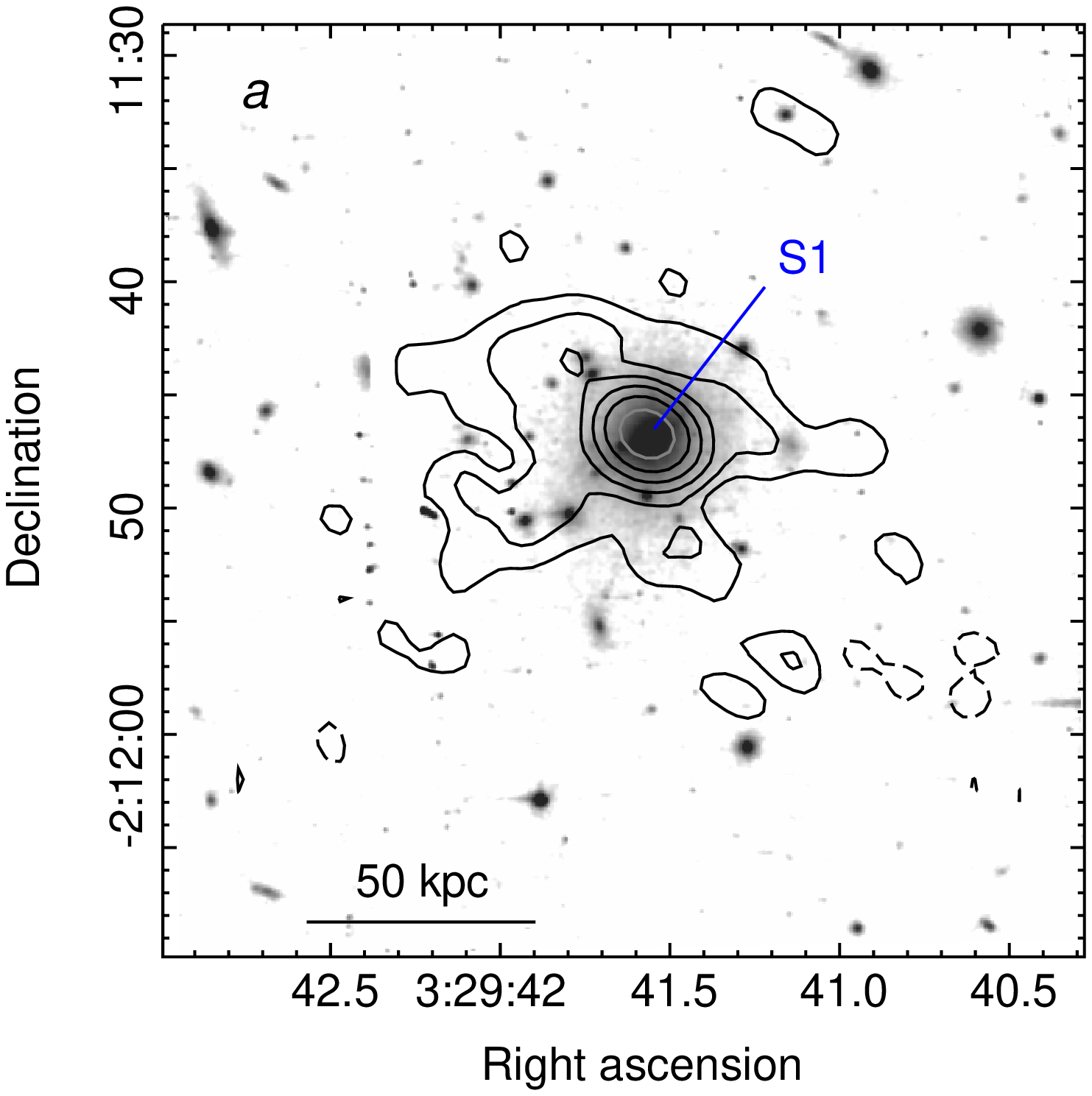}
\includegraphics[width=8cm]{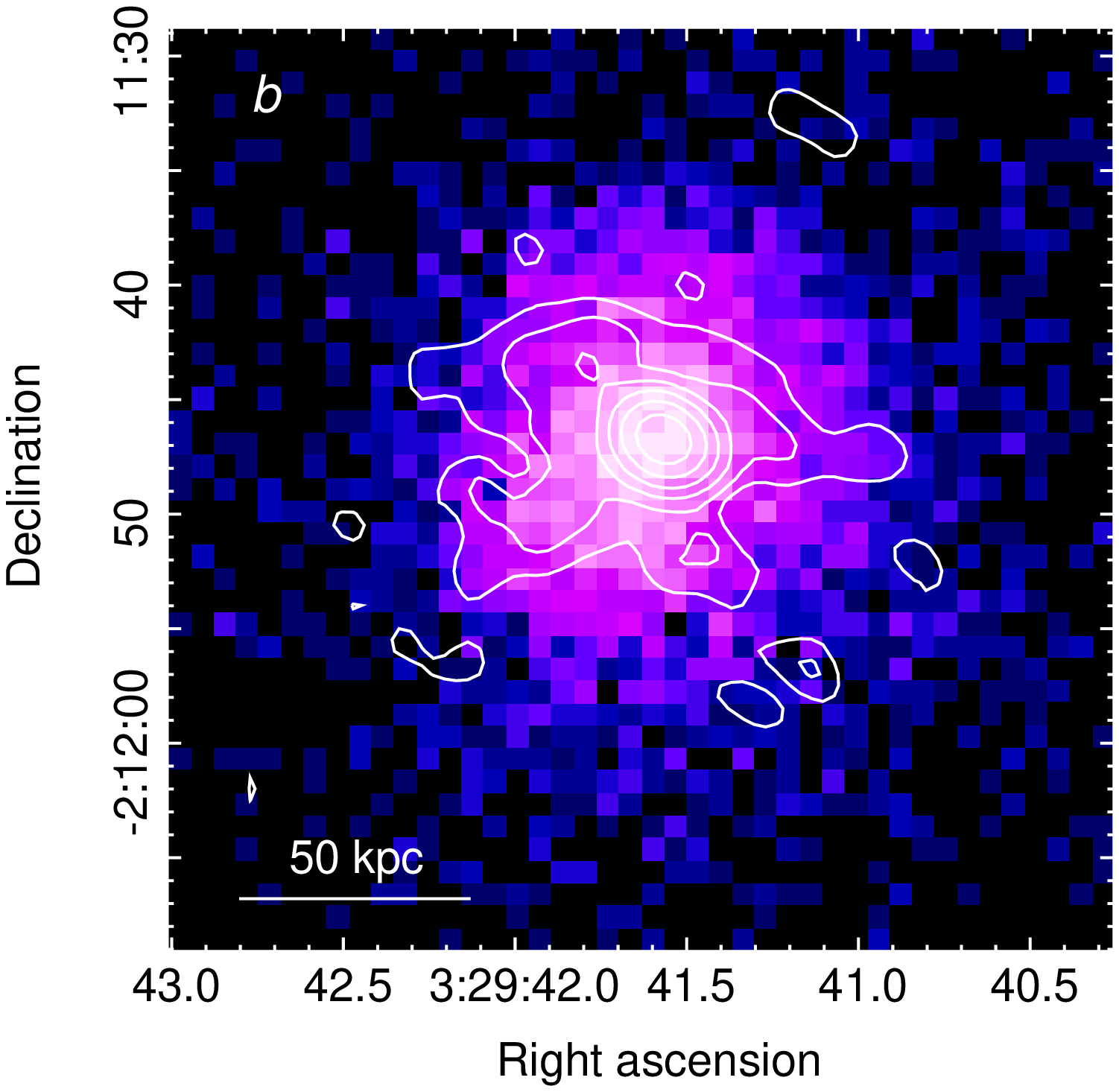}
\smallskip
\caption{MACS\,J0329.6--0211. ({\em a}) \gmrt\ contours 
  at 1.28 GHz of the central unresolved radio galaxy (S1) and surrounding
  minihalo, overlaid on a {\em HST} ACS image  (ID. 12452). The radio image has a restoring beam of
  $2^{\prime\prime}.7\times2^{\prime\prime}.1$, in p.a. $61^{\circ}$ and rms
  noise level of $1\sigma=23$ $\mu$Jy beam$^{-1}$. Contours are spaced by a
  factor of 2, starting from $+3\sigma$. Contours at $-3\sigma$ are shown as
  dashed.  ({\em b}) \gmrt\ 1.28 GHz contours, same as in ({\em a}),
  overlaid on the {\em Chandra} combined X-ray image in the 0.5-4 keV band.}
\label{fig:macs0329}
\end{figure*}
%
%
%

\subsection{MACS\,J0159.8--0849}

MACS\,J0159.8--0849 is a relaxed, cool-core cluster at redshift $z =0.405$,
which is host to a candidate minihalo (G14a). Using the {\em GMRT} at 1.39
GHz, we imaged the cluster center at $2^{\prime\prime}$ resolution. Our
image is shown in Figure ~\ref{fig:0159}.  As a reference, the first contour
of the {\em VLA} 1.4 GHz image of the minihalo from G14a is reported in
magenta.

The radio source associated with the BCG (S1) is unresolved in our new image
at 1.39 GHz.  Its flux density is $S_{\rm 1.39 \, GHz}= 34.0\pm1.7$ mJy
(Tab.~3), which is consistent with $35.0\pm1.8$ mJy measured by the {\em
  VLA} at 1.4 GHz. In the surroundings of the BCG, we detect only hints of
the faint diffuse emission seen in the much higher sensitivity {\em VLA}
image ($1\sigma=15$ $\mu$Jy beam$^{-1}$). To enhance the extended emission,
we made images at slightly lower resolution (down to $\sim
5^{\prime\prime}$, same as the {\em VLA} image), using values of ROBUST $>1$
and tapering the $uv$ data to weight down the long baselines. However, due
to the higher noise in these images ($1\sigma\sim 40$ $\mu$Jy beam$^{-1}$),
we were not able to map the diffuse emission around S1. On the other hand,
integration over the area enclosed by the {\em VLA} isocontour in Figure
~\ref{fig:0159} reveals an excess of $\sim 2$ mJy after subtraction of the
flux of S1. Such flux excess is only $\sim 20\%$ less than the minihalo flux
density measured on the {\em VLA} image ($2.4\pm0.2$ mJy), thus confirming
the presence of an extended component, that is not well imaged by our
shallow {\em GMRT} observation.

Even though higher-sensitivity images are needed to adequately map the
minihalo, the absence in Figure ~\ref{fig:0159} of bright jets and/or lobes
on the scale of the minihalo suggests that the central radio galaxy and
outer extended emission are not morphologically connected. As a further
note, the BCG is still unresolved at the sub--arcsecond resolution of a {\em
  VLA}--A configuration observation at 8.5 GHz (G14a). This indicates that
any possible extended lobes/jets associated with S1 must be smaller than
$\sim 2$ kpc.

\subsection{MACS\,J0329.6--0211}

MACS\,J0329.6--0211 is a distant ($z=0.45$) relaxed cluster with a bright,
low-entropy cool core (Cavagnolo et al.\ 2009). Figure ~\ref{fig:macs0329}
presents our new {\em GMRT} 1.28 GHz image at a resolution of
$2^{\prime\prime}.7\times2^{\prime\prime}.1$, overlaid on the optical {\em
  Hubble Space Telescope} ({\em HST}) and X-ray {\em Chandra}\/ images.  The
diffuse minihalo, first reported by G14a, is clearly detected in the cluster
core region.  It has a size of $\sim50$ kpc in radius, which is comparable
to its extent in the G14a {\em VLA} 1.4 GHz image at $5^{\prime\prime}$
resolution. No large-scale jets or plumes connect the diffuse emission to
central radio galaxy (S1), which is unresolved, thus limiting its size to be
less than 15 kpc.  Observations at sub--arcsecond resolution are needed to
reveal the details of its radio structure at smaller scales. We measure a
flux density of $S_{\rm 1.28 \, GHz}=3.9\pm0.2$ mJy for the BCG and $S_{\rm
  1.28 \, GHz}=4.5\pm0.4$ mJy for the minihalo (Tab.~3).

%
%
%
\begin{figure*}
\centering
\includegraphics[width=12cm]{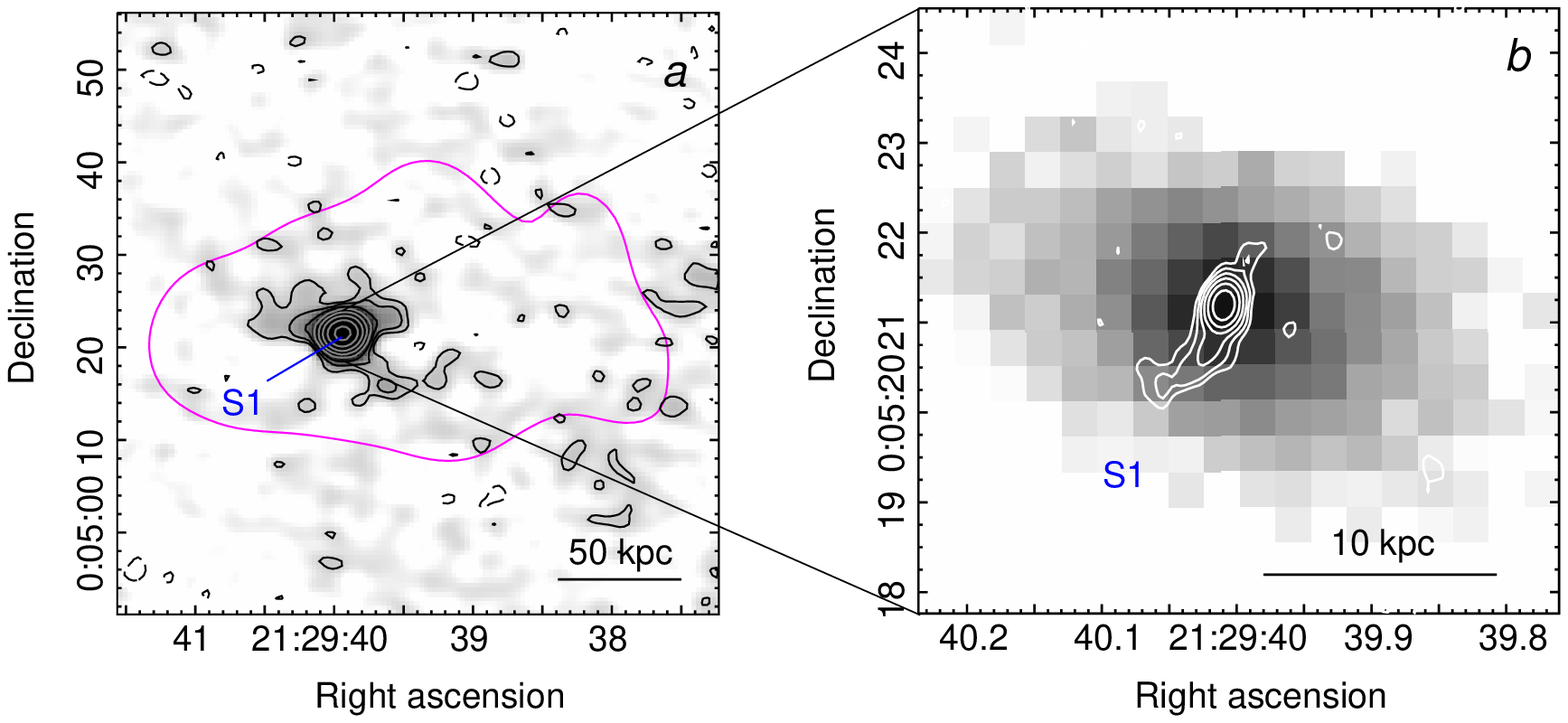}
\hspace{-1cm}
\includegraphics[width=6cm]{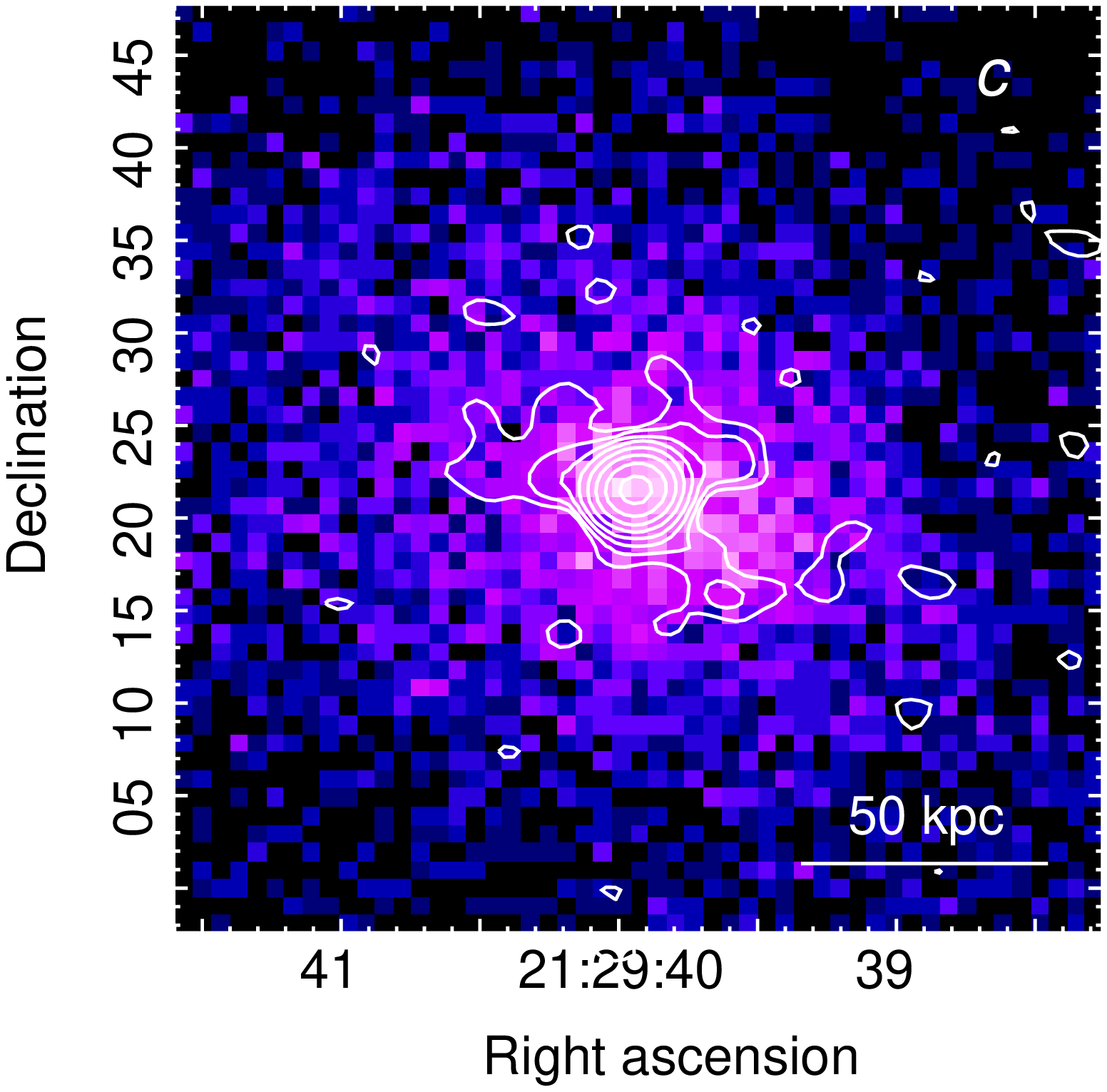}
\smallskip
\caption{RXC\,J2129.6+0005. ({\em a}) \gmrt\ image at 1.30 GHz 
  (greyscale and black and grey contours) of the central radio galaxy (S1)
  and innermost part of the minihalo. The restoring beam is
  $2^{\prime\prime}.6\times2^{\prime\prime}.1$, in p.a. $82^{\circ}$ and rms
  noise level is $1\sigma=45$ $\mu$Jy beam$^{-1}$. Contours are spaced by a
  factor of 2 from 0.12 mJy beam$^{-1}$. Contours at $-0.12$ mJy beam$^{-1}$
  are shown as dashed. The first contour from the {\em GMRT} 610 MHz image
  of the minihalo from K15 (their Figure 2) is reported in magenta. The
  contour level is $0.33$ mJy beam$^{-1}$, for a
  $11^{\prime\prime}.6\times10^{\prime\prime}.6$ beam. ({\em b}) {\em VLA}
  contours at 8.46 GHz of S1, overlaid on the r-band SDSS image. The radio image
  has a restoring beam of $0^{\prime\prime}.33\times0^{\prime\prime}.23$, in
  p.a. $-9^{\circ}$. The rms noise level is $1\sigma=22$ $\mu$Jy
  beam$^{-1}$. Contours are spaced by a factor of 2 from 0.06 mJy
  beam$^{-1}$.  ({\em c}) \gmrt\ 1.30 GHz contours, same as in ({\em
    a}), overlaid on the {\em Chandra} X-ray image in the 0.5-4 keV band.}
\label{fig:2129}
\end{figure*}

\subsection{RXC\,J2129.6+0005}\label{sec:2129}

RXC\,J2129.6+0005 is a relaxed cool-core cluster at $z=0.235$. A central
minihalo was reported by K15, based on a {\em GMRT} observation at 610 MHz.
In Figure \ref{fig:2129}{\em a}, we present our new higher-resolution {\em
  GMRT} image at 1.30 GHz.  The image detects only the innermost $r\sim 35$
kpc portion of the diffuse source seen at lower frequency (magenta contour).
The extended emission encompasses a compact source associated with the BCG
(S1). This latter is still point-like in a 4.86 GHz image at
$1^{\prime\prime}$ resolution, obtained from reanalysis of archival {\em
  VLA} data (Table~2; image not shown here). The source structure is finally
unveiled at 8.46 GHz (Figure \ref{fig:2129}{\em b}), which resolves the
emission into a bright, compact component (possibly the core), coincident
with the optical peak in the SDSS\footnote{Sloan Digital Sky Survery.}
image, and a fainter $\sim 4$ kpc jet pointing south-east. Hints of a much
fainter counter jet are visible on the opposite side of the putative core. A
comparison between the radio and X-ray {\em Chandra} images in Figure
~\ref{fig:2129}{\em c} shows that the diffuse emission detected at 1.30 GHz
is centrally located and permeates the brightest part of the cluster core,
as seen in other minihalo clusters.

The flux densities of S1 are summarized in Table~3, along with the
measurements at lower frequencies by K15. Its spectral index is
$\alpha=0.77\pm0.02$ between 235 MHz and 8.46 GHz. For the minihalo, we
measure a flux density of $S_{\rm 1.30 \, GHz}=2.5\pm0.2$ mJy in the region
enclosed by the magenta contour in Figure \ref{fig:2129}{\em a}. A flux of
$S_{\rm 1.4 \, GHz}=2.4\pm0.2$ mJy, consistent with our measurement, was
estimated by K15 from a comparison between the NVSS\footnote{NRAO VLA Sky Survey, Condon et al. (1998).}
and FIRST\footnote{Faint Images of the Radio Sky at Twenty-cm, Becker et al. (1995).} images. The
minihalo has an inferred spectral index of $\alpha=1.2\pm0.1$ between 235
MHz and 1.4 GHz.

%
%
\begin{figure*}
\centering
\includegraphics[width=15cm]{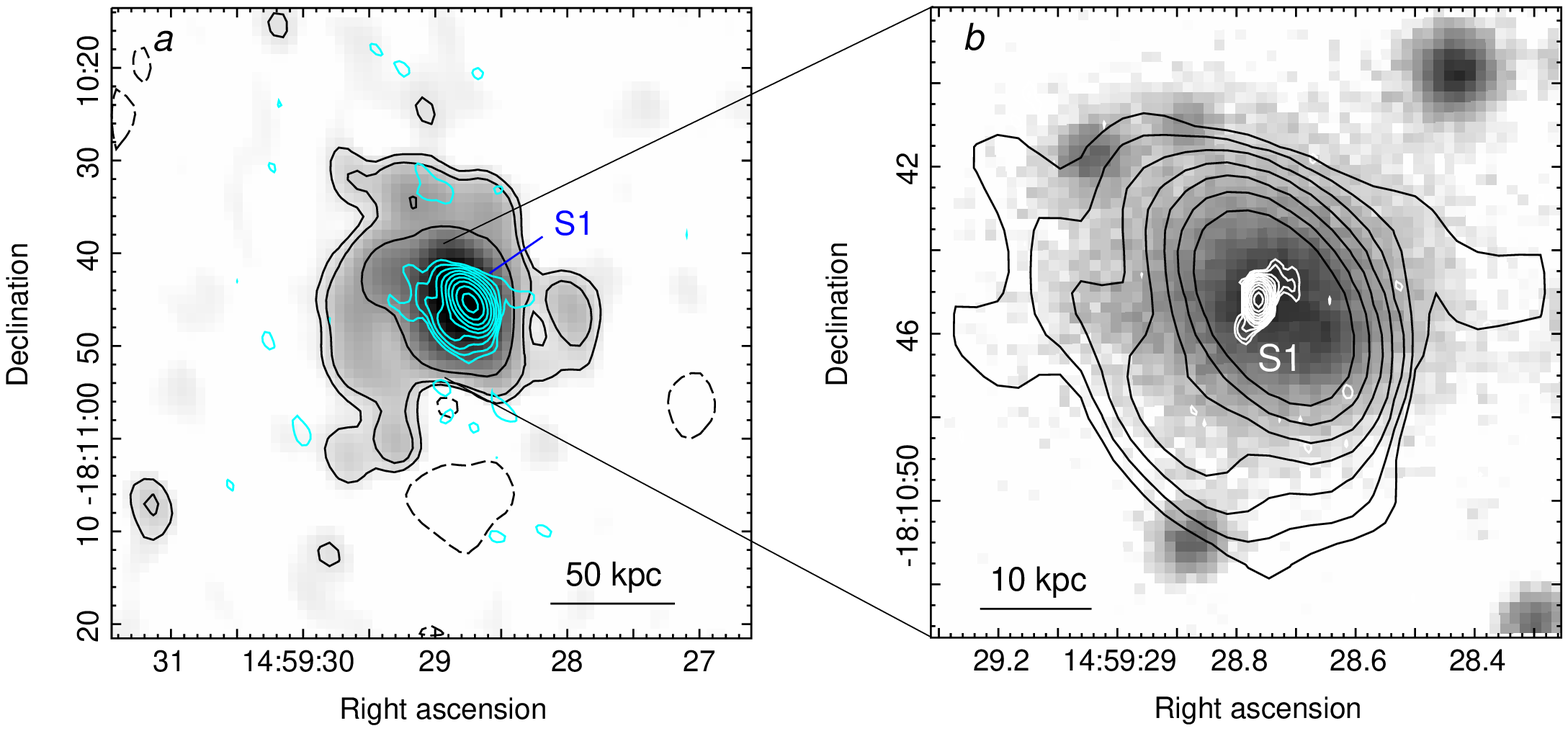}
\includegraphics[width=8cm]{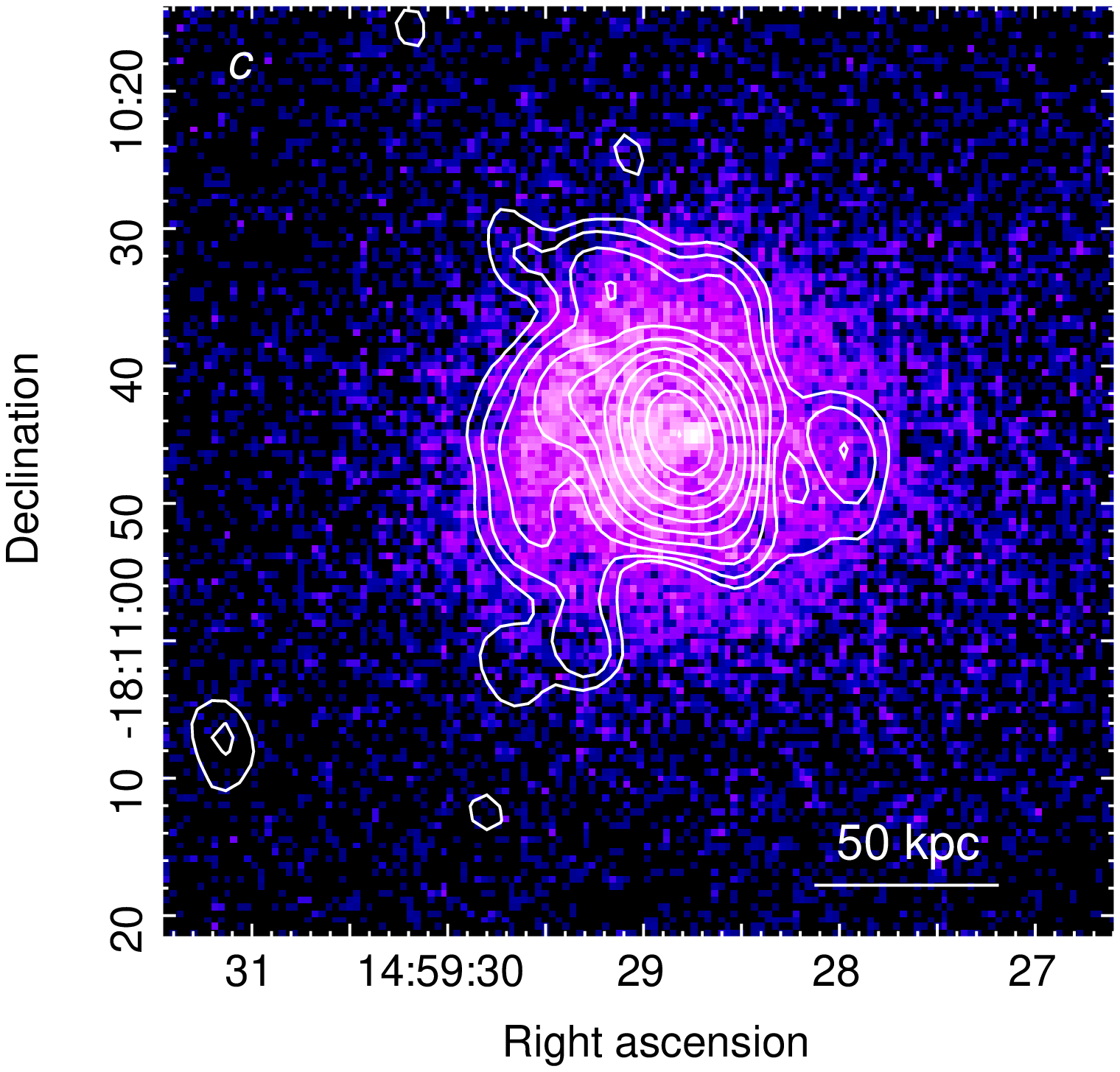}
\hspace{-1cm}\includegraphics[width=8cm]{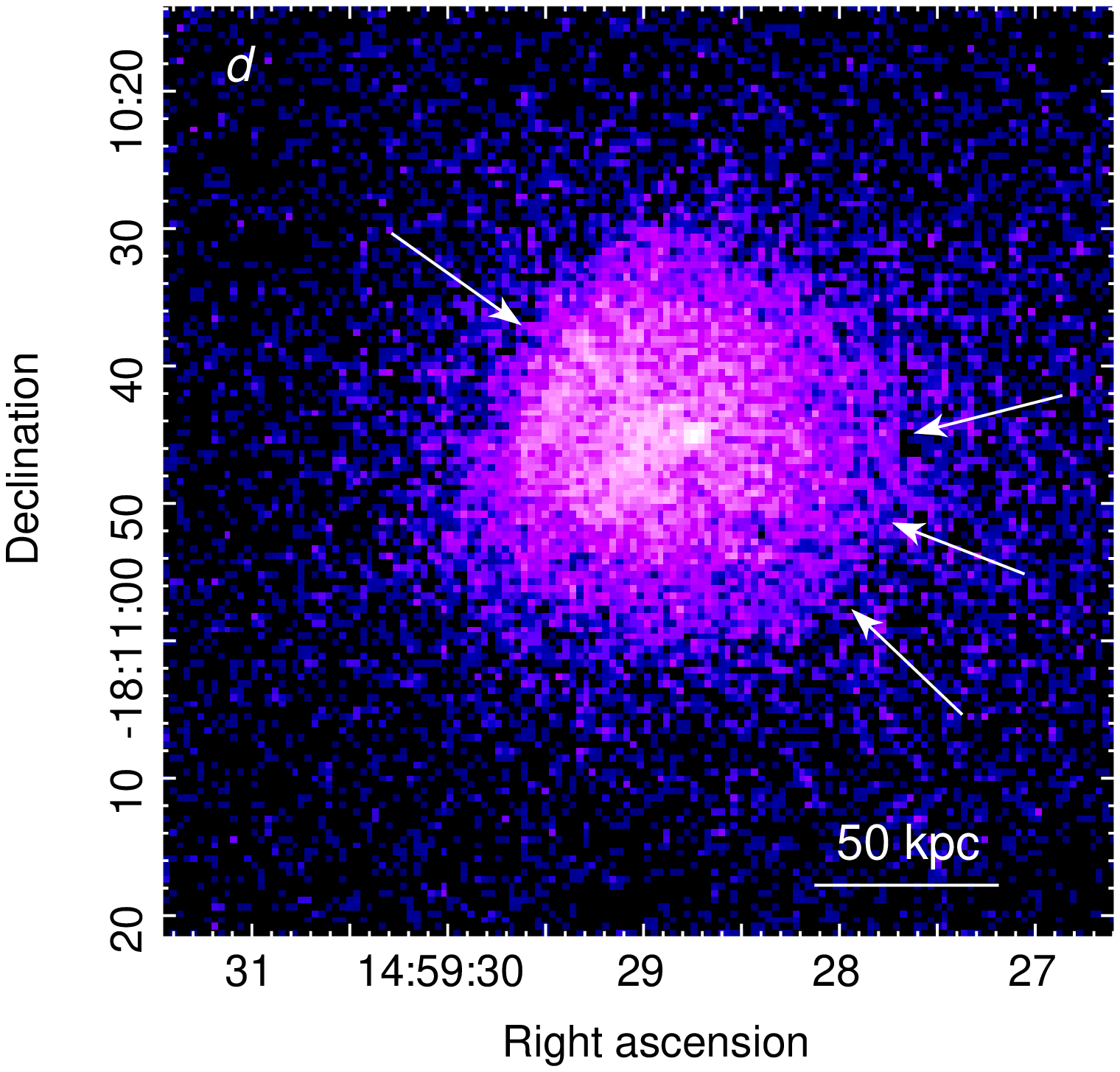}
\smallskip
\caption{{\footnotesize AS\,780. ({\em a}) {\em GMRT} 1.30 GHz contours (cyan), 
overlaid on the {\em GMRT} image of the minihalo at 610 MHz (grey scale 
and black contours).  The radio images have been  restored with a 
$3^{\prime\prime}\times2^{\prime\prime}$ beam at 1.30 GHz and 
$6^{\prime\prime}\times4^{\prime\prime}$ at 610 MHz.  Cyan contours 
start at $+3\sigma$=0.25 mJy beam$^{-1}$ and then scale by a factor of 2.
Black contours are spaced by a factor of 2 between $3\sigma=0.2$ mJy
beam$^{-1}$ and 1.6 mJy beam$^{-1}$. Contours at the $-3\sigma$ level are
shown as dashed. ({\em b}) {\em VLA}--A configuration contours at 8.46 GHz (white) 
of the cluster central radio galaxy S1 on the optical Pan-STARRS-1 $r$ image. 
The angular resolution is $0^{\prime\prime}.4\times0^{\prime\prime}.3$. White contours 
start at $+3\sigma$=0.09 mJy beam$^{-1}$ and then scale by a factor of 2. 
Black contours show the radio emission detected by the {\em GMRT} at 1.30 GHz
({\em a}) and are spaced by a factor of 2 between $+3\sigma=0.25$ mJy
  beam$^{-1}$ and 16 mJy beam$^{-1}$. ({\em c}) {\em GMRT} 610 MHz contours from ({\em a}), 
superposed to the {\em Chandra} X-ray image in the 0.5-4
  keV band. Radio contours start at $+3\sigma$=0.2 mJy beam$^{-1}$ and then
  scale by a factor of 2. ({\em d}) {\em Chandra} X-ray image, same as in
  ({\em c}). Arrows mark the position of two possible cold fronts.}}
\label{fig:as780}
\end{figure*}
%
%

\subsection{AS\,780}\label{sec:as780}

AS\,780 is a relaxed cluster at $z=0.236$ with a regular X-ray morphology on
large scales. The X-ray surface brightness profile peaks at the BCG, and
the gas temperature declines towards the center (e.g., G17), as typically
observed in other cool-core clusters. Hints of diffuse radio emission around
the central radio galaxy were reported at 610 MHz by Venturi et al.\ (2007).
We have obtained new images at 610 MHz by reanalyzing the Venturi et al.
data and compared them to higher-resolution images at 1.30 GHz and 8.46 GHz,
as well as to optical (from Pan-STARRS-1\footnote{Panoramic Survey Telescope
  and Rapid Response System, Chambers et al.\ (2016).})  and X-ray (from {\em
  Chandra}) images of the cluster central region (Figure \ref{fig:as780}).
The 8.46 GHz image (white contours in panel {\em b}) shows an extended
($\sim 8$ kpc) core-dominated and double-jet radio source, coincident with
the dominant optical galaxy and central X-ray point source in the {\em
  Chandra} image (panel {\em d}). The source is embedded in a region of
diffuse emission at 1.30 GHz (black contours), which corresponds to the
innermost part of the minihalo detected at 610 MHz (panel {\em a}). The size
of the minihalo is $\sim 50$ kpc in radius at 610 MHz. As indicated by a
comparison between the radio and X-ray emission (panels {\em c} and {\em
  d}), the minihalo is cospatial with the bright X-ray cool core and is
edged by a pair of possible cold fronts, symmetric with respect to the
cluster center.

The flux densities of the central radio galaxy S1 at all frequencies are
summarized in Table 3. Its spectrum is inverted with $\alpha=-0.17\pm0.04$ between
610 MHz and 8.46 GHz. No high-resolution images at 1.4 GHz (e.g., from FIRST) 
are available for this source, therefore we estimate its radio luminosity 
at 1.4 GHz by extrapolating the 1.3 GHz flux density with $\alpha=-0.17\pm0.04$. We 
obtain $S_{\rm \, BCG, \,1.4 \, GHz} = 102.3\pm0.3$ and $P_{\rm \,BCG, \,1.4 \, GHz} 
= (17\pm1)\times10^{24}$ W Hz$^{-1}$ (Table 4).

The minihalo has a flux of $S_{\rm \,MH, \,610 \, MHz}=34\pm2$ mJy. At 1.30 GHz, where only 
part of the minihalo is detected, we measure $S_{\rm \,MH, \,1.30 \, GHz}=11\pm1$ mJy, which 
has to be considered a lower limit to the actual flux of the minihalo at this frequency. 
This translates into a spectral index $\alpha < 1.45$ between 610 MHz and 1.30 GHz 
and an estimated flux density $S_{\rm \, MH, \,1.4 \, GHz} >10$ mJy, corresponding to a 
limit of $P_{\rm \,MH, \,1.4 \, GHz}>1.7\times10^{24}$ W Hz$^{-1}$ in radio power
(Table 4).

%
%
\begin{figure*}
\centering
\includegraphics[width=7.5cm]{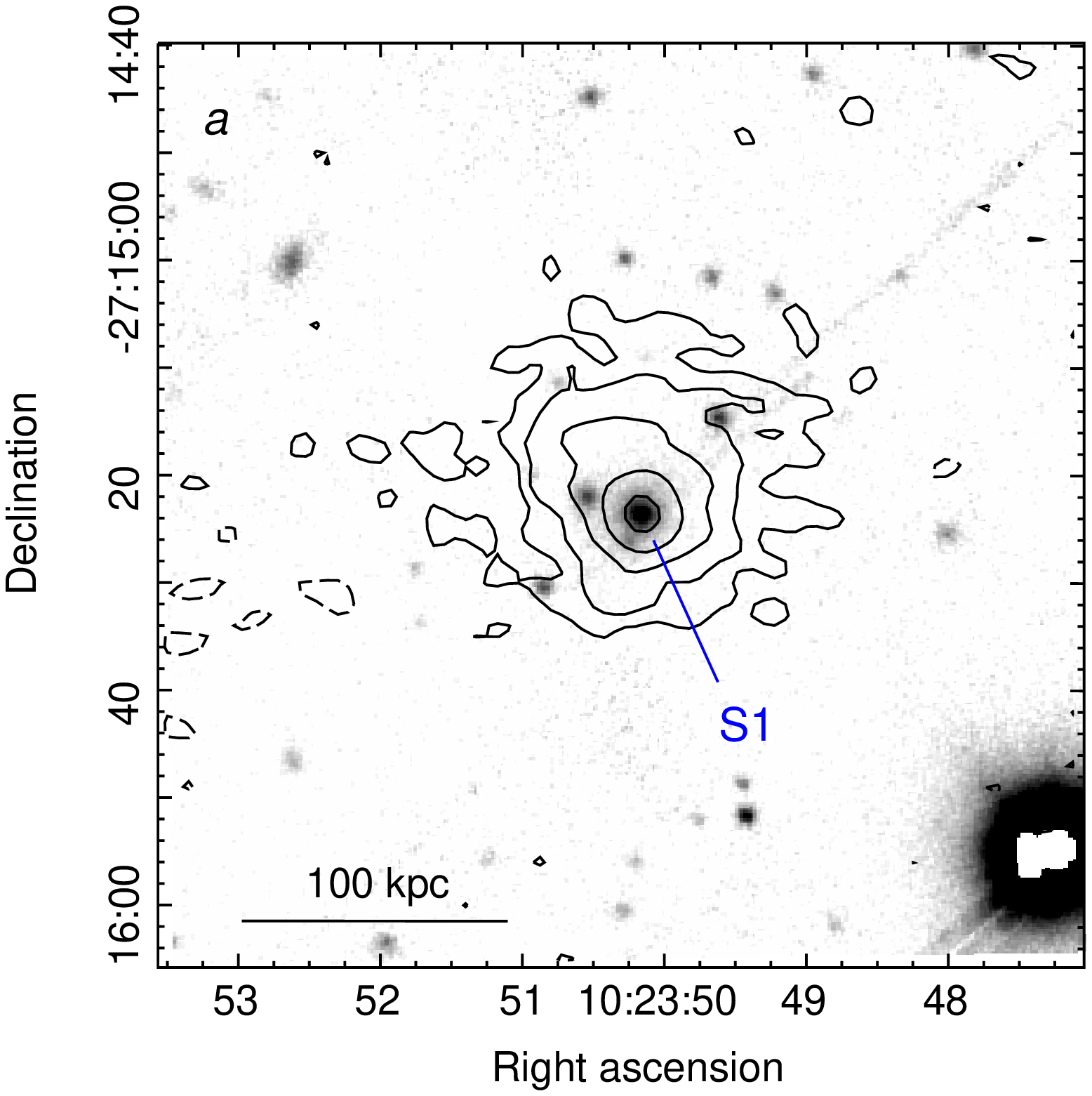}
\includegraphics[width=7.5cm]{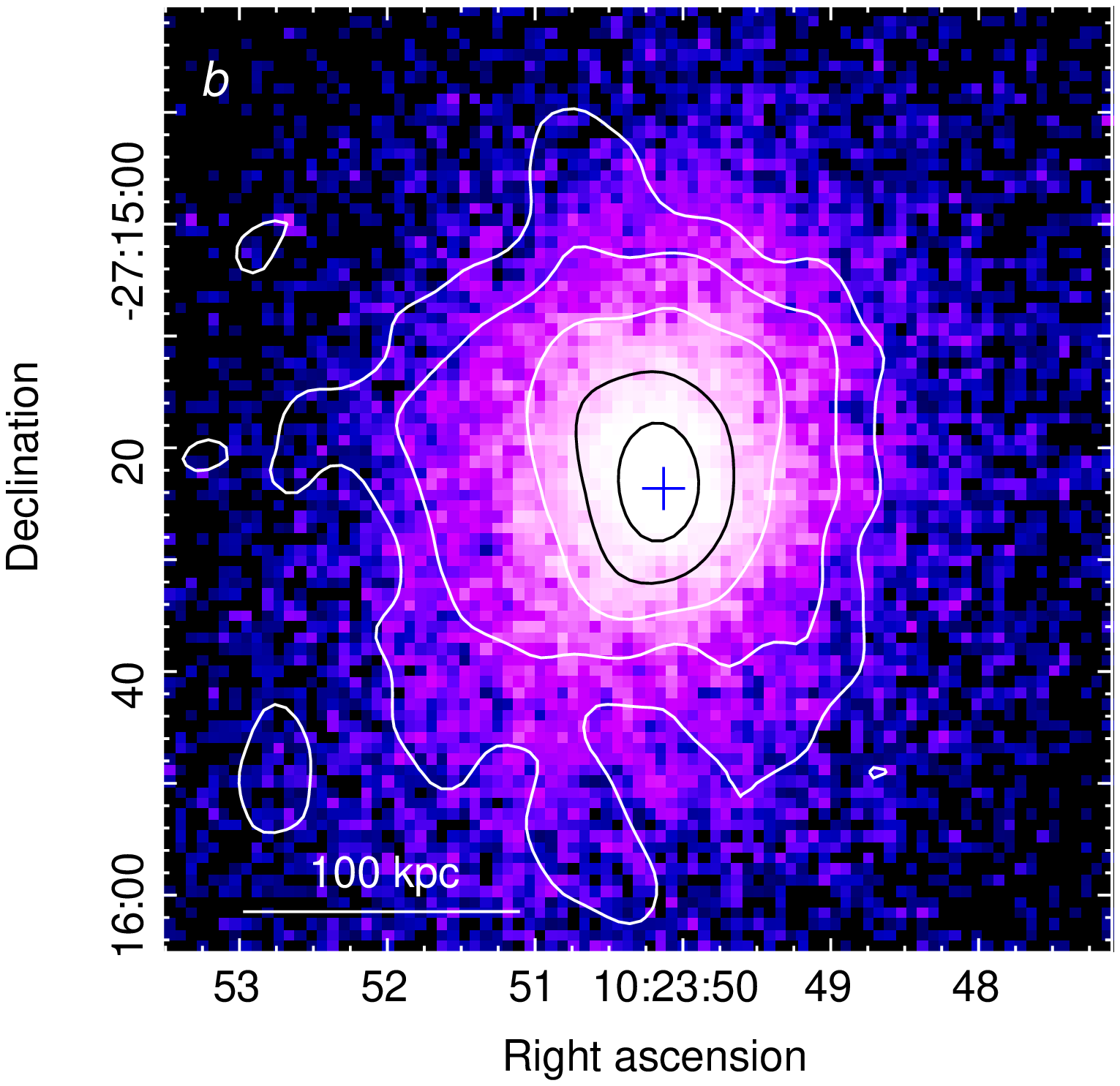}
\vspace{-0.5cm}
\smallskip
\caption{A\,3444. ({\em a}) {\em VLA} 1.4 GHz BnA--configuration 
  image, restored with a $5^{\prime\prime}$ circular beam and overlaid on
  the optical Pan-STARRS-1 $r$ image. The rms noise is $1\sigma=35$ $\mu$Jy
  beam$^{-1}$. Contours start at 0.09 mJy beam$^{-1}$ and then scale by a
  factor of 2. Contours at $-0.09$ mJy beam$^{-1}$ are shown as dashed. S1
  is the central radio galaxy. ({\em b}) {\em VLA} 1.4 GHz combined
  DnC+BnA--configuration contours of the minihalo and central radio galaxy
  S1 (black cross), overlaid on the {\em Chandra} image in the 0.5-4 keV
  band. The radio image has been restored with a $11^{\prime\prime}$
  circular beam. The rms noise is $1\sigma=45$ $\mu$Jy beam$^{-1}$.
  Contours start at 0.13 mJy beam$^{-1}$ and then scale by a factor of 2. No
  contours at a level of $-0.13$ mJy beam$^{-1}$ are present in the part of
  the image shown.}
\label{fig:a3444}
\end{figure*}
%
%

\subsection{A\,3444}\label{sec:a3444}

A\,3444 is a relaxed cluster at $z=0.254$ with a central, low-entropy cool
core (G17). A possible central minihalo was reported in this cluster by
Venturi et al.\ (2007), based on {\em GMRT} images at 610 MHz. In Figure
~\ref{fig:a3444}, we present our {\em VLA} images at 1.4 GHz, compared to
the cluster optical and X-ray emission. The BnA image ({\em a}) shows a
central compact radio source (S1), fully enveloped by a roundish minihalo
containing no jets. The unresolved source is associated with the BCG and its
size is $< 20$ kpc. In a higher-resolution image from recent {\em GMRT}
observations at 1.28 GHz (Giacintucci et al. in preparation), the source is still
pointlike, thus implying that its size must be even smaller ($\sim 8$ kpc or
less). The minihalo size reaches a radius of $\sim 120$ kpc in the combined
DnC+BnA image in Figure ~\ref{fig:a3444}{\it b}.
The flux density of S1 is $S_{\rm 1.4 \, GHz}=1.41\pm0.07$ mJy and its
radio power $P_{\rm \,BCG,  \,1.4 \, GHz}=(0.28\pm0.01)\times10^{24}$ W Hz$^{-1}$. 
The surrounding minihalo has $S_{\rm 1.4 \, GHz}=12.1\pm0.9$ mJy and
$P_{\rm \,MH, \,1.4 \, GHz}=(2.38\pm0.17)\times10^{24}$ W Hz$^{-1}$.
A detailed
spectral study of this minihalo, based on the {\em VLA} data presented here
and new, deep {\em GMRT} observations at 1.28 GHz and 610 MHz will be
presented in a future paper (Giacintucci et al. in preparation), 
where spatial correlation between
the radio and X-ray emission is also investigated.

\section{New minihalo detections}

Our analysis of {\em GMRT} data of the clusters A\,907, A\,2667 and
PSZ1\,G139.61+24.20 (Table 2) led to the detection of large-scale diffuse
emission in their cool cores, which we classify as minihalos. In the
following sections, we describe these newly discovered minihalos and
summarize their properties.

\subsection{A\,907}\label{sec:a907}

\begin{figure*}
\centering
\includegraphics[width=6.5cm]{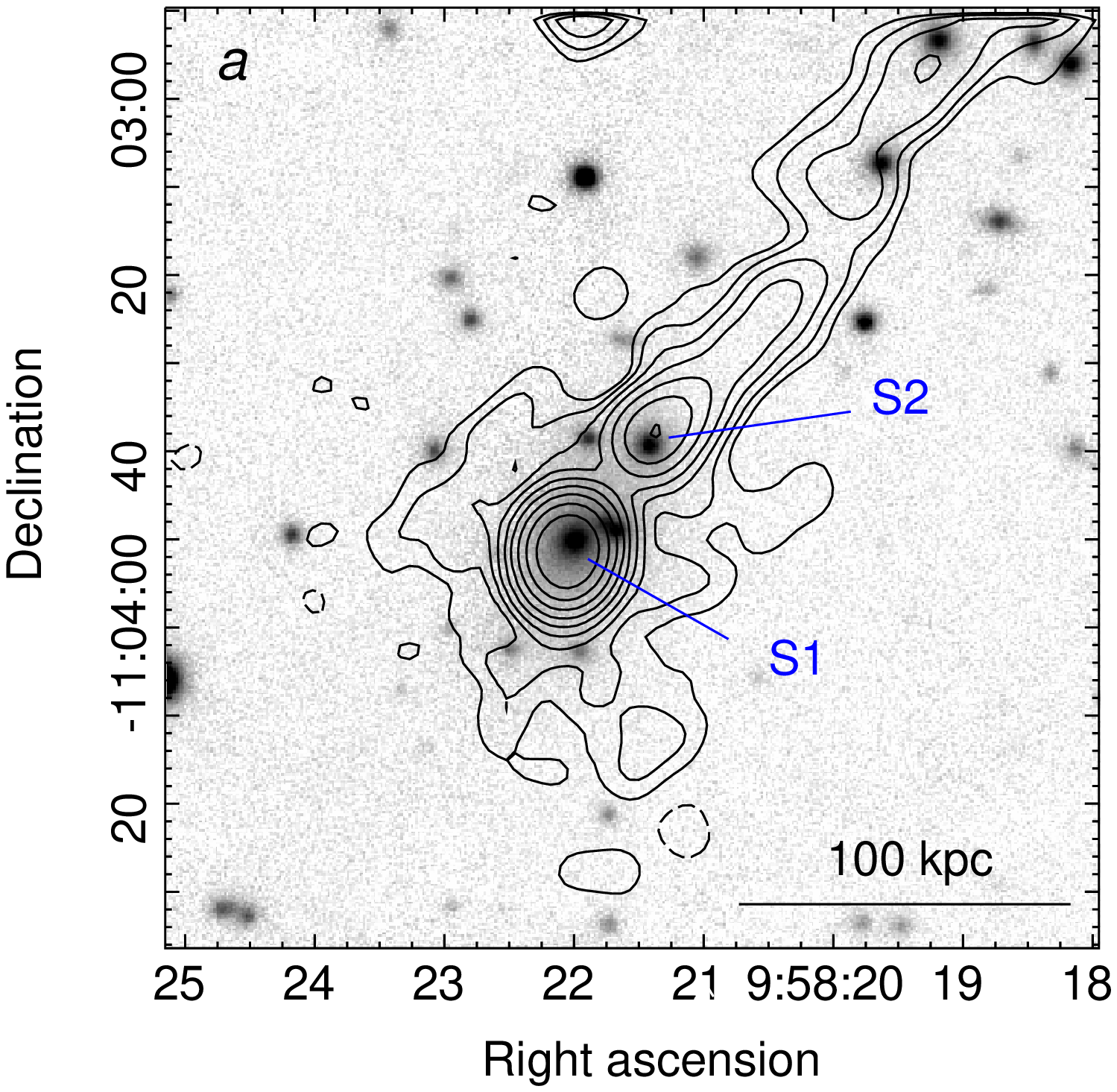}
\hspace{-1cm}\includegraphics[width=6.5cm]{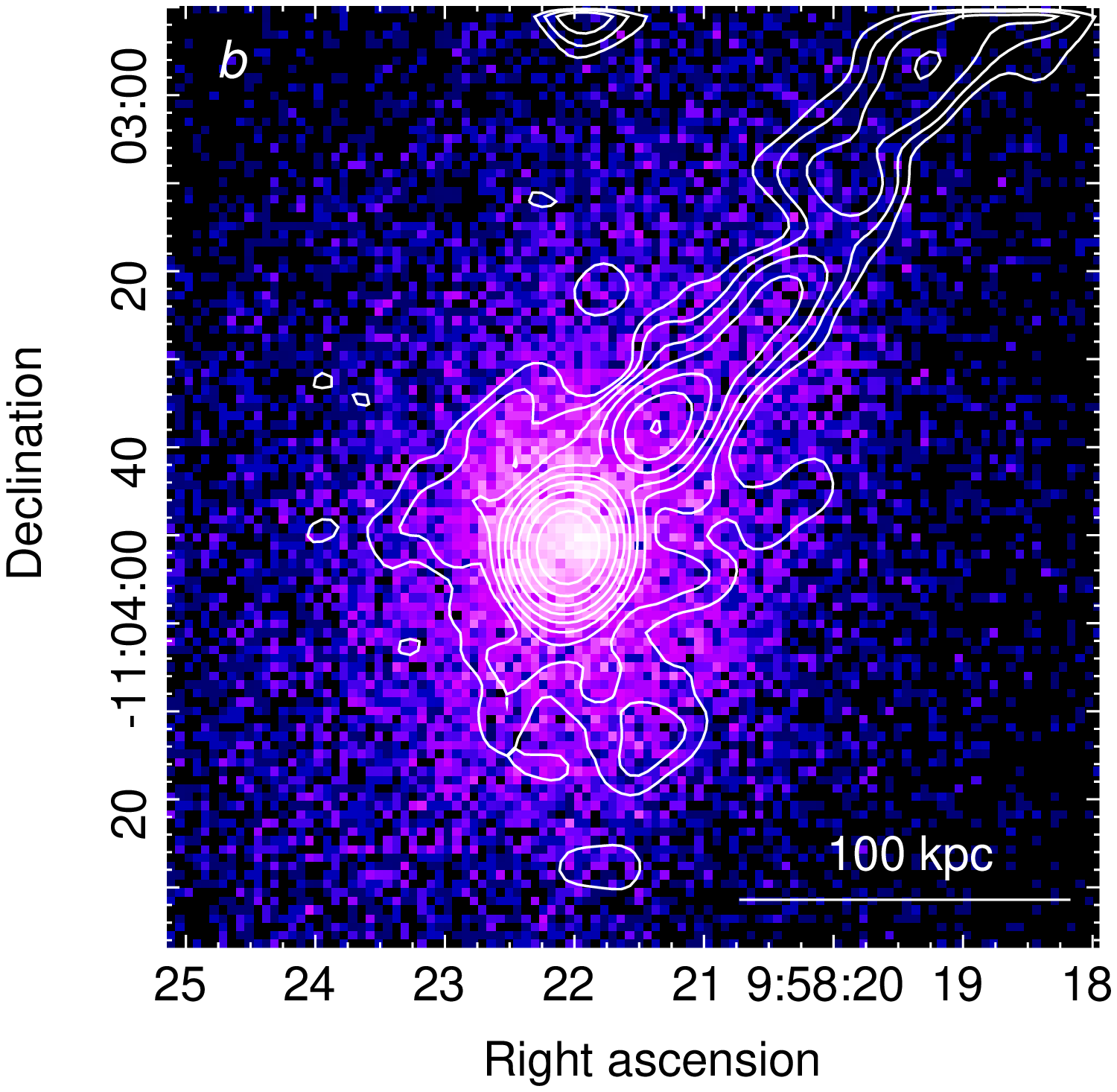}
\hspace{-1cm}\includegraphics[width=6.5cm]{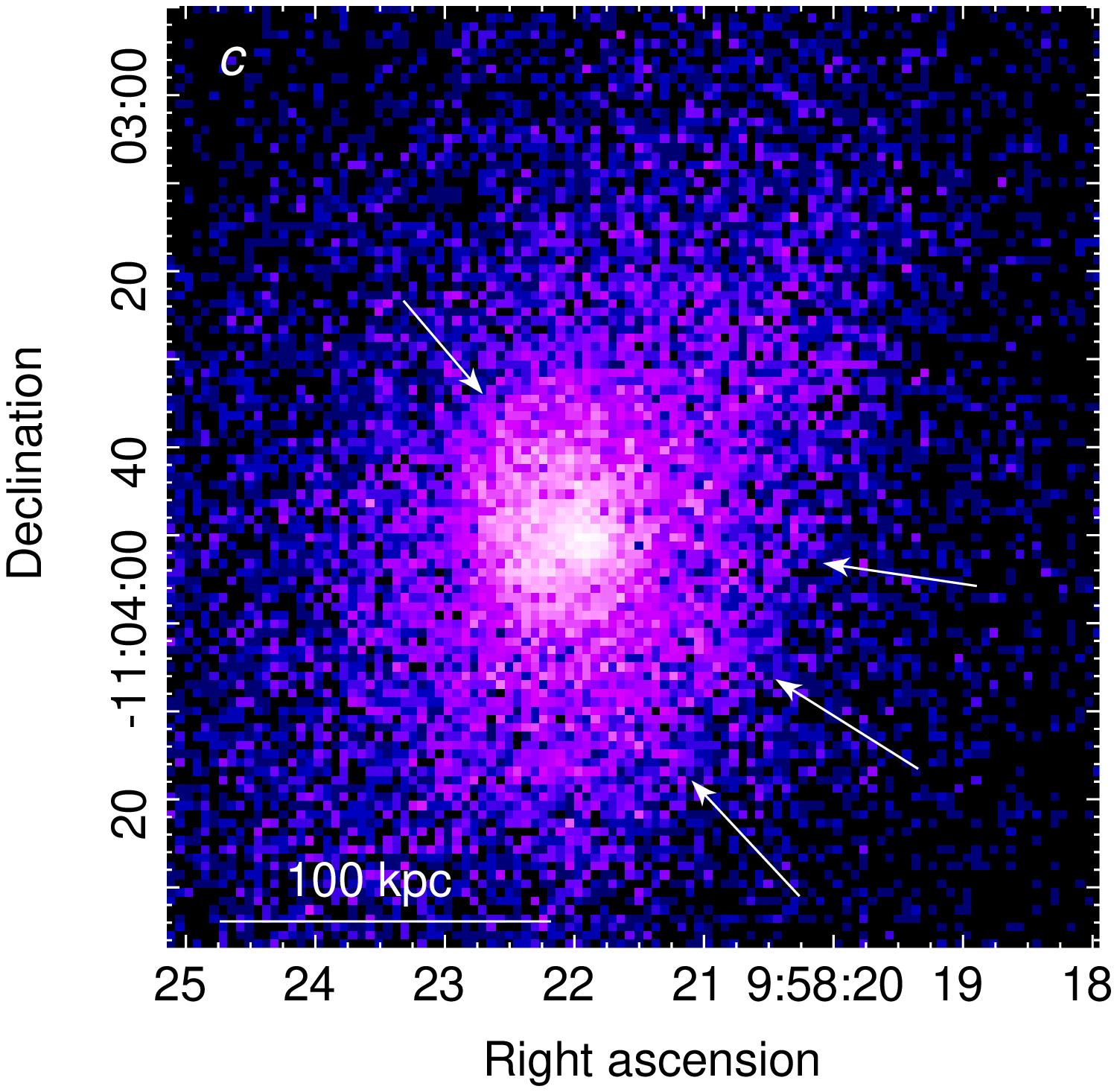}
\smallskip
\caption{A\,907. ({\em a}) {\em GMRT} 610 MHz contours, overlaid 
  on the optical Pan--STARRS-1 $r$--band image (gray scale). The radio image
  has been restored with a $5^{\prime\prime}$ circular beam.  The rms noise
  is $1\sigma=40 \, \mu$Jy beam$^{-1}$ and contours are spaced by a factor
  of 2 starting from 0.12 mJy beam$^{-1}$.  Contours at $-0.12$ mJy
  beam$^{-1}$ are shown as dashed. S1 is associated with the cluster central
  galaxy and S2 is an extended, head-tail radio galaxy.  ({\em b}) {\em GMRT}
  contours at 610 MHz, same as in ({\em a}), overlaid on the {\em Chandra}
  X-ray combined image in the 0.5-2 keV band. ({\em c}) {\em Chandra} X-ray
  combined image, same as in ({\em b}). Arrows mark the position of two
  possible X-ray cold fronts.}
\label{fig:a907}
\end{figure*}

We have reanalyzed an archival {\em GMRT} observation at 610 MHz (Table 2)
of this relaxed, cool-core cluster at $z=0.153$. In Figure
~\ref{fig:a907}{\em a}, we present a 610 MHz image, overlaid as contours an
optical $r$-band image from Pan-STARRS1.  The figure shows a radio point
source (S1) at the center of a region of diffuse emission, about 70 kpc in
radius, that we classify as a minihalo. The source S1 is identified with the
cluster dominant galaxy. An extended head-tail source (S2) is also visible
in the image. No redshift information is available for this object, however
its radio morphology, typical of dense environments, suggests that it is a
cluster member.  Panel {\em b} shows a comparison of the radio and {\em
  Chandra} X-ray emission. The minihalo fills a large fraction of the
central cool core and appears to be bounded by a pair of apparent X-ray cold
fronts, as it is often observed in minihalo clusters with similar sloshing
features in their cores.

We measure a flux density of $S_{\rm 610 \, MHz}=117\pm9$ mJy for the
central galaxy S1 and $S_{\rm 610 \, MHz}=42\pm3$ mJy for S2. The minihalo
has $S_{\rm 610 \, MHz}=34.9\pm5.6$ mJy (Table 3).  No high-resolution
images at 1.4 GHz are available for this cluster. An estimate of the flux
and radio luminosity at 1.4 GHz can be obtained by extrapolating the 610 MHz
fluxes using a spectral index of $\alpha=1.0\div1.3$ for the minihalo (e.g.,
G14a) and $\alpha=0.6\div0.8$ for the central active radio galaxy (e.g.,
Condon 1992). We obtain $S_{\rm 1.4 \, GHz}=13.5\pm2.8$ mJy for the
minihalo, which corresponds to a radio power of $P_{\rm \,MH, \,1.4 \,
  GHz}=(0.9\pm0.2)\times10^{24}$ W Hz$^{-1}$.  For the BCG, we estimate
$S_{\rm 1.4 \, GHz}=65\pm7$ mJy and $P_{\rm \,BCG, \,1.4 \,
  GHz}=(4.1\pm0.4)\times10^{24}$ W Hz$^{-1}$ (Table 4).

%
%
%

%
%
%
\begin{figure*}
\centering
\includegraphics[width=6.5cm]{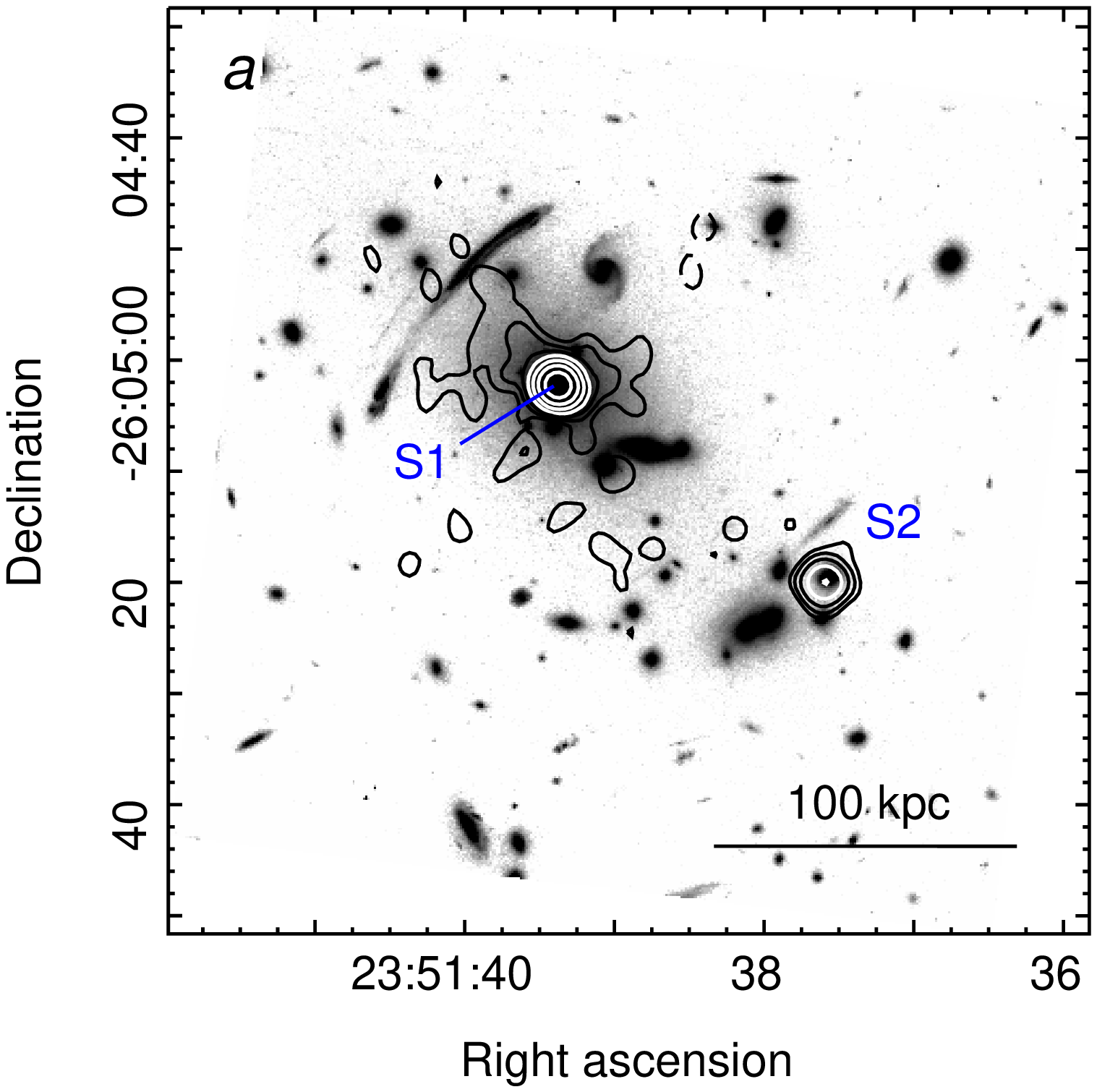}
\hspace{-1cm}\includegraphics[width=6.5cm]{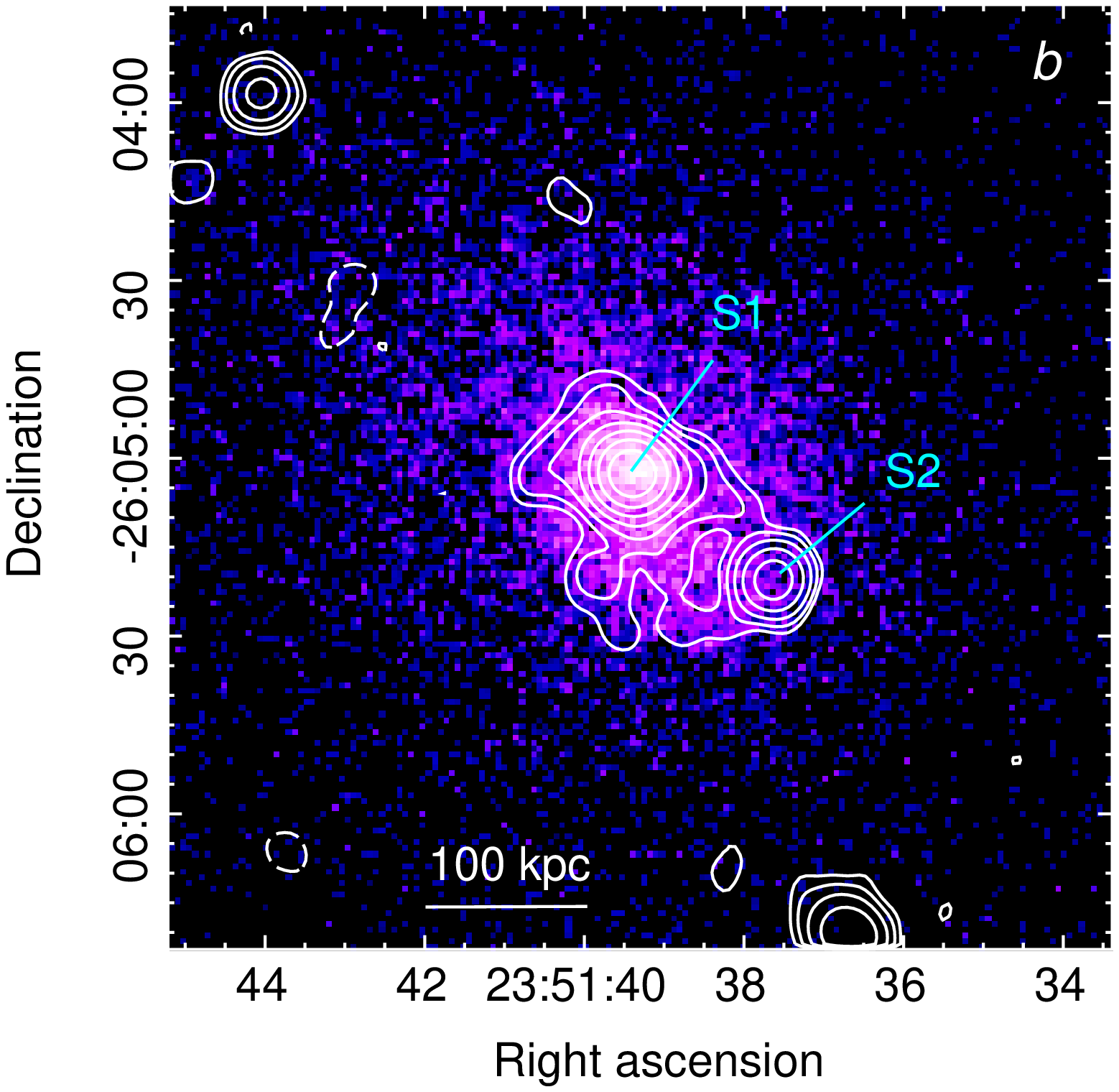}
\hspace{-1cm}\includegraphics[width=6.5cm]{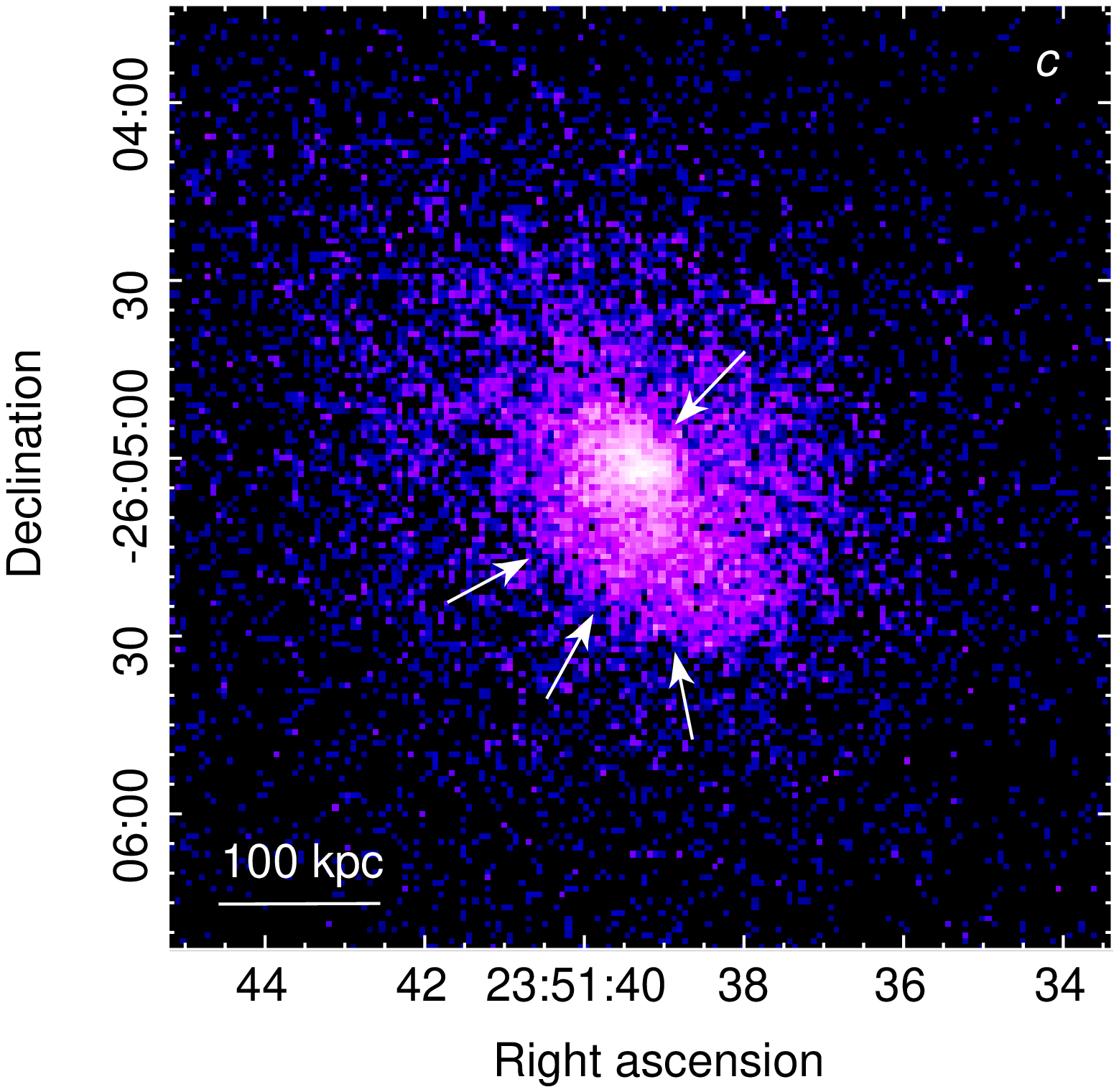}
\smallskip
\caption{A\,2667. ({\em a}) {\em GMRT} 1.15 GHz contours, overlaid on the
  optical {\em HST} ACS image (gray scale). The radio image has been
  restored with a circular $2^{\prime\prime}.5$ beam. The rms noise is
  $1\sigma=35 \, \mu$Jy beam$^{-1}$.  Contours are scaled by a factor of 2,
  starting from 0.12 mJy beam$^{-1}$.  Contours at $-0.12$ mJy beam$^{-1}$
  are shown as dashed.  Individual radio galaxies are labeled. ({\em b})
  {\em GMRT} contours at 610 MHz, overlaid on the {\em Chandra} X-ray image
  in the 0.5-4 keV band.  The radio image has been restored with a circular
  $7^{\prime\prime}$ beam.  The rms noise is $1\sigma=53 \, \mu$Jy
  beam$^{-1}$. Contours are scaled by a factor of 2, starting from 0.16 mJy
  beam$^{-1}$. Contours at $-0.16$ mJy beam$^{-1}$ are shown as dashed.
  Individual radio galaxies are labeled.  ({\em c}) {\em Chandra} X-ray
  image, same as in ({\em b}). White lines indicate two surface
  brightness edges, which may be cold fronts.}
\label{fig:a2667}
\end{figure*}

\subsection{A\,2667}\label{sec:a2667}

A\,2667 is a relaxed cluster at $z=0.23$ with a bright, cool core (Cavagnolo
et al.\ 2009).  We imaged the central region of the cluster using two
high-resolution observations at 1.15 GHz from the {\em GMRT} archive (Table
2). We also reanalyzed {\em GMRT} data at 610 MHz from Venturi et al.
(2008). Our images are presented in Figure ~\ref{fig:a2667}, compared with
optical and X-ray images. Panel {\em a} shows an overlay of the radio
contours from the combined 1.15 GHz data on the {\em HST} ACS image,
revealing a bright unresolved radio source (S1), coincident with the cluster
central galaxy, and a second point source (S2), associated with a member
galaxy ($z=0.234$; Covone et al.\ 2006). 
ATCA\footnote{Australia Telescope Compact Array.} observations at 4.8 GHz
and 8.6 GHz find S1 to be still unresolved at $\sim 1^{\prime\prime}$
resolution (Hogan et al.\ 2015), thus limiting the spatial scale of any
possible jets and/or lobes associated with the galaxy to be $\sim 4$ kpc or
less.  Diffuse radio emission, that we classify as a minihalo, is detected
around S1.  Panel {\em b} presents the radio contours at 610 MHz, overlaid
on the {\em Chandra} image, which is also shown in panel {\em c} with arrows
marking the position of two possible X-ray cold fronts. The minihalo covers
an area of about 70 kpc in radius and appears to be contained within the
sloshing region defined by the X-ray fronts.

The flux densities of S1 and S2 at 1.15 GHz and 610 MHz are summarized in
Table 3.  Their spectral indices are $\alpha=0.5\pm0.1$ and
$\alpha=0.7\pm0.1$, respectively.  We measure the minihalo flux density by
subtracting the fluxes of S1 and S2 form the total emission. We obtain
$S_{\rm 1.15 \, GHz}=8.3\pm0.7$ mJy and $S_{\rm 610 \, MHz}=15.3\pm1.2$ mJy.
A spectral index $\alpha= 1.0\pm0.2$ is inferred for the minihalo. Using
this spectral index, we estimate a minihalo flux density at 1.4 GHz of
$S_{\rm 1.4 \, GHz}=7.1\pm0.6$ mJy and a luminosity of $P_{\rm \,MH, \,1.4
  \, GHz}=(1.1\pm0.1)\times10^{24}$ W Hz$^{-1}$ (Table 4).  For the central
galaxy, we estimate $S_{\rm 1.4 \, GHz}=13.4\pm0.7$ mJy and $P_{\rm \,BCG,
  \,1.4 \, GHz}=(2.1\pm0.1)\times10^{24}$ W Hz$^{-1}$ using
$\alpha=0.5\pm0.1$.

%
%
\begin{figure*}
\centering
\includegraphics[width=6.5cm]{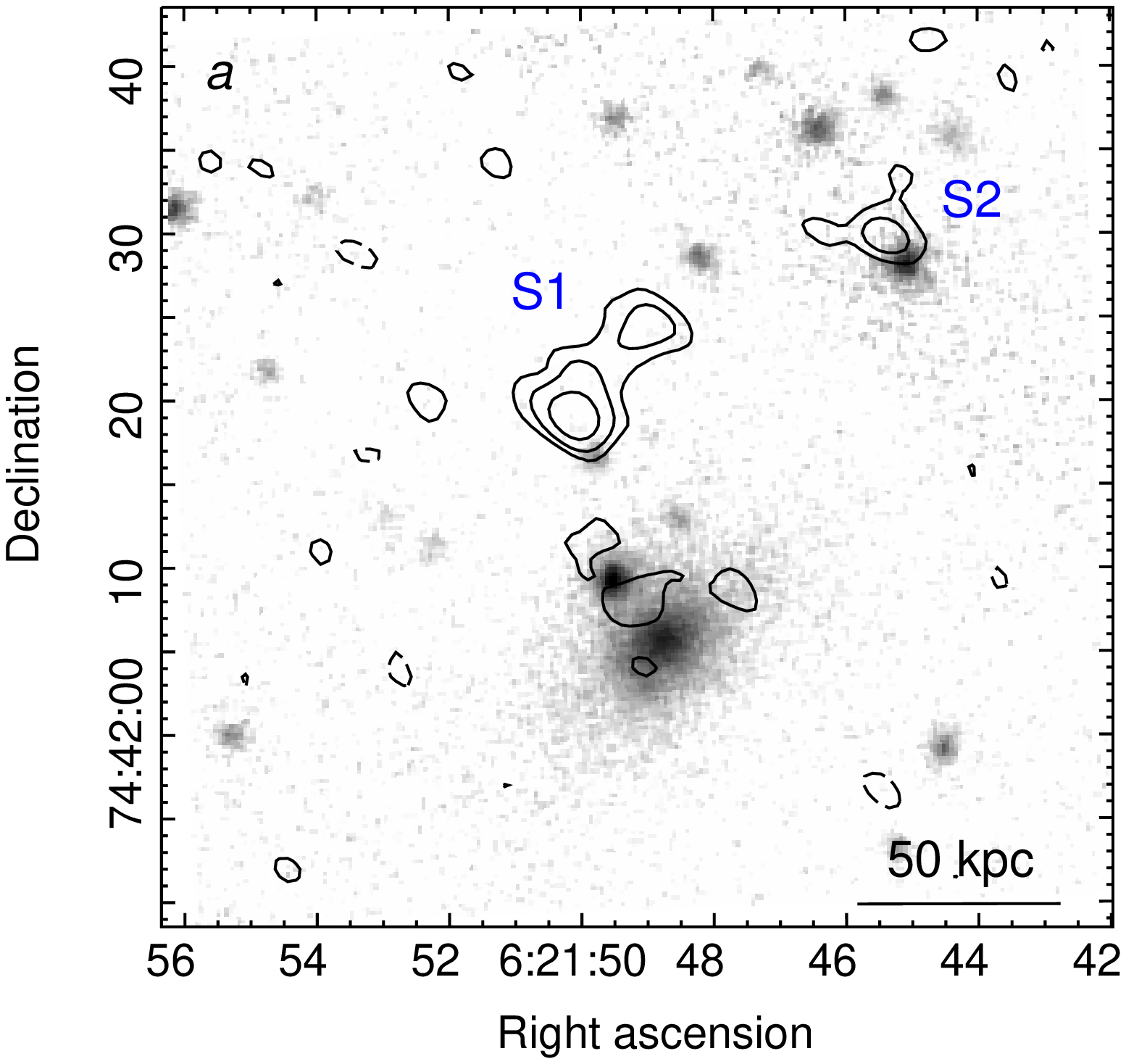}
\hspace{-1cm}\includegraphics[width=6.5cm]{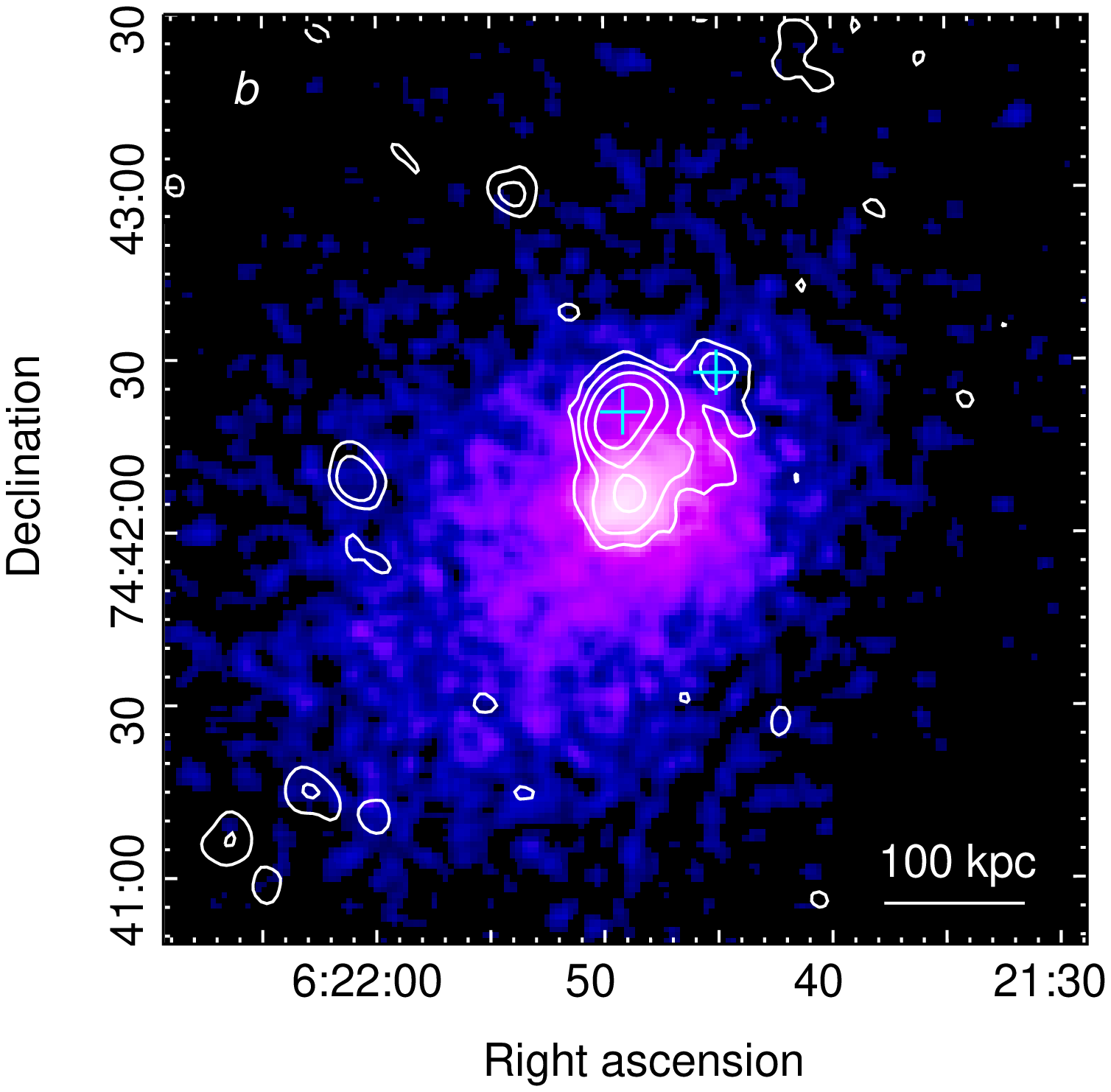}
\hspace{-1cm}\includegraphics[width=6.5cm]{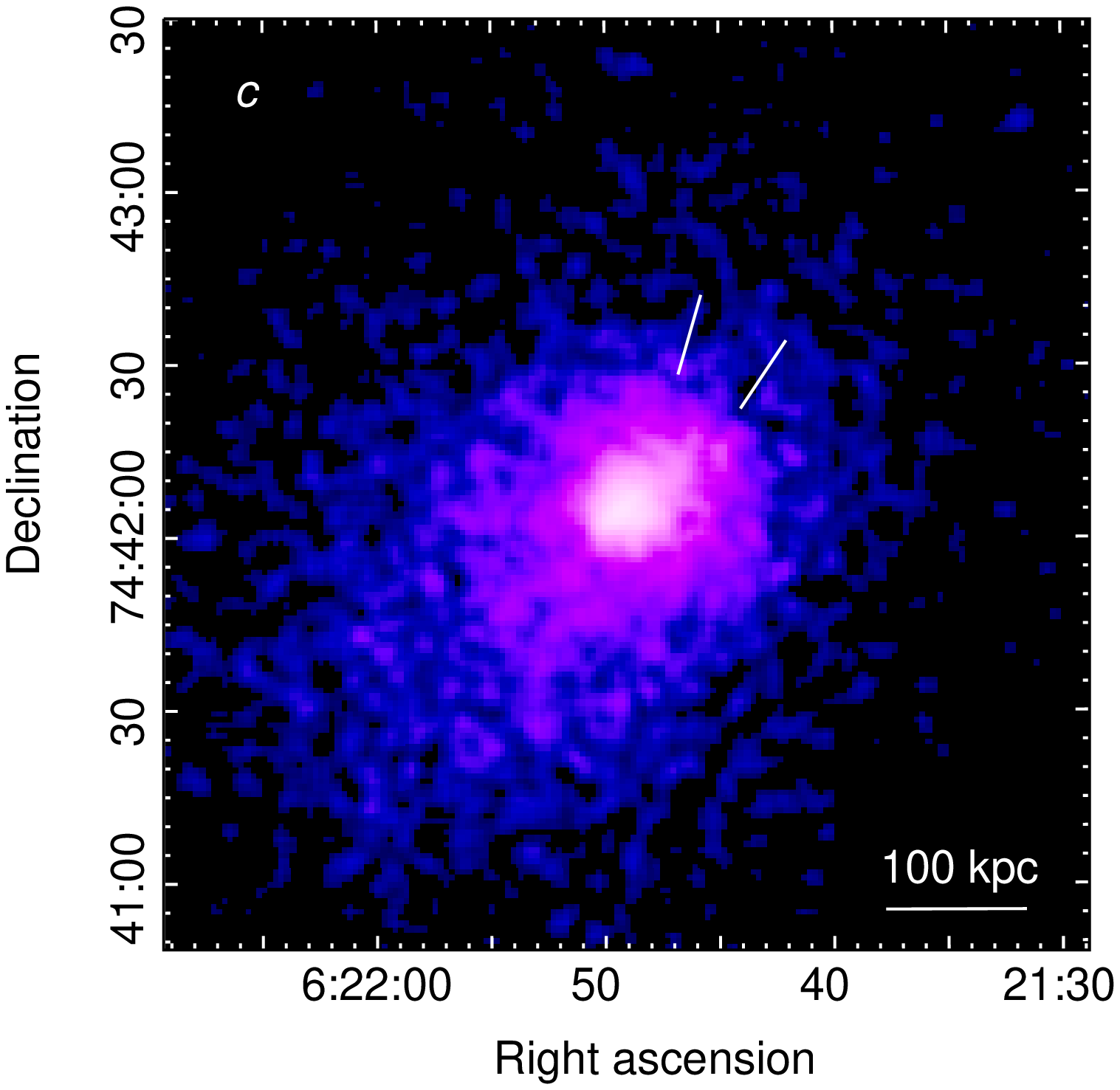}
\smallskip
\caption{PSZ1\,G139.61+24.20. ({\em a}) {\em GMRT} 1.28 GHz contours
  at the resolution of $3^{\prime\prime}\times2^{\prime\prime}$, in p.a.
  $53^{\circ}$, overlaid on the $r$-band Pan--STARRS1 optical image. The rms
  noise of the radio image is $1\sigma=15 \, \mu$Jy beam$^{-1}$. Contours
  are scaled by a factor of 2, starting from 0.05 mJy beam$^{-1}$. Contours
  at $-0.05$ mJy beam$^{-1}$ are shown as dashed. S1 is an extended radio
  source with no optical identification on the POSS-II image.  S2 is a radio
  source with an optical counterpart.  ({\em b}) Contours from the combined
  image at 610 MHz, restored with a $7^{\prime\prime}$ circular beam,
  overlaid on a {\em Chandra} smoothed X-ray image in the 0.5-4 keV band.
  The rms noise of the radio image is $1\sigma=30 \, \mu$Jy beam$^{-1}$.
  Contours are scaled by a factor of 2, starting from 0.08 mJy beam$^{-1}$.
Crosses mark the position of the radio galaxies S1 and S2 ({\em a}). ({\em c}) 
{\em Chandra} X-ray image, same as in ({\em b}). White lines mark the position 
of an X-ray cold front (Savini et al.\ 2018).}
\label{fig:pg139}
\end{figure*}

\subsection{PSZ1\,G139.61+24.20}\label{sec:pg139}

PSZ1\,G139.61+24.20 is a massive, cool-core cluster at $z=0.27$, that was
discovered by {\em Planck} (Planck Collaboration et al.\ 2014). 
We observed the cluster with the {\em GMRT} at
1.28 GHz for approximately 5 hours and two times at 610 MHz for a total of
$9.4$ hours (Table 2).
A high-resolution image at 1.28 GHz is shown as contours in Figure
~\ref{fig:pg139}{\em a}, compared to an optical $r$-band image from
Pan-STARRS1.  A 610 MHz image at $7^{\prime\prime}$ resolution is presented
in Figure ~\ref{fig:pg139}{\em b}.  At 1.28 GHz, hints of extended emission
are visible at the location of the central optical galaxy, which lacks a
clear radio counterpart. An extended double radio source (S1) is detected at
$\sim 15^{\prime\prime}$ from the center. The source has no visible optical
counterpart in the Pan-STARRS1 image, thus suggesting that it is a
background, high-redshift radio galaxy. A second radio source (S2), located
to the North-West, is associated with an optical galaxy with no redshift
information.

At 610 MHz, the double source S1 blends with a central diffuse source, with
a size of $r\sim 50$ kpc, that we classify as a minihalo (Figure
~\ref{fig:pg139}{\em b}).  An X-ray cold front, whose position is marked by
arrows in panel {\em c}, was reported by Savini et al.\ (2018), indicating
that the central cool gas is sloshing.  The position of the cold front is
roughly coincident with the northwestern boundary of the minihalo at 610
MHz, suggesting that its radio emission may be confined here by the front.

The minihalo has been also detected by Savini et al.\ (2018) at 610 MHz
(using the same {\em GMRT} observations analyzed here) and at 144 MHz with
LOFAR.  The LOFAR image revealed the existence of a larger-scale diffuse
component, which surrounds the cool core and its central and more compact
minihalo. This second component is undetected at higher frequencies,
implying a very steep radio spectrum ($\alpha > 1.7$), and has been
interpreted as an ultra-steep spectrum radio halo, similar to those found in
unrelaxed clusters without a central cool core (e.g., Brunetti et al.\ 2008,
Macario et al.\ 2010). Savini et al.\ argued that a recent cluster collision
with a smaller subcluster may have perturbed the cluster on the large scale,
without destroying its central cool core, and generated a large scale radio
halo. The same gravitational interaction may also have induced gas sloshing
in the cool core and ensuing minihalo. This suggests that both a minihalo
and a giant radio halo could at times co-exist in the same cluster (Storm et al. \ 2015, 
Venturi et al.\ 2017; Savini et al.\ 2018, 2019; van Weeren et al.\ 2019).

The flux densities of S1 and S2 are summarized in Table 3, along with their
spectral indices. For the minihalo, we estimated a flux density of $S_{\rm
  610 \, MHz}=2.3\pm0.8$ mJy.  This value was obtained by subtracting the
flux of S1 and S2, measured on the full $5^{\prime\prime}$-resolution image
(not shown here), from the total emission in Figure ~\ref{fig:pg139}{\em b}.
At 1.28 GHz, we used a $5^{\prime\prime}$-resolution image (not shown here)
to measure the minihalo flux and obtained $S_{\rm 1.28 \, GHz}=0.65\pm0.08$
mJy.  A flux density of 12 mJy (with 15\% uncertainty) is reported at 144
MHz by Savini et al.\ (2018).  A spectral index of $\alpha=1.33\pm0.08$ is
inferred between 144 MHz and 1.28 GHz, consistent with the index in the
144-610 MHz range (Savini et al.\ 2018).

The flux density of the minihalo at 1.4 GHz, extrapolated from the
lower-frequency measurements with $\alpha=1.33\pm0.08$, is $S_{\rm 1.4 \,
  GHz}=0.6\pm0.1$ mJy. The corresponding luminosity at 1.4 GHz is $P_{\rm
  \,MH, \,1.4 \, GHz}=(1.3\pm0.2)\times10^{23}$ W Hz$^{-1}$ (Table 4).  For
the BCG, the non-detection at 1.28 GHz at a $3\sigma$ level of 0.035 mJy
implies an upper limit to its 1.4 GHz radio power of $P_{\rm \,BCG, \,1.4 \,
  GHz} < 7\times10^{21}$ W Hz$^{-1}$ (adopting $\alpha=0.6\div0.8$). 
We note that, to date, this is the
only non-detection of radio emission from the BCG in a cluster that hosts a
radio minihalo.

\section{{\em CHANDRA}\/ X-ray analysis}

We use high angular resolution {\em Chandra} observations to map the X-ray
emitting gas in the cluster cores and compare its distribution to the radio
minihalo images (Figs.\ \ref{fig:2129}--\ref{fig:a907},
\ref{fig:a1835}--\ref{fig:2a0335}). In particular, we want to examine
correlations of the minihalo radio luminosity at 1.4 GHz and the X-ray
luminosity of the whole clusters and of their cores (\S\ref{sec:disc}). In
the following sections, we describe the analysis of the {\em Chandra}\/
observations that we have used. They are listed in Table \ref{tab:xray}
along with clean exposure times and the adopted Galactic absorption column
density $N_H$ (discussed below).

\subsection{Data reduction}
\label{sec:xray}

The Level-1 ACIS event files from the archive were reprocessed following the
standard procedure \citep[as described in][]{2005ApJ...628..655V} using the
\chandra\ Calibration Database (CALDB) 4.6.3. We excluded the background
flare intervals and modeled the background as described in
\cite{2003ApJ...586L..19M}. To model the detector + sky background, we used
the CALDB blank-sky datasets appropriate for the date of each observation,
projected to the sky and normalized using the ratio of the observed to
blank-sky count rates in the source-free 9.5--12 keV energy band.  Following
\cite{2000ApJ...541..542M}, we also subtracted the ACIS readout artifact
(using the contributed tool make\_readout\_bg), particularly important in
the presence of sharp X-ray brightness peaks in clusters that we study here.
We used images in the 0.5--4 keV and 2--7 keV energy bands to detect the
unrelated point sources and small-scale extended sources. These sources were
masked out from the spectral analysis.

For each cluster, we obtained a background-subtracted, exposure-corrected
image in the 0.5--4 keV band. For clusters with multiple ACIS observations,
we first coadded the individual background-subtracted images and then
divided the counts images by the sum of the corresponding exposure maps.

To derived the X-ray temperatures and luminosities, we extracted a spectrum
of each cluster region of interest and generated the instrument responses
(ARF and RMF) using calibration files version N0008 for the telescope
effective area, N0006 for the CCD quantum efficiency, and N0009 for the ACIS
time-dependent contamination model. Background spectra were extracted for
the same regions from the blank-sky datasets, normalized as described in
\S\ref{sec:xray}. The cluster X-ray emission was modeled with an absorbed
single-temperature APEC model in the the 0.7--7 keV energy band, with the
heavy element abundance (same for all elements above He) relative to solar
free to vary. The absorption column density $N_H$ was
initially allowed to vary; if the best-fit $N_H$\/ value was found to be
consistent with that from the Leiden/Argentine/Bonn (LAB) radio survey of
Galactic HI \citep{2005A&A...440..775K}, we then fixed it at the database
value for subsequent analysis. Deviations from the radio survey values can
be either real (because of nonuniformity of the column density on the
cluster scale or additional molecular hydrogen) or caused by the residual
inaccuracies of the ACIS contamination model. Using the best-fit $N_H$\/
models such complications away and serves our purpose of deriving the
cluster intrinsic X-ray luminosity and temperature. The choice of the lower
energy boundary of 0.7 keV further limits the importance of such deviations.
The adopted $N_H$ values are listed in Table \ref{tab:xray}.

For several clusters, we combined multiple {\em Chandra}\/ observations by
fitting their spectra separately but simultaneously, with the X-ray model
parameters tied together but using separate detector responses.  For
RXC\,J1504.1-0248, for which the observations had the same pointing position
and roll angle and thus the same responses, we instead coadded the spectra.

\startlongtable
\begin{deluxetable*}{lcccr}
\tablecaption{{\em Chandra} observations analyzed in this work
\label{tab:xray}}
\tablehead{
\colhead{Cluster name} & \colhead{Observation} & \colhead{Detector} & \colhead{Exposure} & \colhead{$N_{H}$}  \\
\colhead{} & \colhead{ID} & \colhead{(ACIS)}  &  \colhead{(ksec)} &  \colhead{($10^{20}$ cm$^{-2}$)}   \\
}
\startdata
A\,478                   &  1669                    & S     & 36.3              & $31.03^{+0.18}_{-0.18}$\phantom{0} \\
MACS\,J0159.8--0849      &  6106                    & I     & 35.3              & $2.06^\star$                      \\
MACS\,J0329.6--0211      &  3582                    & I     & 19.7              & $4.64^\star$                      \\
ZwCl\,3146               &  9371                    & I     & 38.1              & $2.46^\star$ \\
RX\,J1532.9+3021         & 1665,14009               & I,S   & 9,88              & $2.30^\star$ \\ 
A\,2204                  &   7940                   & I     & 76.7              & $8.23^{+0.45}_{-0.39}$\phantom{0} \\
Perseus                  &   11715                  & I     & 70.5              & $13.6^\star$ \\
A\,1835                  &   6880                   & I     & 119               & $2.04^\star$ \\
Ophiuchus                &   16627                  & I     & 35.0              & $35.7^{+0.26}_{-0.25}$\phantom{0}\\
A\,2029                  &   4977,6101              & S,I   & 76.7,9.3          & $3.25^\star$  \\
RBS\,797                 &   2202,7902              & I,S   & 11,7 38.4         & $2.27^\star$ \\
RXC\,J1504.1--0248       &   17197,17669,17670      & I     & 29.8,28.5,44.6    & $14.35^{+4.6}_{-4.6}$\phantom{0} \\
RX\,J1347.5-1145         &   14407                  & I     & 60.1              & $4.60^\star$ \\ 
RXC\,J1720.1+2637        &   4361                   & I     & 22.0              & $3.36^\star$ \\
MS\,1455.0+2232          &   4192                   & I     & 91.2              & $3.18^\star$ \\ 
2A\,0335+096             &   7939                   & S     & 49.8              & $23.18^{+0.27}_{-0.24}$\phantom{0} \\
A\,907                   &   3185,3205              & I     & 47.7, 35.8        & $5.45^\star$ \\
A\,2667                  &   2214                   & S     &  9.3              & $1.73^\star$ \\ 
RX\,J2129.6+0005         &   9370                   & I     & 29.8              & $3.64^\star$\\
AS\,780                  &   9428                   & S     & 39.4              & $7.39^\star$ \\
A\,3444                  &   9400                   & S     & 36.3              & $5.55^\star$ \\ 
PSZ1\,G139.61+24.20      &   15139                  & I     & 17.9              & $13.11^{+1.7}_{-1.6}$\phantom{0} \\
Phoenix                  &   16545                  & I     & 59.9              & $1.52^\star$ \\
\enddata
\tablecomments{Column 1: cluster name; Column 2: observation identification number; Column 3: {\em Chandra} ACIS detector;
Column 4: total clean exposure; Column 5: Galactic absorption column density adopted in this paper; values marked 
with $^\star$ are from LAB (\S\ref{sec:xray}).
}
\end{deluxetable*}



\startlongtable
\begin{deluxetable*}{lrrrrrrrr}
\tabletypesize{\scriptsize} 
\tablewidth{0pt}
\tablecaption{Properties of cluster hosts
\label{tab:hosts}\label{tab:lum}}
\tablehead{
\colhead{Cluster name} & \colhead{$M_{500}$} & \colhead{$R_{500}$} & \colhead{$T_{\rm X, \,R_{500}}$} & \colhead{$T_{\rm X, \,0.15R_{500}}$}
& \colhead{$T_{\rm X, \,70 \, kpc}$} & \colhead{$L_{\rm X, \,R_{500}}$} & \colhead{$L_{\rm X,\, 0.15R_{500}}$} & \colhead{$L_{\rm X, \,70 \, kpc}$}  \\
\colhead{} & \colhead{($10^{14}$ $M_{\odot}$)}   & \colhead{(Mpc)}    & \colhead{(keV)} &  \colhead{(keV)} & \colhead{(keV)} & 
\colhead{($10^{44}$ erg s$^{-1}$)} & \colhead{($10^{44}$ erg s$^{-1}$)} & \colhead{($10^{44}$ erg s$^{-1}$)}   \\
}
\startdata
%
$\diamond$ $\dagger$ A\,478                     & $7.1^{+0.3}_{-0.4}$\phantom{0}     &  1.3  & $6.4\pm0.7$\phantom{0}    & $6.2\pm0.6$\phantom{0}  &  $5.2\pm0.5$\phantom{0}    & $23.3\pm3.0$\tablenotemark{a}   &   $19.5\pm2.3$\phantom{0}  & $7.8\pm0.9$\phantom{0}  \\ 
\phantom{000}MACS\,J0159.8--0849                & $6.9^{+0.9}_{-1.0}$\phantom{0}     &  1.2  & $8.5\pm0.5$\phantom{0}    & $7.5\pm0.4$\phantom{0}  &  $5.9\pm0.3$\phantom{0}   & $40.6\pm4.4$\phantom{0}  &   $21.1\pm2.2$\phantom{0}  & $9.6\pm1.0$\phantom{0}   \\ 
$\star$ \phantom{0} MACS\,J0329.6--0211\tablenotemark{b}   & $4.9\pm0.7$\phantom{0}            &  1.0  & $5.1\pm0.3$\phantom{0}    & $5.1\pm0.3$\phantom{0}   &  $4.3\pm0.3$\phantom{0}   & $27.2\pm2.7$\phantom{0}  &   $15.7\pm1.7$\phantom{0}  & $9.5\pm1.0$\phantom{0}  \\
$\star$ \phantom{0} ZwCl\,3146                  & $6.7\pm0.8$\phantom{0}            &  1.2  & $6.5\pm0.1$\phantom{0}   & $5.9\pm0.1$\phantom{0}   &  $4.7\pm0.1$\phantom{0}    & $45.6\pm4.6$\phantom{0}  &   $28.9\pm3.0$\phantom{0}  & $13.8\pm1.4$\phantom{0}   \\
$\star$ \phantom{0} RX\,J1532.9+3021            & $4.7\pm0.6$\phantom{0}            &  1.0  & $6.0\pm0.3$\tablenotemark{c}          & $5.1\pm0.1$\phantom{0}  &  $4.5\pm0.1$\phantom{0}    &  $41.6\pm4.5$\tablenotemark{c}         &   $27.8\pm2.8$\phantom{0}  &  $15.9\pm1.6$\phantom{0} \\
$\diamond$ $\dagger$ A\,2204                    & $8.0\pm0.4$\phantom{0}            &  1.3  & $6.6\pm0.2$\phantom{0}   & $6.3\pm0.2$\phantom{0}  &  $5.1\pm0.1$\phantom{0}       &  $35.0\pm3.8$            &   $25.1\pm2.7$\phantom{0}  & $13.1\pm1.3$\phantom{0}  \\
$\star$ $\dagger$ Perseus                       & $6.1\pm0.6$\phantom{0}            &  1.3  & $6.4\pm0.1$\tablenotemark{d}          & $6.4\pm 0.2$\phantom{0}   &  $4.1\pm0.1$\phantom{0}    &  $14.9\pm1.5$\phantom{0} &  $10.4\pm1.0$\tablenotemark{e}   &   $4.4\pm0.5$\phantom{0}  \\ 
\phantom{000}A\,1835                           & $8.5^{+0.5}_{-0.6}$\phantom{0}       &  1.3  & $7.7\pm0.1$\phantom{0}  & $6.9\pm0.1$\phantom{0}   &  $5.6\pm0.1$\phantom{0}    & $56.5\pm5.7$\phantom{0}  &   $34.5\pm3.5$\phantom{0} & $16.8\pm1.7$\phantom{0}  \\
$\star$ $\dagger$ Ophiuchus                     & $12.4\pm1.2$\phantom{0}           &  1.6  & $10.3\pm0.2$\tablenotemark{d}         &  $9.4\pm0.1$\phantom{0}  &  $8.8\pm1.1$\phantom{0}    & $19.7\pm2.0$\tablenotemark{f}         & $12.9\pm1.3\phantom{0}$   & $3.0\pm0.4$\phantom{0}  \\
$\diamond$ $\dagger$  A\,2029                   & $6.8\pm0.2$\phantom{0}            &  1.3  & $8.1\pm0.2$\tablenotemark{g}          &  $7.5\pm0.1$\tablenotemark{h}         &   $6.6\pm0.1$\tablenotemark{h}          & $24.3\pm2.4$ \phantom{0}  & $15.0\pm1.5$\tablenotemark{h}           & $6.3\pm0.6$\tablenotemark{h}          \\
\phantom{000}RBS\,797                          & $6.3^{+0.6}_{-0.7}$\phantom{0}       &  1.2  & $6.7\pm0.3$\tablenotemark{i}          & $6.1\pm0.1$\tablenotemark{j}          &  $5.3\pm0.1$\tablenotemark{j}           & $41.9\pm5.4$\tablenotemark{i}          & $32.2\pm3.2$\tablenotemark{j}            & $18.9\pm1.9$\tablenotemark{j}        \\
\phantom{000}RXC\,J1504.1--0248                & $7.0\pm0.6$\phantom{0}             &  1.3  & $5.9\pm0.4$\phantom{0}    & $5.6\pm0.4$\phantom{0}   &   $4.9\pm0.3$\phantom{0}  &  $68.4\pm7.0$\phantom{0}  & $52.1\pm5.6$\phantom{0}   & $31.3\pm3.3$\phantom{0}    \\
\phantom{000}RX\,J1347.5--1145                 & $10.6^{+0.7}_{-0.8}$\phantom{0}      &  1.3  & $12.8\pm0.5$\phantom{0}   & $12.7\pm0.5$\phantom{0} &  $9.8\pm0.4$\phantom{0}   &  $147.3\pm15.4$\phantom{0} & $99.0\pm10.0$\phantom{0}  & $40.9\pm4.3$\phantom{0}   \\
$\diamond$ $\dagger$  RX\,J1720.1+2638          & $6.3\pm0.4$\phantom{0}            &  1.2  & $6.6\pm0.1$\phantom{0}  &  $6.1\pm0.2$\phantom{0} &  $5.1\pm0.1$\phantom{0}   &  $17.2\pm1.7$\phantom{0}   &  $12.9\pm1.3$\phantom{0}  &  $6.2\pm0.6$\phantom{0}   \\
$\star$ \phantom{0} MS\,1455.0+2232            & $3.5\pm0.4$\phantom{0}             & 1.0   & $4.9\pm0.1$\phantom{0}     & $4.8\pm0.1$\phantom{0}  & $4.6\pm0.1$\phantom{0}    &  $21.1\pm2.2$\phantom{0}   & $14.7\pm1.5$\phantom{0}  & $8.6\pm0.9$\phantom{0}   \\ 
\phantom{0} $\dagger$ 2A\,0335+096              & $2.3^{+0.2}_{-0.3}$\phantom{0}      & 0.9   & $3.6\pm0.1$\tablenotemark{k}          &  $2.5\pm0.1$\phantom{0} &  $2.3\pm0.1$\phantom{0}  & $4.4\pm0.5$\phantom{0}      & $3.4\pm0.3$\phantom{0}   & $2.3\pm0.2$\phantom{0}  \\ 
$\diamond$ $\dagger$  A\,907                    & $5.2\pm0.5$\phantom{0}            & 1.2   & $5.8\pm0.1$\phantom{0}  & $5.4\pm0.1$\phantom{0}   & $4.8\pm0.1$\phantom{0}   &  $12.1\pm1.6$\phantom{0}   &  $7.0\pm0.7$\phantom{0}   & $2.9\pm0.3$\phantom{0} \\ 
$\diamond$ $\dagger$  A\,2667                   & $6.8\pm0.5$\phantom{0}            & 1.2   & $6.4\pm0.2$\phantom{0}  & $5.5\pm0.2$\phantom{0}  & $4.6\pm0.2$\phantom{0}    &  $28.8\pm4.1$\phantom{0}  &  $16.0\pm1.7$\phantom{0}  &  $6.2\pm0.7$\phantom{0}  \\
$\diamond$ $\dagger$  A\,3444                   & $7.6^{+0.5}_{-0.6}$\phantom{0}     & 1.3    & $6.2\pm0.2$\phantom{0}  & $5.7\pm0.2$\phantom{0}  & $5.0\pm0.2$\phantom{0}    &  $28.3\pm4.0$\phantom{0}    & $18.5\pm1.9$\phantom{0} & $8.9\pm0.9$\phantom{0}  \\
\phantom{000}RX\,J2129.6+0005                  & $4.2\pm0.6$\phantom{0}             & 1.1   & $6.4\pm0.2$\phantom{0}     & $6.1\pm0.2$\phantom{0}  & $5.1\pm0.2$\phantom{0}    &  $21.1\pm2.1$\phantom{0}     & $11.2\pm1.1$\phantom{0}  & $5.5\pm0.5$\phantom{0} \\
$\diamond$ $\dagger$  AS\,780                   & $7.7\pm0.6$\phantom{0}            & 1.3   & $5.2\pm0.2$\phantom{0}   & $4.7\pm0.1$\phantom{0} & $3.9\pm0.1$\phantom{0}    &  $30.0\pm6.2$\phantom{0}    & $10.9\pm1.1$\phantom{0}   & $5.9\pm0.6$\phantom{0} \\ 
\phantom{000}PSZ1\,G139.61+24.20               & $7.1\pm0.6$\phantom{0}             & 1.2   & $6.9\pm0.5$\phantom{0}      & $7.2\pm0.6$\phantom{0}  & $6.2\pm0.5$\phantom{0}   &  $25.2\pm2.7$\phantom{0}   & $12.6\pm1.4$\phantom{0}  & $5.6\pm0.6$\phantom{0} \\ 
\phantom{000}Phoenix                           & $12.6\pm2.0$\tablenotemark{l}                   & 1.3  & $10.6\pm0.4$\phantom{0}     & $10.0\pm0.3$\phantom{0}  & $8.8\pm0.4$\phantom{0}  &  $143.7\pm15.0$\phantom{0}   & $112.2\pm11.3$\phantom{0}  & $62.6\pm6.5$\phantom{0} \\
\enddata 
\tablecomments{Column 1: cluster name. Column 2: total mass within $R_{500}$ from
Planck Collaboration et al.\ (2014). For clusters marked with $^\star$, 
$M_{500}$ is from G17 and is estimated from the $M_{\rm 500}-T_X$ relation 
of Vikhlinin et al.\ (2009) using core-excised ($r>70$ kpc) temperatures.  
Column 3: $R_{500}$, derived from $M_{500}$. Columns 4--6: cluster temperature 
measured within a radius corresponding to $R_{500}$, $0.15R_{500}$ and $70$ kpc. 
For the clusters marked by a $\diamond$, the total temperature was measured 
within radius smaller than $R_{500}$. A\,478: $r=315$ kpc; A\,2204: $r=360$ kpc; A\,2029:
$r=390$ kpc; RX\,J1720.1+2638: $r=860$ kpc; A\,907: $r=930$ kpc; A\,2667:
$r=620$ kpc; A\,3444: $r=600$ kpc; A\,S780: $r=600$ kpc. Columns 7--9:
bolometric X-ray luminosity measured within a radius corresponding to
$R_{500}$, $0.15R_{500}$, and $70$ kpc. For the clusters marked by a
$\dagger$, the $R_{500}$ luminosity is from MCXC (see \S\ref{sec:lum}).}
\tablenotetext{a}{ The MCXC luminosity was increased by factor 1.34 to account for the
different $N_H$ used in our {\em Chandra}\/ analysis.}
\tablenotetext{b}{ From ObsId 3582.}
\tablenotetext{c}{ From ObsId 1665.} 
\tablenotetext{d}{ Ikebe et al.\ (2002).}
\tablenotetext{e}{ See \S\ref{sec:lum}.}
\tablenotetext{f}{ The MCXC luminosity was increased by factor 1.25 to account for the different $N_H$ used in our {\em Chandra}\/ analysis.}
\tablenotetext{g}{ From ObsId 6101.}
\tablenotetext{h}{ From ObsId 4977.}
\tablenotetext{i}{ From ObsId 2202.}
\tablenotetext{j}{ From ObsId 7902.}
\tablenotetext{k}{ David et al.\ (1993).}
\tablenotetext{l}{ McDonald et al.\ (2015).}
\end{deluxetable*}

\subsection{X-ray luminosities}
\label{sec:lum}

For each cluster, we define 3 interesting regions: (1) the whole cluster,
$r \le R_{500}$, (2) the core region, $r \le 0.15 R_{500}$ (which is in the
$r=140-240$ kpc range for our clusters), and (3) the central coolest core
region, $r \le 70$ kpc, which is within the cooling radius of all our clusters. 
$R_{500}$ was derived from the total cluster masses $M_{500}$ estimated from the 
Planck Sunyaev-Zeldovich (SZ) signal, which are
available for 17 out of 23 clusters (Table 6). For the remaining clusters
(marked by a $\star$), we used a mass estimate from G17, which is based 
on the $M_{\rm 500}-T_{\rm  X}$\/ relation of \cite{2009ApJ...692.1033V}, 
using X-ray measured, core-excised ($r> 70$ kpc) temperatures.

For these three regions, we first determined an X-ray luminosity in the
0.1--2.4 keV band for each cluster. For 12 clusters in the sample, \chandra\ 
covered $R_{500}$, and we determined the luminosity by fitting the
spectrum in the respective region and converting its normalization to
luminosity (in the XSPEC package). For 11 clusters marked with $\dagger$ in
Table \ref{tab:lum}, $R_{500}$ was outside the {\em Chandra}\/ field of
view, so we adopted the 0.1--2.4 keV $R_{500}$ luminosities from MCXC 
(Meta-Catalog of X-Ray Detected Clusters of Galaxies, Piffaretti et al.\ 2011), 
which is mostly based on the {\em ROSAT}\/ All Sky Survey (RASS). 
For A\,478 and Ophiuchus, we corrected the MCXC luminosities by factors 1.34 and 1.25,
respectively, to account for the different $N_H$ derived from our {\em
  Chandra}\/ fits, which were factors 2.0 and 1.8 higher than the database
values adopted for the RASS catalogs. This makes the luminosities internally
consistent with those for the central regions that were derived directly
from \chandra\ observations. For Perseus, even the $0.15 R_{500}$ radius is
not fully covered by one \chandra\ ACIS-I observation, so we used the
\chandra\ mosaic from Aharonian et al.\ (2016) to derive the luminosity
within $0.15 R_{500}$.

To convert the 0.1--2.4 keV luminosities to the more physically meaningful
bolometric ones, we need gas temperatures, which we have derived from
\chandra\ spectral fits in the same regions for most clusters. The
temperature fits are summarized in Table \ref{tab:lum}; their errors include
the statistical uncertainty and a 10\% uncertainty on the value of $N_H$.
For Perseus, Ophiuchus and 2A\,0335+096, \chandra\ coverage is too small to
probe an average temperature within $R_{500}$, so we used average cluster
temperatures from the literature instead, which come from instruments with a
larger field of view (see Table \ref{tab:lum}). For Perseus and Ophiuchus,
whose total mass is estimated from the $M_{\rm 500}-T_{\rm X}$\/ relation
(the 2A\,0335+096 mass comes from Planck), this should result in slightly 
underestimated masses and $R_{500}$, because the cool core biases the average 
temperature low. This will not affect our conclusions in any significant way. 
For further 8 clusters, marked with $\diamond$ in Table \ref{tab:lum}, 
the \chandra\ field of view is smaller than $R_{500}$ but covers enough of it 
to derive a reliable temperature for the region, considering the steeply 
declining radial brightness profiles and the relatively small contribution 
of the missing emission to the spectrum. For these clusters, we use \chandra\ 
temperatures fitted in smaller regions given in the notes for Table \ref{tab:lum}. 
For Perseus, for simplicity, we used one ACIS-I pointing (Table \ref{tab:xray}) 
out of many available to derive a temperature for $r<0.15 R_{500}$, 
as it almost covers the whole region. The statistics of that data subset is 
more than sufficient. The above simplifications are adequate because we are 
using the temperatures only for the luminosity conversion.

For each cluster region, we converted the 0.1--2.4 keV luminosities to the
bolometric luminosities (Table \ref{tab:lum}) using a conversion factor
appropriate for the corresponding temperature (Boehringer et al.\ 2004).
The luminosity errors in Table \ref{tab:lum} include the uncertainty on the
temperature and a $10\%$ {\em ROSAT}--{\em Chandra}\/ flux cross-calibration
uncertainty, added in quadrature.  The latter was evaluated by comparing the
$R_{500}$ 0.1--2.4 keV luminosities derived from {\em Chandra} and those
determined from {\em ROSAT} for the 10 clusters in our sample with both the
{\em Chandra}\/ and the MCXC measurement. We found that our {\em Chandra}
luminosities are on average $6\%$ higher than the RASS ones, with a scatter
of $\sim 10\%$.

\section{Discussion}
\label{sec:disc}

In this paper, we have presented new {\em GMRT}\/ and {\em VLA}\/ radio
images of 8 cool-core clusters. Our images confirm the presence of radio
minihalos in 3 of these clusters that had minihalo candidates, confirm
previously-known minihalos in 2 clusters, and discover new minihalos in 3
clusters. For consistency with the present analysis and previous work (G14a, G17),
we have also reanalyzed {\em VLA}\/ 1.4 GHz archival observations of 7
clusters with known minihalos (Appendix A).
Below, we briefly summarize our main radio results.

\begin{enumerate}

\item Our {\em GMRT}\/ images of MACS\,J0159.8-0849 (1.4 GHz),
MACS\,J0329.6-0211 and RXC\,J2129.6+0005 (1.3 GHz) detail the radio emission
in the cluster center on angular scales $2''$ and up. This corresponds to a
physical size of $\gax 10$ kpc (7 kpc for RXC\,J2129.6+0005). We do not
detect any well-defined jets, lobes or tails from the central AGN on this
scale, which is considerably smaller than that of the surrounding diffuse
minihalo seen in the lower-resolution images ($R_{\rm MH}>70$ kpc; G14a,
K15). We thus confirm the presence of minihalos in these clusters (no
morphological connection between the diffuse radio emission and the AGN).

\item We have presented new {\em GMRT}\/ images at 1.28 GHz and 610 MHz of
the minihalo in AS\,780 (Venturi et al.\ 2017). The minihalo, which extends
$\sim 50$ kpc in radius, appears to be confined within a pair of prominent
sloshing cold fronts seen in the {\em Chandra}\/ image. Our reanalysis of an
archival {\em VLA}\/ observation at 8.46 GHz shows the cluster central
galaxy to be a core-dominated, double-lobe radio source of 8 kpc in length 
with no morphological connection to the much larger minihalo.

\item We have presented {\em VLA}\/ images at 1.4 GHz of the minihalo in
A\,3444, first reported by Venturi et al.\ (2007). The diffuse emission
covers the central $r\sim120$ kpc region of the cluster core and encloses a
weak compact source ($\lax 8$ kpc) associated with the BCG.  In a
forthcoming paper (Giacintucci al. in preparation), 
we will present new, deep {\em GMRT}\/ observations at 1.28 GHz
and 610 MHz, study the radio spectral properties of the minihalo and
investigate possible spatial correlation between the radio and X-ray
emission.

\item Using {\em GMRT}\/ observations, we have detected new minihalos in the
cores of A\,2667, A\,907 and PSZ1\,G139.61+24.20. For the latter, a recent
\lofar\ observation at 144 MHz has shown the central minihalo to be
surrounded by a larger-scale and fainter diffuse component, with a much
steeper radio spectrum (Savini et al.\ 2018) --- perhaps another example of
minihalo-like emission inside a giant radio halo as possibly seen 
in A\,2319 (Storm et al.\ 2015), A\,2142 (Venturi et al. \ 2017) and
RX\,J1720.1+2638 (Savini et al. 2019).

\end{enumerate}

With the new minihalo detections and candidate verifications presented
in this work, the total number of confirmed minihalos has now increased to
23 (Table 4). This allows us to revisit some correlations between
non-thermal emission and thermal properties of the minihalo cluster
hosts. We will consider some other correlations in the forthcoming work.

\subsection{Radio minihalos and properties of the cluster hosts}

A possible intrinsic relation between the non-thermal properties and global
thermal properties of clusters with a central minihalo has been suggested by
early works, based on a small number of minihalos ($\lax 10$, e.g., Cassano
et al.\ 2008, Kale et al.\ 2013). More recently, K15 and Gitti et al.\ (2018)
used a larger sample of minihalos to investigate the distribution of
minihalo clusters in the $P_{\rm \,MH, \,1.4 \, GHz}$--$L_{\rm X}$ plane
and reported a positive correlation. The physical origin of this
correlation might be a correlation of the radio power with the total
cluster mass, as found for the giant radio halos (e.g., Cassano et al.\ 2013), because the
cluster mass and X-ray luminosity correlate. However, in G14a,
we did not find evidence for a correlation with the mass for 14
minihalos with available masses from the {\em Planck}\/ SZ observations.
We also did not find a strong correlation between the radio power and
the cluster core-excised X-ray temperature, which is a better proxy
for the cluster total mass than the X-ray luminosity (e.g., Vikhlinin et al.\ 2009).

With our new minihalo detections, we can explore these possible correlations
further. Our statistical analysis will include 22 confirmed minihalos,
excluding AS\,780, which has only a lower limit for the minihalo 1.4
GHz luminosity, but we show it in the plots. To evaluate the presence of any
possible correlation and its strength, we use the Spearman's test, which
computes a rank correlation coefficient, $r_s$, whose value can
range from --1 (perfect anticorrelation) through 0 (no correlation) to
1 (perfect correlation). The test also 
returns the probability {\em probrs}\/ to get the observed $r_s$ value
by chance under the null hypothesis of no correlation (false positive).  In
Table 7, we summarize the values of $r_s$ and {\em probrs}\/ found for
the relations analyzed in the following paragraphs.

\begin{figure*}
\centering
\includegraphics[width=6.5cm,draft=false]{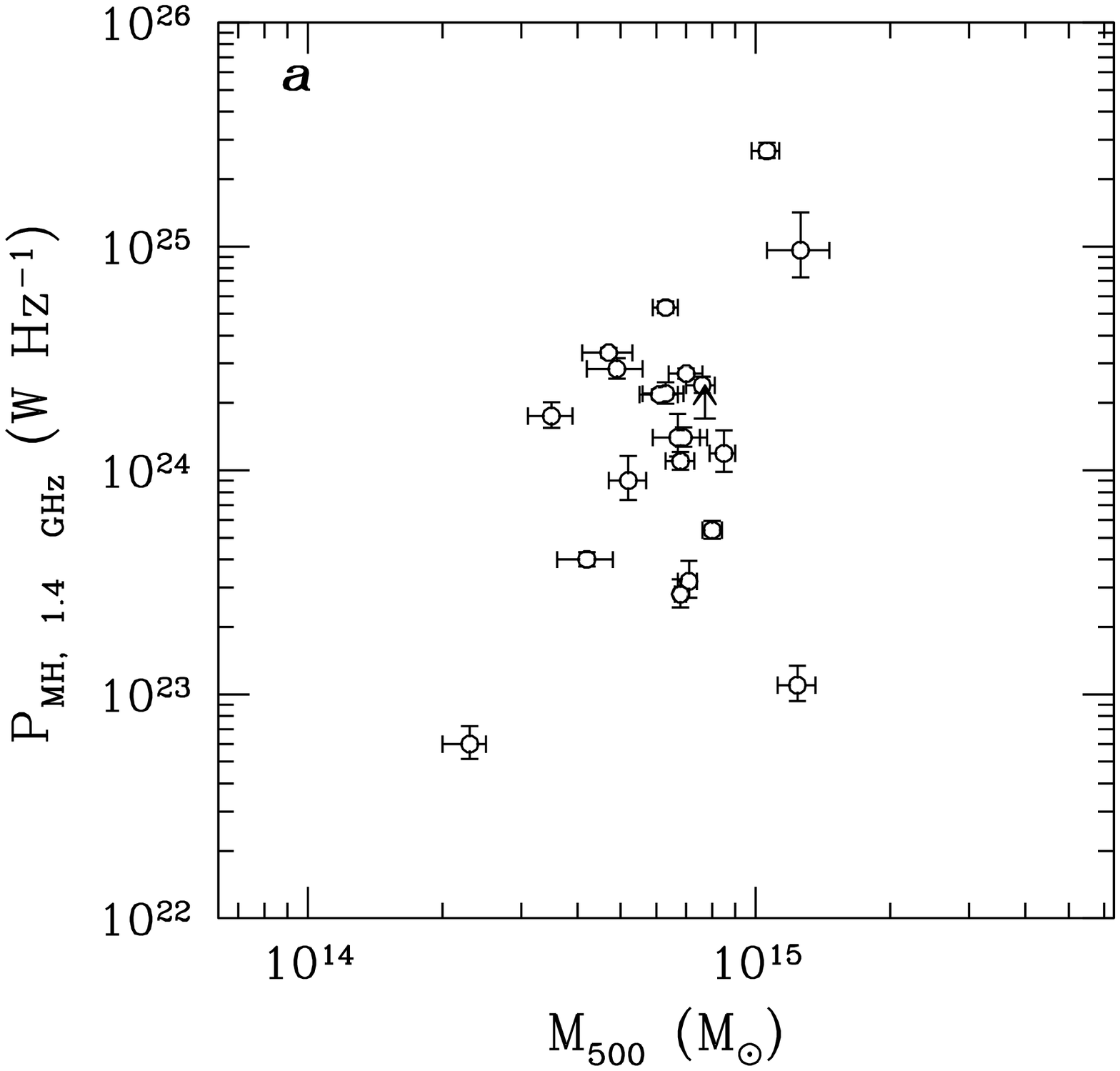}
\includegraphics[width=6.5cm,draft=false]{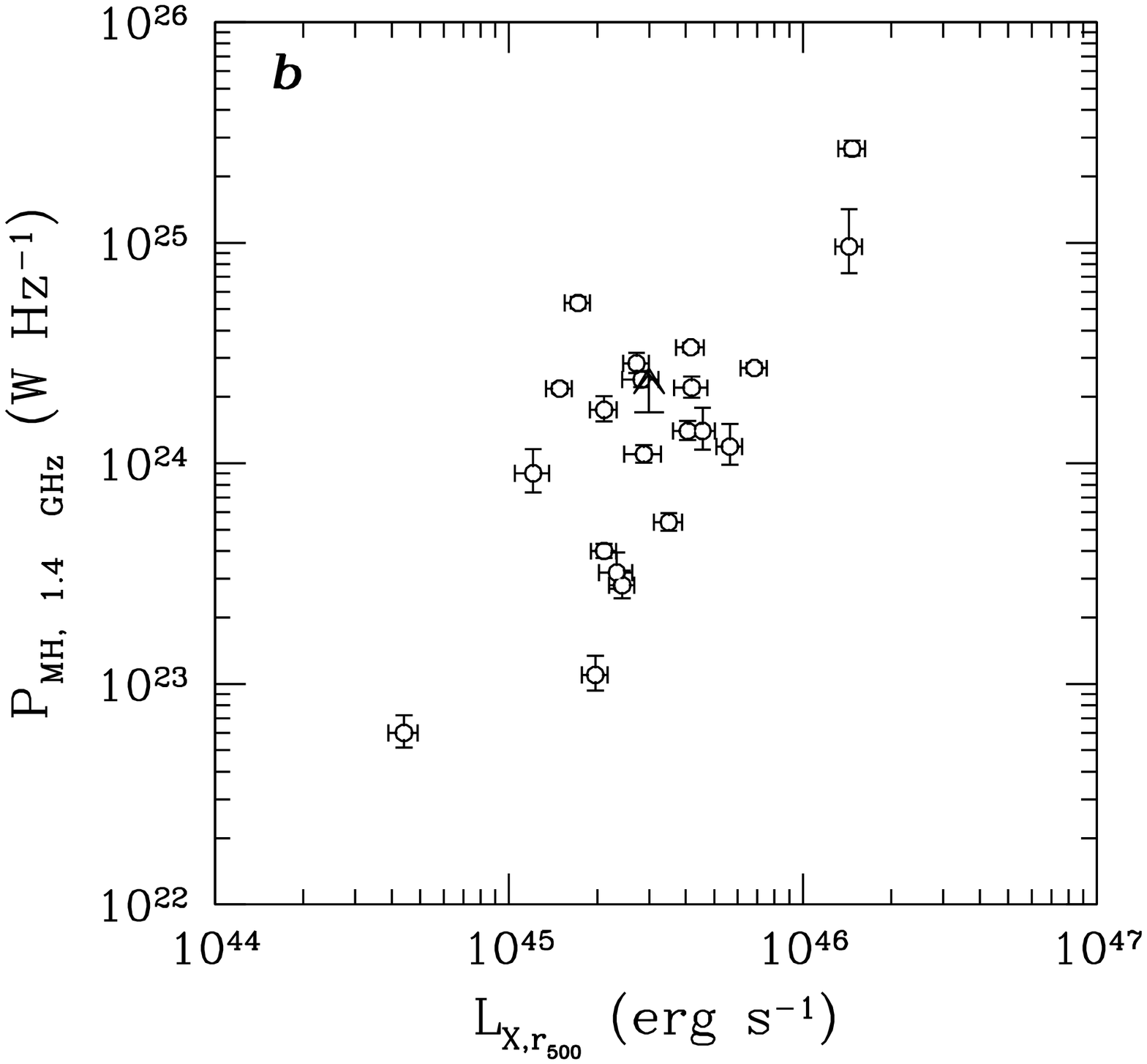}
\includegraphics[width=6.5cm,draft=false]{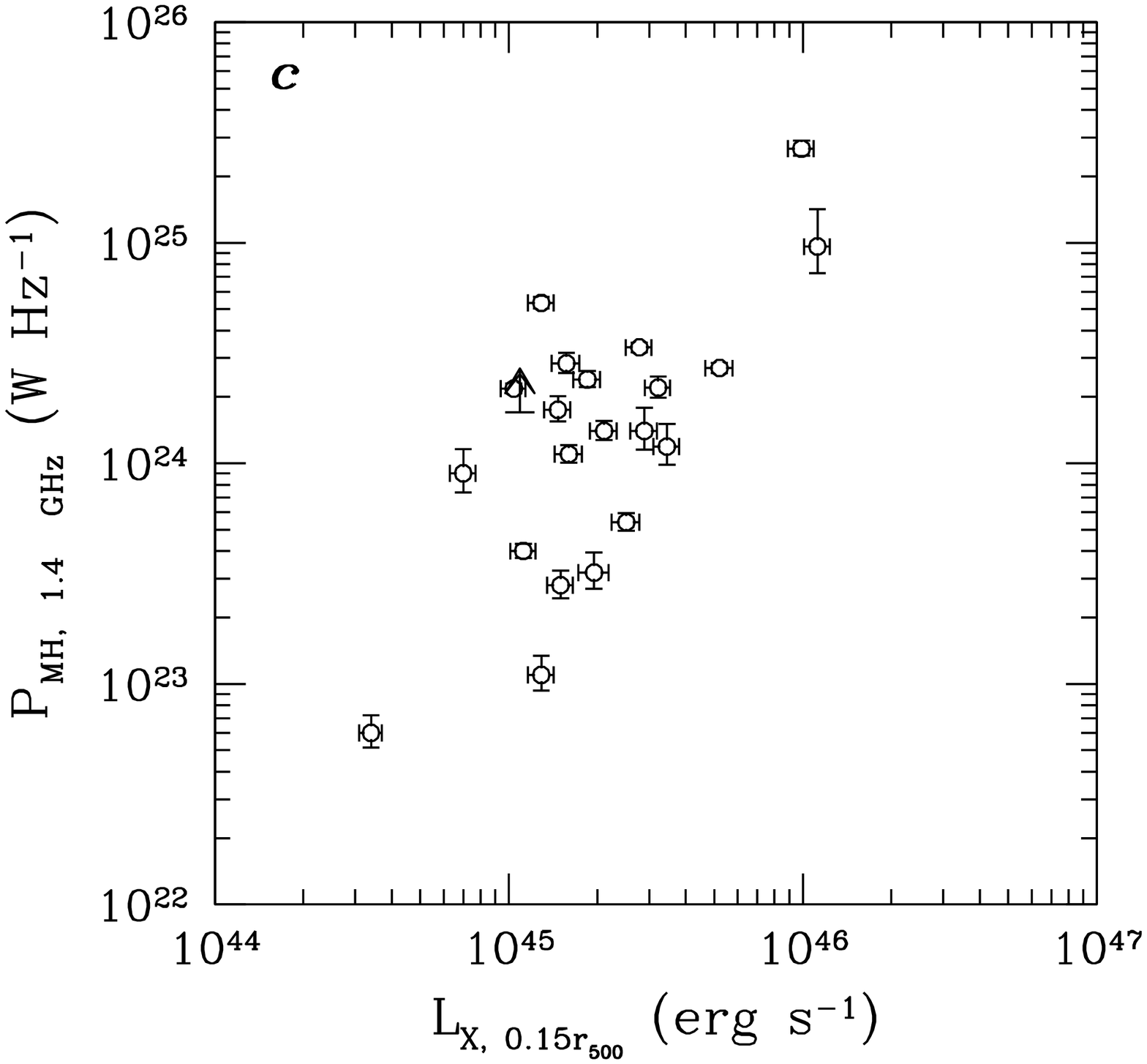}
\includegraphics[width=6.5cm,draft=false]{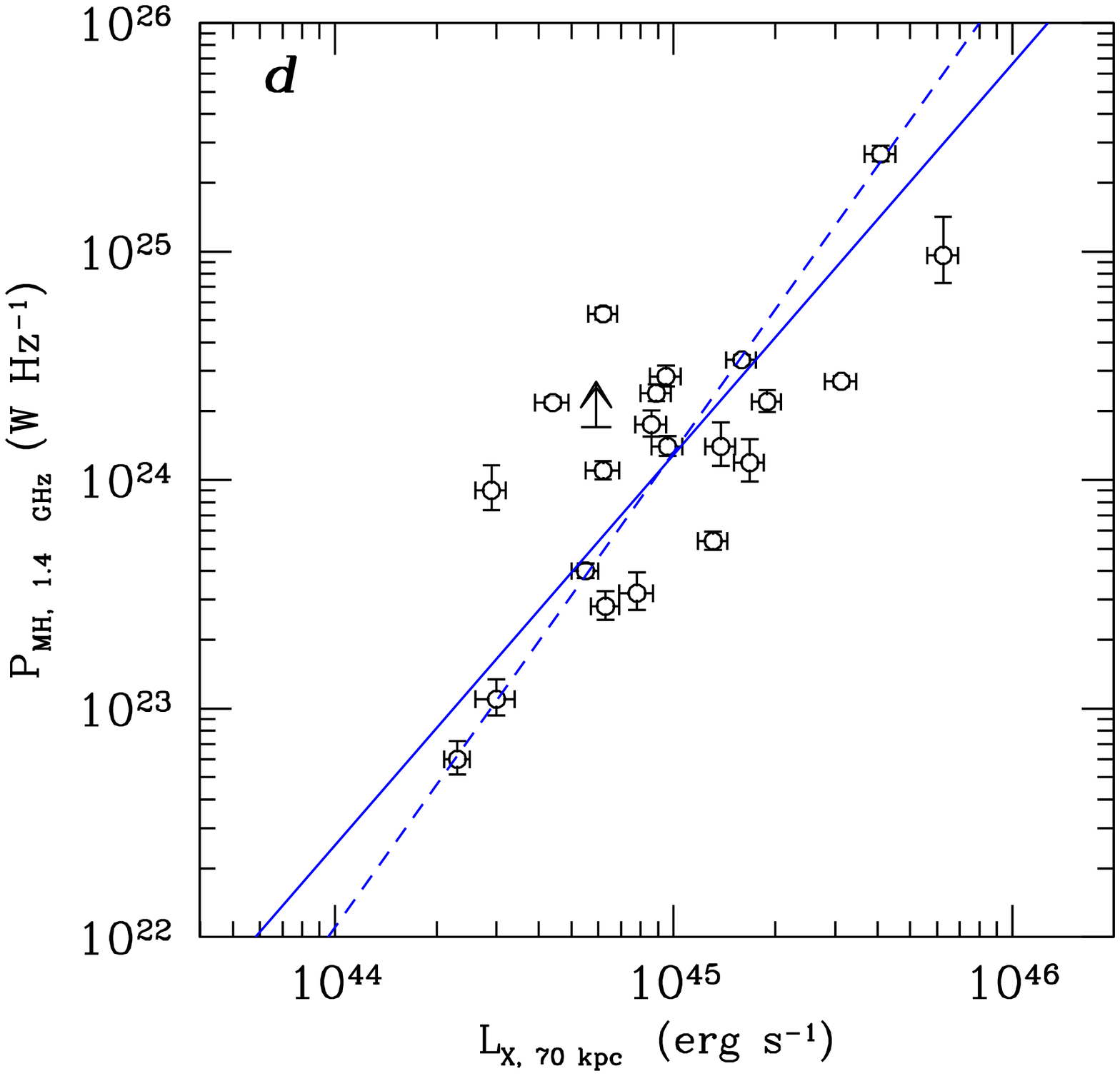}
\includegraphics[width=6.5cm,draft=false]{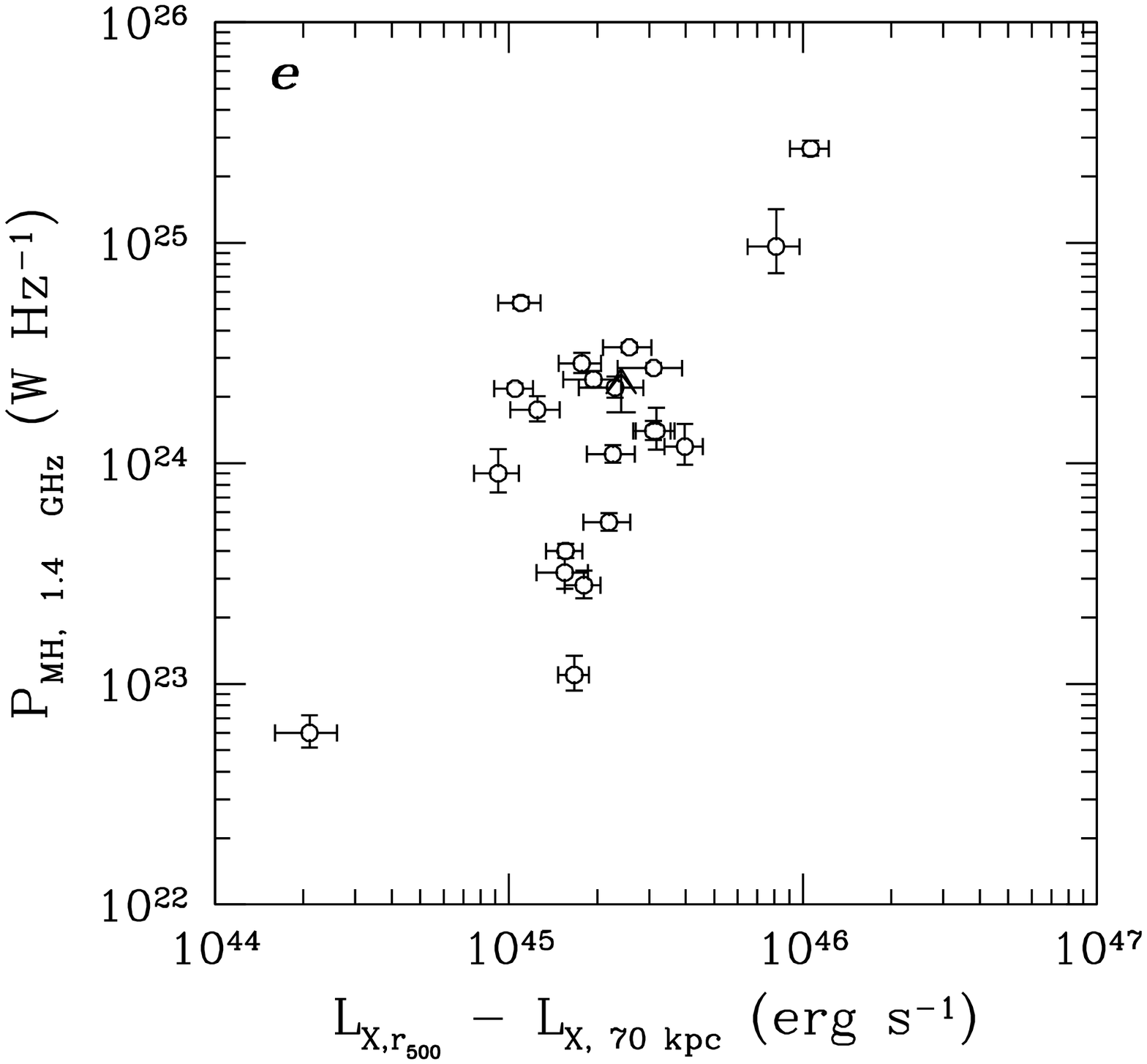}
\includegraphics[width=6.5cm,draft=false]{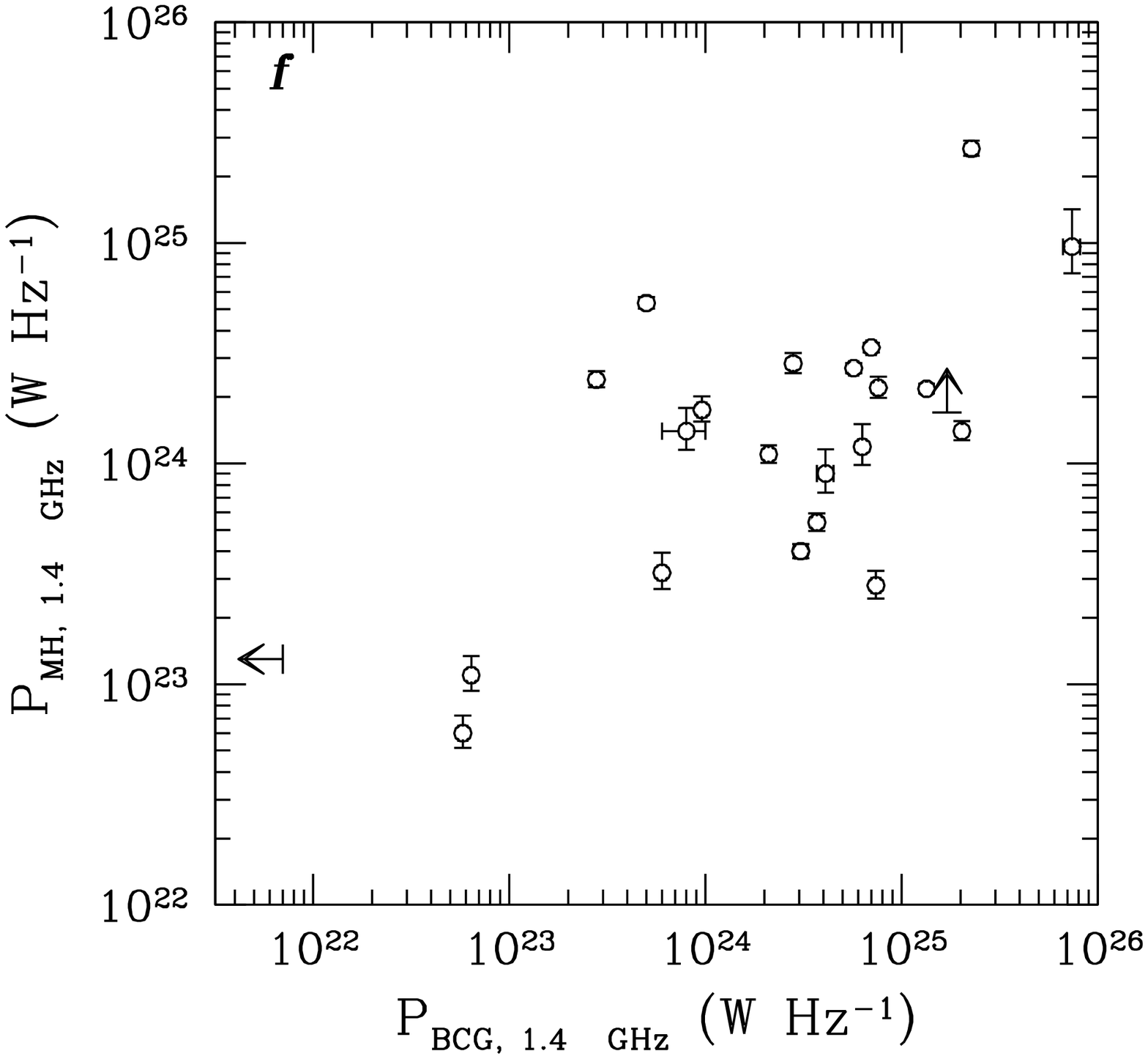}
\smallskip
\caption{ \footnotesize ({\em a}) Comparison between the 1.4 GHz radio luminosity 
of minihalos and total mass within $R_{500}$ of their cluster hosts.
{(\em b,c,d)} Comparison between the 1.4 GHz radio luminosity of minihalos 
and the cluster X-ray bolometric luminosity within $R_{500}$ ({\em a}),
$0.15 R_{500}$ ({\em b}), and $r=70$ kpc ({\em c}). Blue lines in ({\em c}) 
are the best-fit relations derived using the BCES bisector 
(solid line) and orthogonal (dashed line) regression algorithms 
(Eq.\ref{eq:fit}). ({\em e}) The minihalo 1.4 GHz luminosity 
is compared to the core-excised X-ray bolometric luminosity, a proxy for 
the cluster mass. ({\em f}) Comparison between the 1.4 GHz luminosity of 
minihalos and 1.4 GHz luminosity of the BCG.}
\label{fig:corr2}
\end{figure*}

\subsubsection{Radio luminosity--cluster mass}

In Fig.\ \ref{fig:corr2}{\em a}, we compare the radio luminosity 
(Table 4) and the total cluster mass within $R_{500}$ (Table 6) for 
our 23 minihalo clusters. The distribution in this plot is scattered 
and no obvious correlation is visible. The Spearman test gives a rank 
correlation coefficient of $r_s = 0.06$ and $probrs=0.79$, which indicates 
lack of correlation with the cluster mass, consistently with the G14a
finding from a smaller sample. 

\subsubsection{Luminosity relations}

In line with previous studies (e.g., Gitti et al.\ 2018), we find a positive
correlation ($r_s = 0.53$) between the minihalo radio power and cluster
total X-ray bolometric luminosity measured within $R_{500}$
(Fig.\ \ref{fig:corr2}{\em b}). In cool-core clusters, such as the minihalo
hosts, a large fraction of the total X-ray luminosity is found to arise from
the central $0.15R_{500}$ region (e.g., Maughan et al.\ 2012), which
is also the region where the minihalo is located. It is
therefore interesting to check if the correlation seen in Fig.\
\ref{fig:corr2}{\em b}\/ is driven by a correlation with the X-ray
luminosity of the cluster core. In Figs.\ \ref{fig:corr2}{\em c}\/ and {\em d}, 
we compare the minihalo radio
power to the X-ray bolometric luminosity measured within the central
$r=0.15R_{500}$ (which is in the
$r=140-240$ kpc range for our clusters) and the smaller $r=70$ kpc regions, respectively.

The plot with the $0.15R_{500}$ X-ray luminosity ({\em c}) is rather similar
to the plot with the total X-ray luminosity ({\em b}), and we find
a similar rank correlation coefficient, $r_s = 0.56$. This is not
surprising, since for our sample of strongly peaked clusters, the
core/total luminosity ratio, $L_{\rm X,\, 0.15R_{500}}/L_{\rm X, \,
  R_{500}}$, ranges from 0.4 (AS\,780) 
to 0.8 (A\,478), with an average value of 0.6. Thus, the central
$r=0.15R_{500}$ core region is the dominant contributor to the cluster total
luminosity for most clusters in our sample, explaining the similarity of
diagrams {\em b} and {\em c}.

In Fig.\ \ref{fig:corr2}{\em d}, we cut an even smaller region of the core
and plot the radio luminosity vs.\ the X-ray bolometric luminosity
within the central $r=70$ kpc, which is within the cooling radius 
for all our clusters. This correlation is tighter, with a higher correlation 
coefficient, $r_s = 0.67$. The probability of the null hypothesis is less 
than $1\%$. For this relation, we also show the best-fit correlations in the form
\begin{equation}\footnotesize
{\rm log}(P_{\rm \,MH, \,1.4 \, GHz} /10^{24}) = B \times {\rm log}(L_{\rm X, \,70 \, kpc}/10^{44}) +A
\end{equation} \label{eq:fit}
derived by implementing the bivariate correlated error and intrinsic scatter
(BCES) bisector (blue solid line) and orthogonal (blue dashed line) regression
algorithms (Akritas \& Bershady 1996).  The best-fit parameters are
$A=-1.60\pm0.23$ and $B=1.71\pm0.21$ (bisector) and $A=-1.96\pm0.31$ and
$B=2.08\pm0.34$ (orthogonal).

\begin{table}[t]
\label{tab:spearman}
\caption{Spearman's Rank Correlation Coefficient test}
\begin{center}
\begin{tabular}{lcc}
\hline\noalign{\smallskip}
\small
\phantom{0} Relation & $r_s$ & $probrs$ \\ 
\noalign{\smallskip}
\hline\noalign{\smallskip}
\phantom{0} $P_{\rm \,MH, \,1.4 \, GHz}$--$M_{\rm 500}$                                   &  +0.06     &   79\%  \\
\phantom{0} $P_{\rm \,MH, \,1.4 \, GHz}$--$L_{\rm X, \, R_{500}}$                           &  +0.53     &   11\%  \\
\phantom{0} $P_{\rm \,MH, \,1.4 \, GHz}$--$L_{\rm X,\, 0.15R_{500}}$                         &  +0.56     &   6\%   \\ 
\phantom{0} $P_{\rm \,MH, \,1.4 \, GHz}$--$L_{\rm X, \,70 \, kpc}$                          &  +0.67     &   0.7\% \\
\phantom{0} $P_{\rm \,MH, \,1.4 \, GHz}$--($L_{\rm X, \,R_{500}}$-$L_{\rm X, \,70 \, kpc}$)  &  +0.44     &   4\% \\
 $^\star$ $P_{\rm \,MH, \,1.4 \, GHz}$--($L_{\rm X, \,R_{500}}$-$L_{\rm X, \,0.15R_{500}}$)    &  +0.43     &   4\% \\
\phantom{0} $P_{\rm \,MH, \,1.4 \, GHz}$--$P_{\rm \,BCG, \,1.4 \, GHz}$                      &  +0.43     &   5\% \\
\hline\noalign{\smallskip}
\end{tabular}
\end{center}
NOTE. Column 1: relation (Fig.~\ref{fig:corr2}; $^\star$ figure not shown). Column 2: Spearman's rank correlation 
coefficient, a non--parametric measure of the statistical dependence between two variables. Column 3: probability 
of uncorrelated quantities producing datasets that have a Spearman correlation at least as extreme as the one 
computed from these datasets. 
\end{table}


In Fig.\ \ref{fig:corr2}{\em e} we plot the radio luminosity as a
function of the core {\em excised}\/  ($r> 70$ kpc) X-ray bolometric luminosity 
within $R_{500}$, which can be used as a proxy for the cluster total mass. This
relation is weaker, with a correlation coefficient of $r_s = 0.44$, in
agreement with the lack of a strong scaling with the SZ--estimated cluster
mass ({\em a}). A similar diagram (not shown here) is obtained if we excise 
the central $0.15R_{500}$ region instead of the smaller $r=70$ kpc region, 
with similar Spearman correlation coefficient and null-hypothesis probability
(Table 7).

Finally, in Fig.\ \ref{fig:corr2}{\em f} we compare the radio luminosity of
the minihalo to the radio luminosity of the central BCG (Table 4). A possible
trend is visible, in line with previous analysis by Govoni et al. (2009)
and G14a, but it is relatively weak ($r_s = 0.43$, $probs=5\%$\footnote{The test
was done using 21 minihalos: in addition to AS\,780, we also excluded PSZ1\,G139.61+24.20,
which has only a lower limit for the BCG luminosity at 1.4 GHz.}), 
indicating that the AGN activity and minihalo emission are not tightly correlated. 
Interestingly, this correlation is weaker than that with the $r<70$ kpc core 
X-ray luminosity, suggesting that the properties of the core gas may have a more 
direct effect on the minihalos than the AGN output.
\\
\\
\indent Our minihalo sample is not statistically complete. 
Analysis of selection effects is beyond the scope of this paper, however we 
checked for the possibility of the luminosity-luminosity correlations being 
dominated by effects originating from flux selection. The corresponding radio 
flux--X-ray flux diagrams show strong positive correlations reflecting the large 
range of distances, but do not show any obvious flux-selection effects. 
This does not affect our main result of a better correlation of the minihalo
radio power with the cool-core luminosity compared to the total luminosity, 
which is derived from the same sample and thus not driven by selection effects.
\\
\\
\indent Our finding of a significant correlation between the minihalo radio
power and the X-ray luminosity of the cool core is consistent with earlier
findings by Gitti et al.\ (2004, 2012) and Bravi et al.\ (2016), who
considered a correlation with the ``cooling flow power'' --- which by
their derivation is very close to the X-ray luminosity of the cooling region.
Since we know that some physical process in the cluster cores compensates for
X-ray radiative cooling (e.g., McNamara \& Nulsen 2012), that process has
to have the power very similar to the X-ray bolometric luminosity of the core, 
which is what cools the gas. Among the popular processes are the turbulence 
generated by AGN activity (Zhuravleva et al.\ 2014) and cosmic rays generated 
by the AGN (e.g., Guo \& Oh 2008, Pfrommer 2013, Yang \& Reynolds 2016, 
Ruszkowski et al.\ 2017). A further process is gas sloshing, which can 
provide a net heat inflow by bringing the outer, higher-entropy gas into 
the cool core (ZuHone et al.\ 2010). It is possible to make the conjecture that the same 
heating process is responsible for the radio minihalos. However, in addition 
to the minihalo power, any such hypothesis would need to explain other 
properties of the minihalos, in particular, their apparent close relation 
to the sloshing cold fronts (e.g., Mazzotta \& Giacintucci 2008; ZuHone et al.\
2013; Giacintucci et al. 2014b).

\section{Summary and conclusions}

We have presented new \gmrt\ and \vla\ images of 8 cool-core
clusters. We have found new minihalos in A\,2667, A\,907 and
PSZ1\,G139.61+24.20, confirmed the candidates in
MACS\,J0159.8--0849, MACS\,J0329.6--0211 and RXC\,J2129.6+0005, and
presented new radio images of the previously-known minihalos AS\,780
and A\,3444. For consistency with our radio analysis, we have also
reanalyzed archival {\em VLA}\/ 1.4 GHz observations of 7 cool-core clusters
with known minihalos. Combining our new detections and 
confirmations with the previously known minihalos results in a total sample
of 23 confirmed minihalos, the largest sample to date. We revisited
possible correlations between non-thermal emission and thermal properties of
their cluster hosts. We confirmed the lack of a strong relation
between the minihalo radio power and the total cluster mass. Instead, we found
a relatively strong positive correlation between the minihalo radio
power and the X-ray bolometric luminosity of the cool core ($r<70$
kpc). This supplements our earlier finding (G17) that most if not all
cool cores in massive clusters contain radio minihalos. In many cases,
we observe that the minihalo may be confined by the sloshing cold fronts 
in the core gas. Taken together, these findings suggest that the origin 
of the minihalos should be closely related to the properties of the 
cool-core gas.  
\\
\\
\indent {\it Acknowledgements.}

Basic research in radio astronomy at the Naval Research Laboratory is
supported by 6.1 Base funding.  RK acknowledges support through the 
DST-INSPIRE Faculty Award by the Government of India. We thank the staff of 
the GMRT that made these observations possible. GMRT is run by the National Centre for
Astrophysics of the Tata Institute of Fundamental Research. The National
Radio Astronomy Observatory is a facility of the National Science Foundation
operated under cooperative agreement by Associated Universities, Inc.

{}

\appendix

\section{RADIO OBSERVATIONS AND IMAGES}\label{sec:imagesr}

Table \ref{tab:re_obs} lists the {\em VLA} 1.4 GHz archival observations of
7 clusters with known minihalos, that we have reanalyzed in this work. The
data were reduced as described in \S2.2. The radio images are
presented in Figs.~11--17, compared to the X-ray {\em Chandra} and
{\em XMM-Newton} images. In
Table 9, we summarize for each cluster the flux density of the discrete
radio galaxies detected in the area occupied by the minihalo and the
source-subtracted flux density of the minihalos, measured as 
described in \S2.1. A brief description of each individual cluster is given below.

\begin{table*}[t]
\caption{Newly analyzed 1.4 GHz {\em VLA} observations}
\footnotesize
\begin{center}
\begin{tabular}{lrcccccccrc}
\hline\noalign{\smallskip}
\hline\noalign{\smallskip}
Cluster  & Project & Configuration & Bandwidth & Date & Time  & FWHM, PA  &   rms    \\  
  &                  &            &      (MHz)   &   &  (min)    &
  ($^{\prime \prime} \times^{\prime \prime}$, $^{\circ}$)\phantom{00}  & ($\mu$Jy b$^{-1}$) \\
\noalign{\smallskip}
\hline\noalign{\smallskip}
A\,1835             & AI75  & B & 25  & 1998 Jul 13    & 443 &  $5.2\times4.5, -74$      & 15 \\ 
                    & AG729 & C & 25  & 2006 Oct 23,28 & 325 &  $16.2\times13.6, 37$     & 30 \\
                    & AG639 & D & 50  & 2003 Mar 11    & 260 &  $50.5\times45.0, 4$      & 40 \\

Ophiuchus           & AC261 & D & 25  & 1990 Jan 25    &  28 &  $91.6\times39.2, -25$    & 100 \\

A\,2029             & AL77  & B &12.5 & 1984 Jan 15 &  69 &  $4.5\times4.3$, 11       & 60 \\
                    & AG729 & C & 50  & 2006 Dec 26 & 355 &  $15.2\times13.7$, $-15$  & 30 \\
                    & AG639 & D & 50  & 2003 Mar 09 & 293 &  $57.3\times45.2$, 0      & 40 \\

RBS\,797            & AS720 & A & 50  & 2002 Mar 27 & 257 &  $1.6\times1.2,-11$       & 17 \\
                    & AS720 & B & 50  & 2002 Jul 27 &  45 &  $5.2\times3.9, 10$       & 27 \\

RX\,J1347.5--1145   & AS632 & A & 25  & 1998 Apr 22 & 115 &  $1.7\times1.3,-18$       & 35\\
                    & AD511 & C & 50  & 2005 Sep 03 &  94 &  $16.4\times13.0,-1$      & 40 \\

MS\,1455.0+2232     & AL317 & A & 50  & 1994 Mar 12  & 33 &  $1.5\times1.2,33$        & 55 \\
                    & AS823 & C & 50  & 2006 Dec 12  & 31 &  $16.5\times14.6,-7$      & 70 \\

2A\,0335+096        & AM828 & B   & 6.3 & 2005 Apr 29 & 136 & $4.2\times3.9$, $-53$   & 60 \\ 
                    & AS465 & C   & 50  & 1992 Mar 14 & 165 & $15.0\times14.3$, $-39$ & 38 \\
                    & AS465 & D   & 50  & 1992 Jul 14 & 117 & $50.9\times45.2$, 40    & 50 \\
\hline{\smallskip}
\end{tabular}
\end{center}
\label{tab:re_obs}
{\bf Notes.} Column 1: cluster name. Columns 2 and 3: {\em VLA} array configuration 
and project. Columns 4--6: observing bandwidth and date. Column 7: 
observing time. Column 8: radio beam (for ROBUST=0 in IMAGR). Column 9: 
image rms level ($1\sigma$).
\end{table*}


\subsection{Notes on individual clusters}

\begin{figure*}
\centering
\includegraphics[width=8cm]{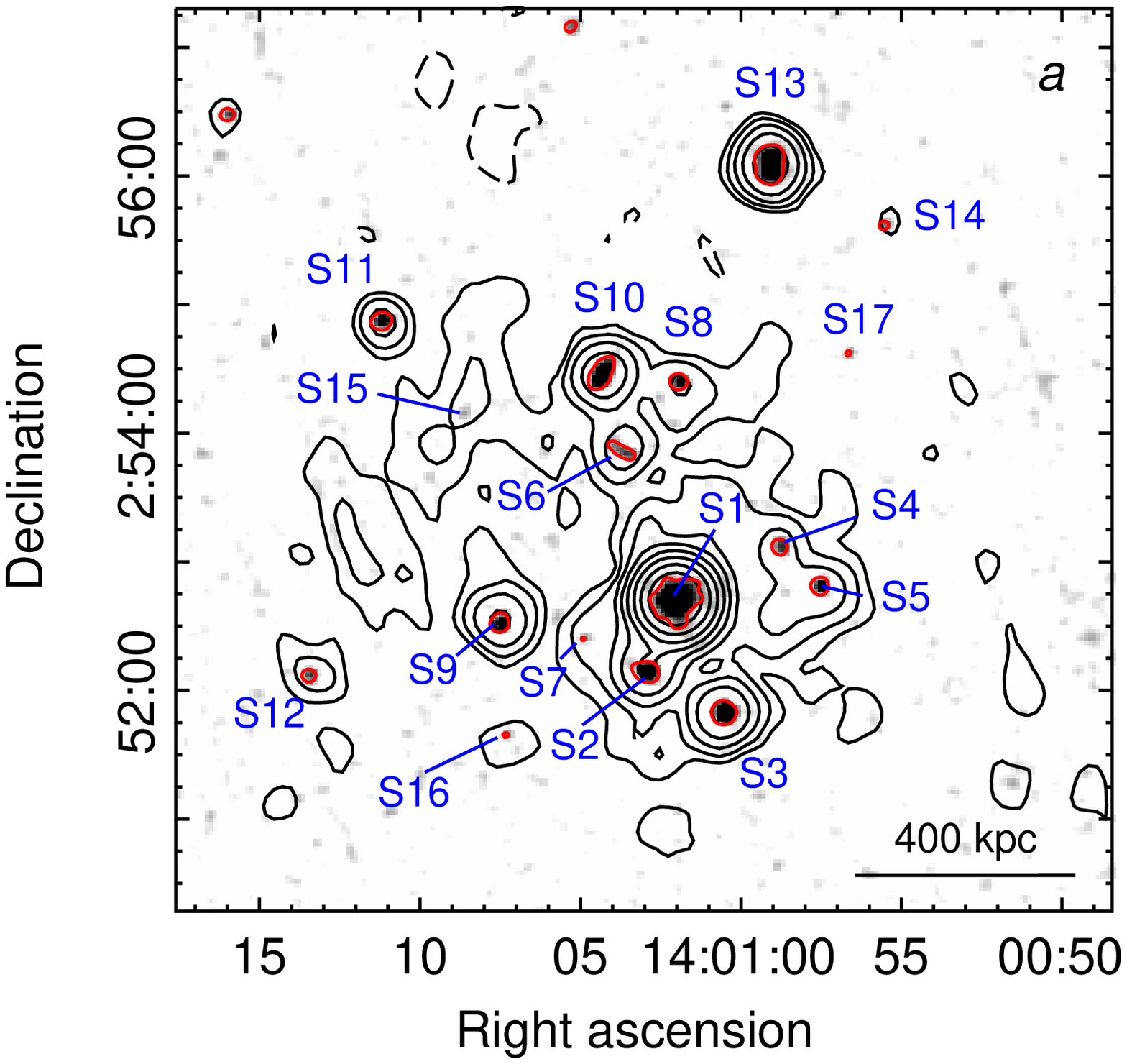}
\includegraphics[width=8cm]{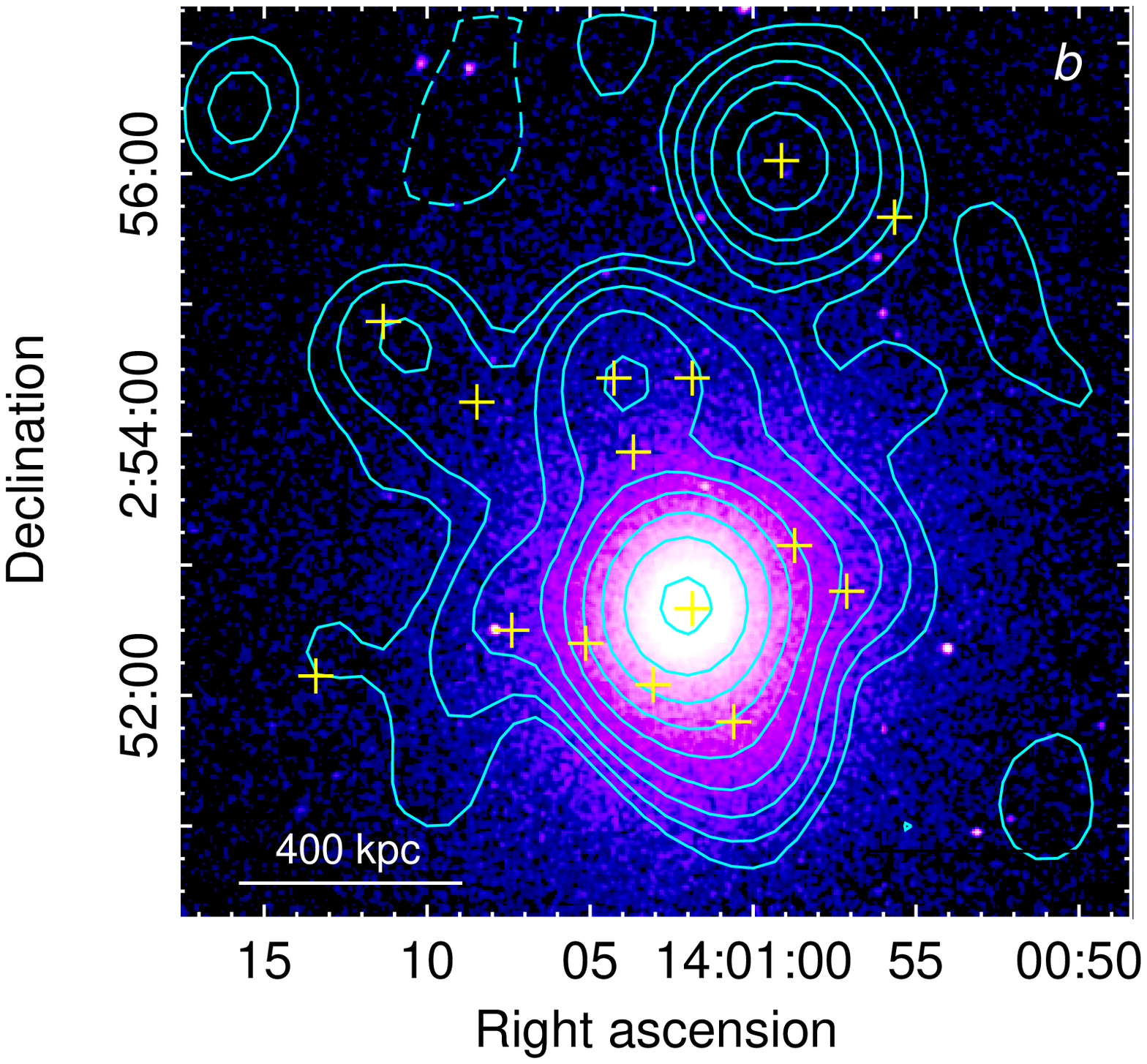}
\smallskip
\caption{A\,1835. ({\em a}) {\em VLA} 1.4 GHz C--configuration contours at
  $20^{\prime\prime}$ resolution (black), overlaid on the 1.4 GHz
  B--configuration image (grey scale and red contours) at $5^{\prime\prime}$
  resolution. The rms noise levels are $1\sigma=30 \, \mu$Jy beam$^{-1}$ and
  $1\sigma=15 \, \mu$Jy beam$^{-1}$, respectively.  Black contours are $-1$
  (dashed), 1, 2, 4, 8, 16, . . . $\times 3\sigma$.  Red contours are at
  $+5\sigma$. Individual discrete radio sources are labeled.  ({\em b}) {\em
    VLA} 1.4 GHz D--configuration contours of the minihalo and embedded
  radio sources (yellow crosses; see panel {\em a}), overlaid on the {\em
    Chandra} X-ray image in the 0.5--4 keV band. The radio image has a
  restoring beam of $51^{\prime\prime}\times50^{\prime\prime}$. The rms
  noise is $1\sigma=40 \, \mu$Jy beam$^{-1}$.  Contours are $-1$ (dashed),
  1, 2, 4, 8, 16, . . . $\times 2.5\sigma$.}
\label{fig:a1835}
\end{figure*}

\smallskip\noindent{\bf A\,1835.} A central minihalo was reported by Govoni
et al.\ (2009).  Our images are presented in Figure ~\ref{fig:a1835}. The
identification of the discrete sources embedded in the minihalo emission was
done using the B--configuration image, shown in ({\em a}). We identified
17 discrete sources above a $5\sigma=0.075$ mJy beam$^{-1}$ level (Table 9),
accounting for a total of $S_{\rm \, 1.4 \, GHz}=44.8$ mJy. 
A global flux of $S_{\rm \,tot, \, 1.4 \, GHz}=50.6\pm2.5$ mJy is
measured on the D--configuration image in the region occupied by the
minihalo (Figure ~\ref{fig:a1835}{\em b}). After subtraction of the flux in
discrete sources, $S_{\rm \,MH, \, 1.4 \, GHz}=6.1\pm1.3$ mJy remain for 
the minihalo, corresponding to a radio luminosity at 1.4 GHz of 
$P_{\rm \,MH, \,1.4 \, GHz} = (1.19\pm0.25)\times10^{24}$ W Hz$^{-1}$. 
Our flux value is in agreement with the flux density reported by 
Govoni et al.\ (2009) using the same C-- and D--configuration observations 
analyzed here. The flux of the BCG is $S_{\rm \,BCG, \, 1.4 \, GHz}=32.4\pm1.6$ mJy
and its radio power $P_{\rm \,MH, \,1.4 \, GHz} = (6.3\pm0.3)\times10^{24}$ W Hz$^{-1}$.

%
%
%
%

\begin{figure*}
\centering
\includegraphics[width=8cm]{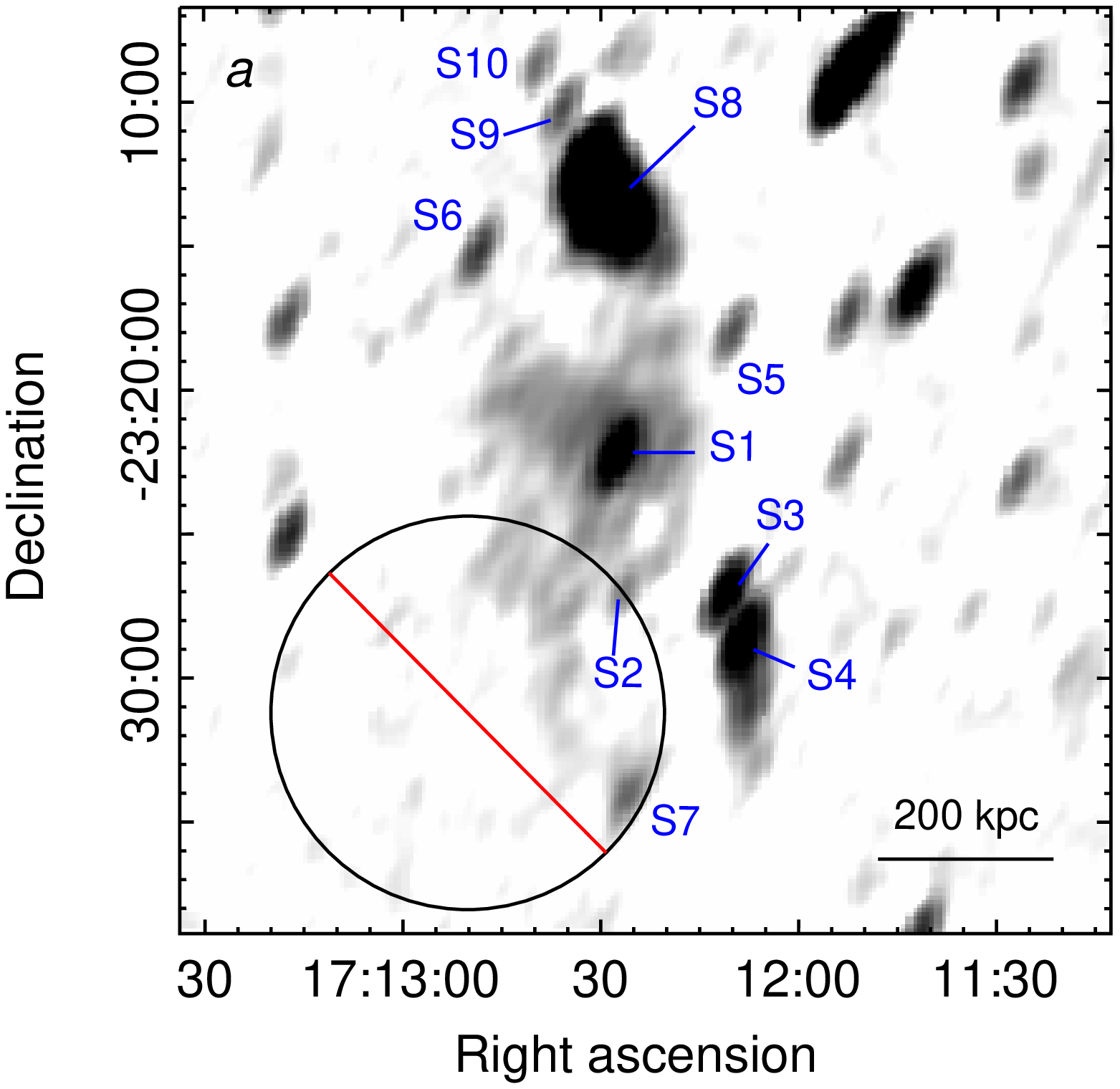}
\includegraphics[width=8cm]{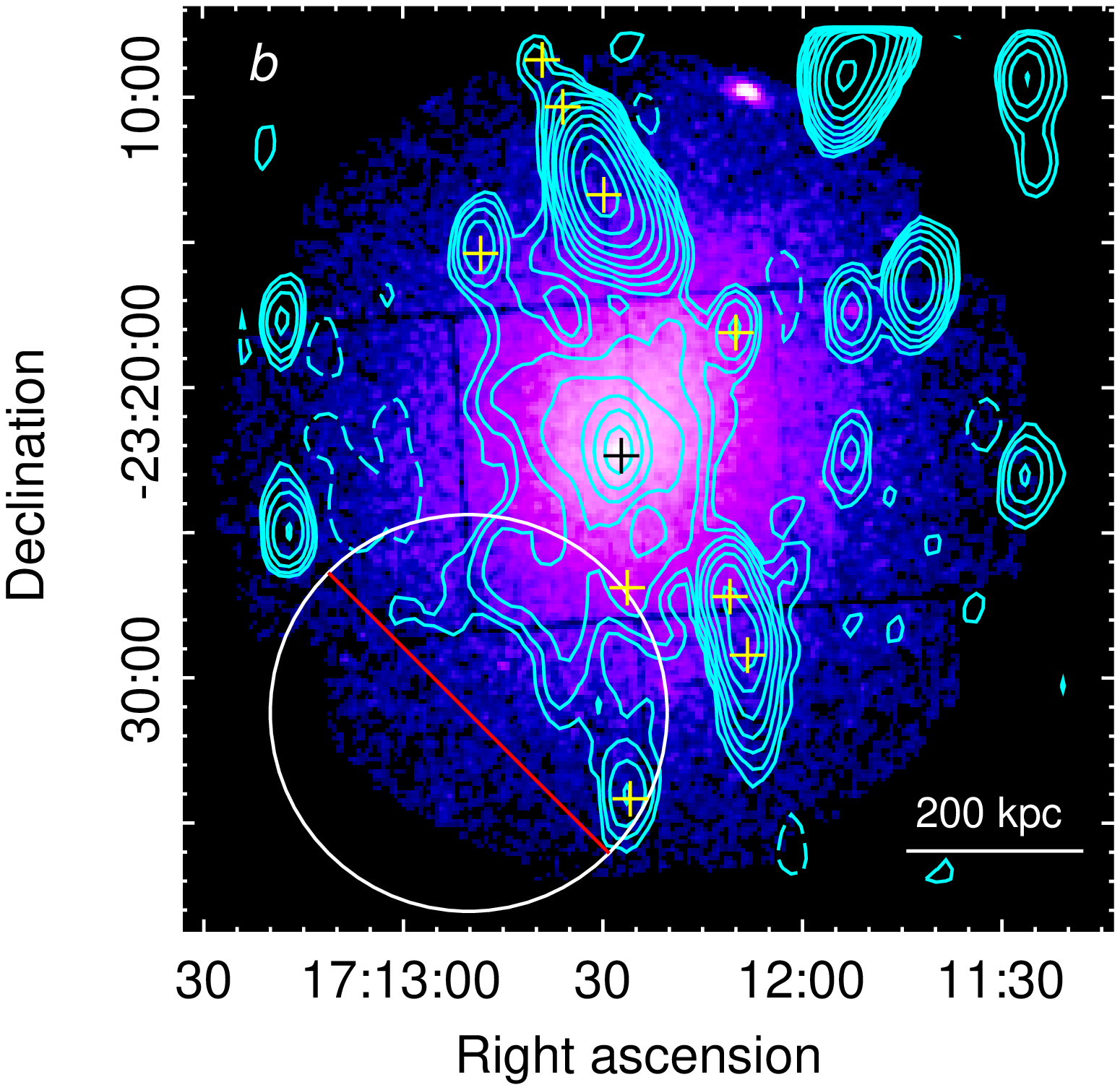}
\smallskip
\caption{Ophiuchus. ({\em a}) {\em VLA} 1.4 GHz D--configuration image. 
  The restoring beam is $92^{\prime\prime}\times39^{\prime\prime}$ and rms
  noise is $1\sigma=100 \, \mu$Jy beam$^{-1}$. The individual radio galaxies
  are labeled. ({\em b}) {\em VLA} 1.4 GHz D--configuration contours of the
  minihalo and radio galaxies (yellow and black crosses; see panel {\em a}),
  overlaid on the {\em XMM-Newton} X-ray image (OBS ID 0505150101).  
The radio image has a
  restoring beam of $110^{\prime\prime}\times60^{\prime\prime}$.  The rms
  noise is $1\sigma=120 \, \mu$Jy beam$^{-1}$.  Contours are $-1$ (dashed),
  1, 2, 4, 8, 16, . . . $\times 2.5\sigma$.  The region associated with the
low-frequency relic lobe that fills a possible X-ray cavity (circle; 
Giacintucci et al.\ 2019) has been excluded to measure the flux density of 
the minihalo.}
\label{fig:oph}
\end{figure*}

\smallskip\noindent{\bf Ophiuchus.}  A central minihalo was reported by
Govoni et al.\ (2009) and Murgia et al.\ (2010). Our images are presented in
Figure ~\ref{fig:oph}.  We identified 10 discrete radio sources in the
minihalo area (Table 9) using an image at $92^{\prime\prime}\times39^{\prime\prime}$
resolution, shown in ({\em a}). These sources account for a total flux
of $S_{\rm \, 1.4 \, GHz}=733.8$ mJy. To measure the minihalo flux density, 
we used a lower resolution image, shown as contours in ({\em b}) on the 
{\em XMM-Newton} X-ray image. We subtracted the contribution from discrete sources.  
We also masked the region marked by a circle, which coincides with an 
X-ray cavity suggested by the {\em Chandra} and {\em XMM-Newton} images 
of the cluster core (Werner et al.\ 2016, Giacintucci et al.\ 2019). 
The cavity is filled by a very steep spectrum relic lobe, likely inflated 
by a previous outburst of central AGN, discovered in {\em GMRT} and {\em MWA} 
images at lower frequencies (Giacintucci et al.\ 2019). The relic lobe is only 
partially visible at 1.4 GHz (Figure ~\ref{fig:oph}{\em b}). We obtained a 
total flux of  $S_{\rm \,MH, \,1.4 \, GHz}=62\pm9$ mJy for the minihalo
($P_{\rm \,MH, \,1.4 \, GHz} = (0.11\pm0.02)\times10^{24}$ W Hz$^{-1}$). 
A higher flux of $106.4^{+10.4}_{-8.9}$ mJy has been
measured by Govoni et al. (2009) using an exponential model to fit the
surface brightness profile of the minihalo. A flux density of $85\pm3$ mJy
is reported by the same authors by integration within a radius of $\sim 230$
kpc. We note, however, that both these measurements include part of the
emission from the newly detected relic lobe. The flux density 
of the BCG is $S_{\rm \,BCG, \, 1.4 \, GHz}=30.6\pm1.5$ mJy
and its radio power $P_{\rm \,MH, \,1.4 \, GHz} = (0.064\pm0.003)\times10^{24}$ W Hz$^{-1}$.

\begin{figure*}
\centering
\hspace{-1cm}
\includegraphics[width=8cm]{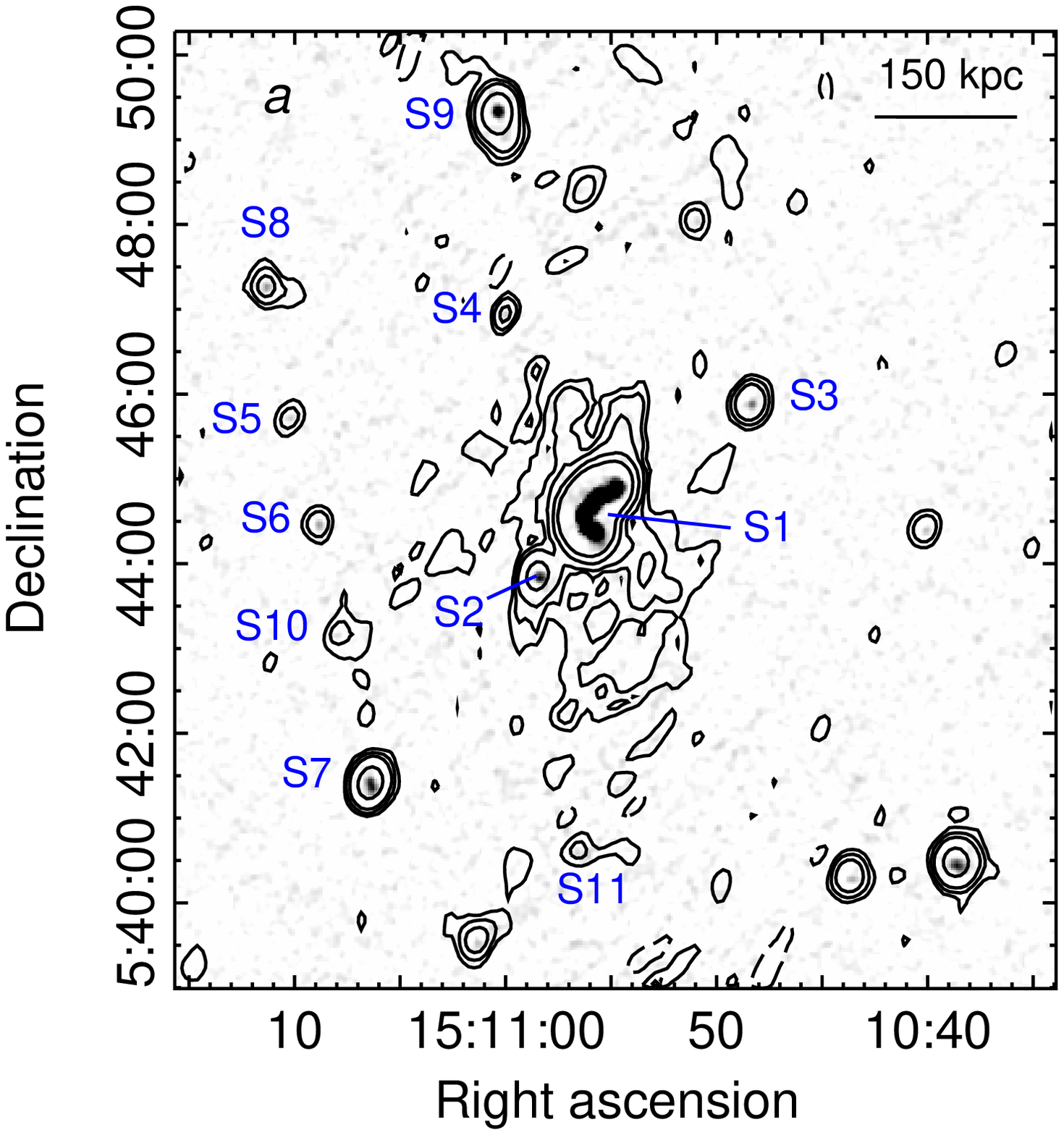}
\includegraphics[width=8cm]{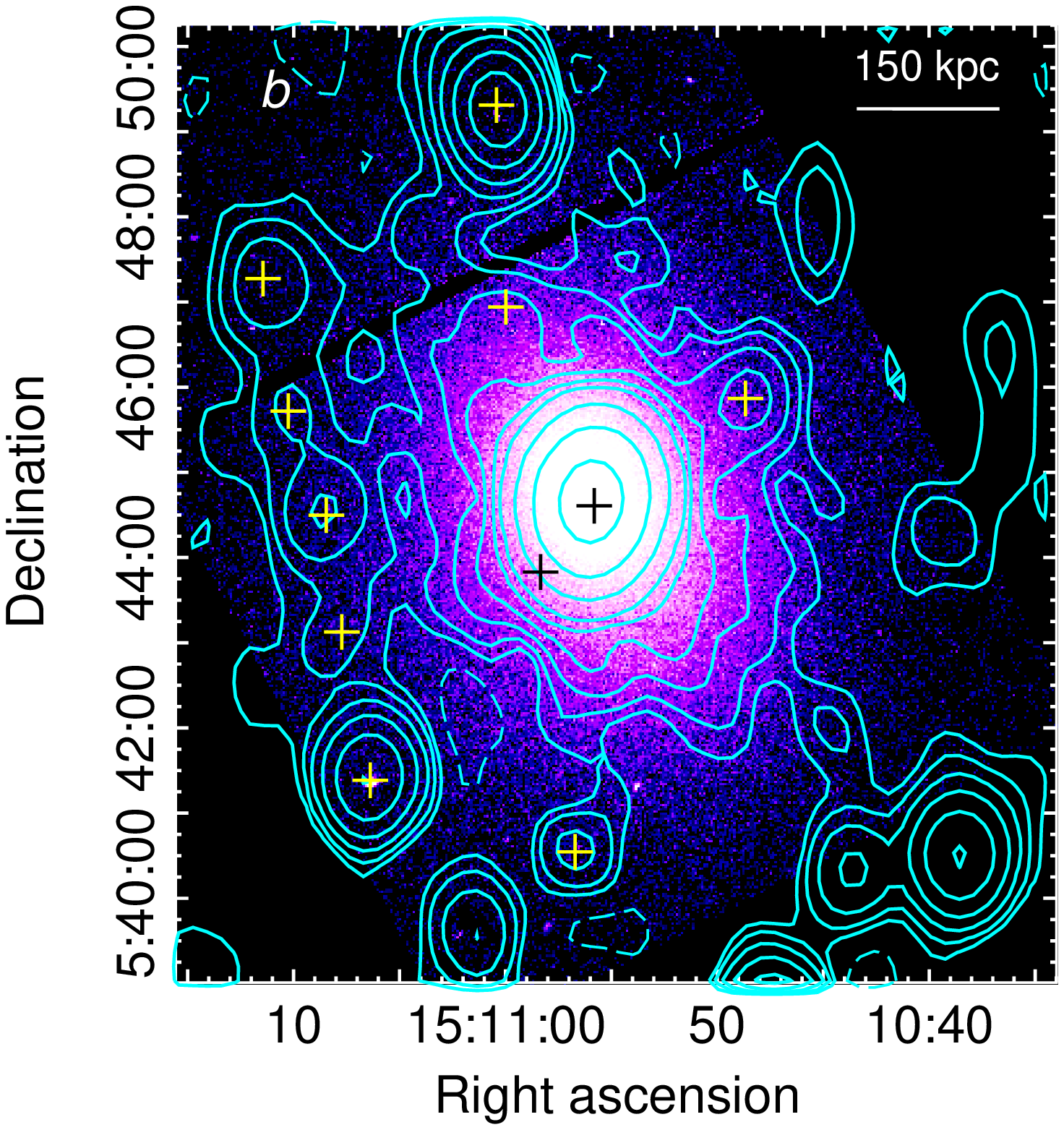}
\smallskip
\caption{A\,2029. ({\em a}) {\em VLA} 1.4 GHz C--configuration 
  contours, overlaid on the B--configuration image.  The restoring beam and
  rms noise are $20^{\prime\prime}\times16^{\prime\prime}$ and $1\sigma=30
  \, \mu$Jy beam$^{-1}$ and $5^{\prime\prime}\times4^{\prime\prime}$ and
  $1\sigma=50 \, \mu$Jy beam$^{-1}$, respectively. Contours are $-0.1$
  (dashed), 0.1, 0.2, 0.4, 1.6, 6.4 mJy beam$^{-1}$. The individual radio
  galaxies are labeled. ({\em b}) {\em VLA} 1.4 GHz D--configuration
  contours of the minihalo and radio galaxies (yellow and black crosses; see
  panel {\em a}), overlaid on the {\em Chandra} image in the 0.5-4 keV band.  
The radio image has a restoring beam of
  $57^{\prime\prime}\times45^{\prime\prime}$.  The rms noise is $1\sigma=40
  \, \mu$Jy beam$^{-1}$.  Contours are $-0.12$ (dashed), 0.12, 0.24, 0.48,
  0.96, 2, 4, 16, 64, 256 mJy beam$^{-1}$.}
\label{fig:2029}
\end{figure*}

\smallskip\noindent{\bf A2029}. A central minihalo was reported by Govoni et
al.\ (2009). Our images are presented in Figure ~\ref{fig:2029}. We
identified 11 discrete radio sources (Table 9), including the central dominant 
wide angle tail (e.g., Taylor et al.\ 1994), using the B-- and C--configuration
images ({\em a}). The discrete source account for a total of $S_{\rm \, 1.4 \, GHz}=531.2$ mJy. 
We used the D-configuration observation to map the minihalo ({\em b}) and measure 
its flux density. Our value of $S_{\rm \,MH, \,1.4 \, GHz}=19.5\pm2.5$ mJy, 
after removal of the flux in discrete sources, is in agreement within the errors
with $18.8\pm1.3$ mJy measured by Govoni et al.\ (2009). The 1.4 GHz 
radio luminosity of the minihalo is $P_{\rm \,MH, \,1.4 \, GHz}= (0.28\pm0.04)\times10^{24}$ W Hz$^{-1}$. The flux density 
of the BCG is $S_{\rm \,BCG, \, 1.4 \, GHz}=513\pm26$ mJy and its radio power 
$P_{\rm \,BCG, \,1.4 \, GHz} = (7.4\pm0.4)\times10^{24}$ W Hz$^{-1}$.

%
%
%
%

%
%
%
%

\begin{figure*}
\centering
\includegraphics[width=8cm]{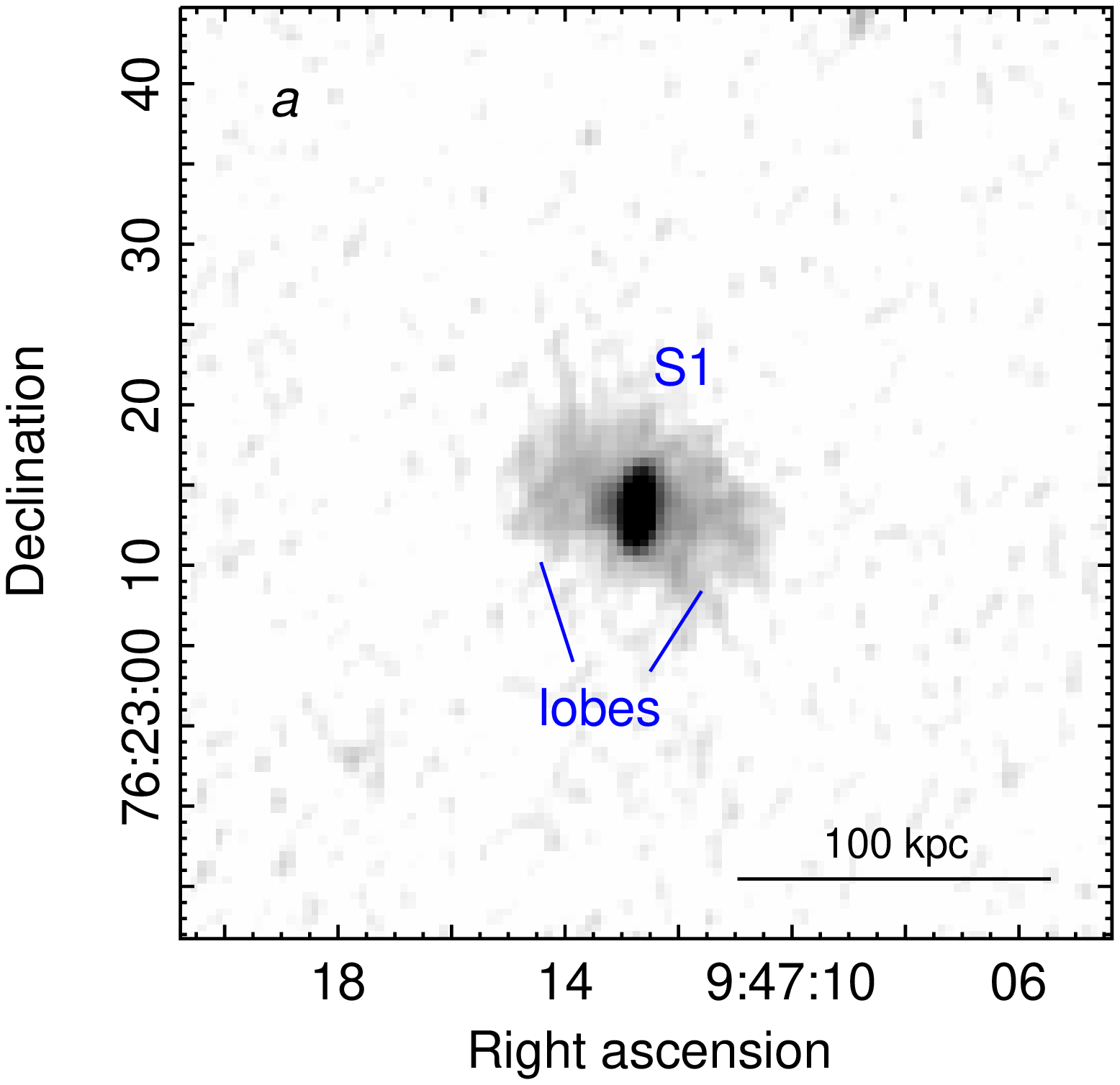}
\includegraphics[width=8cm]{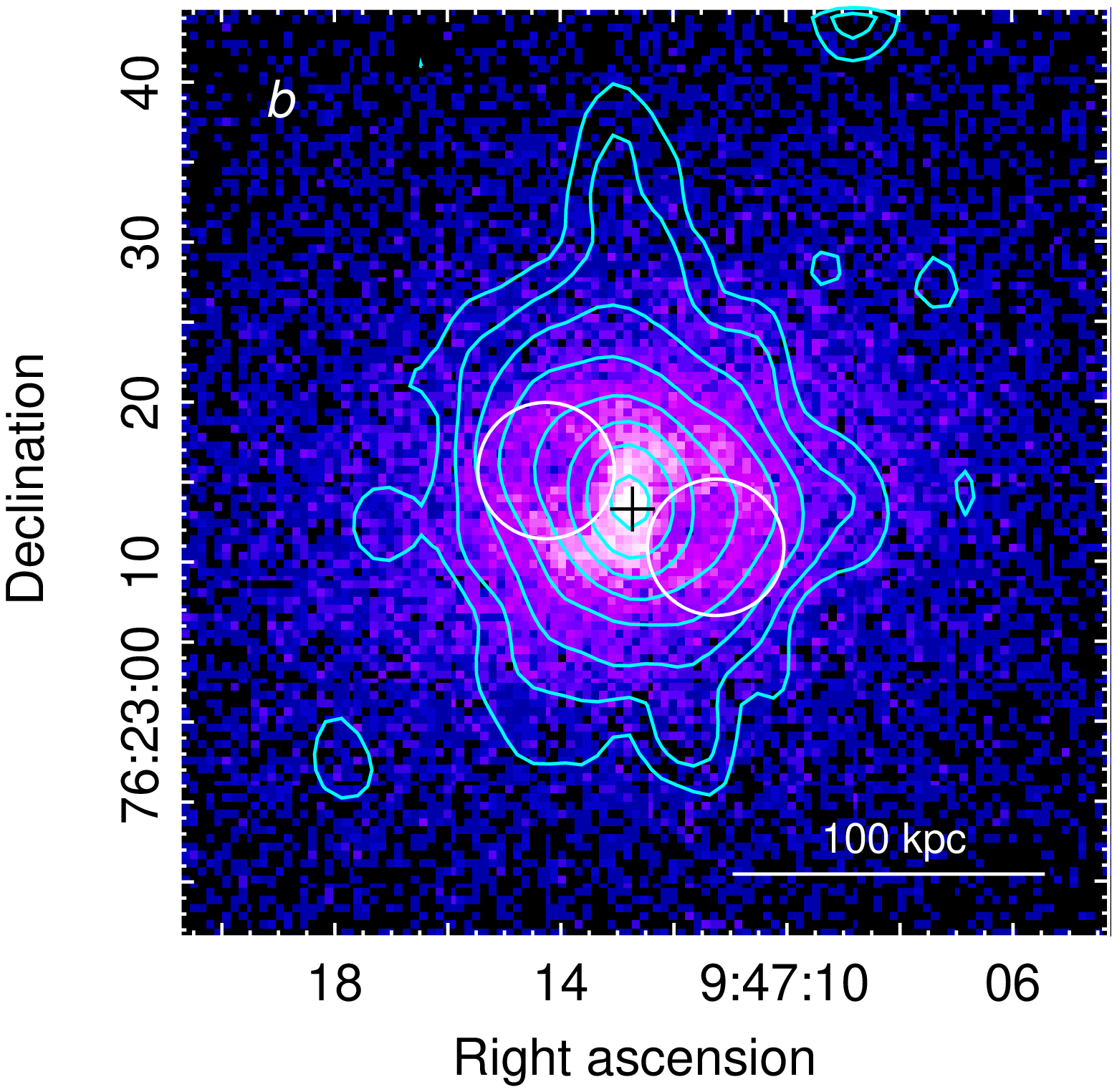}
\smallskip
\caption{RBS\,797. ({\em a}) {\em VLA} 1.4 GHz A--configuration image,
  restored with a $2^{\prime\prime}\times1^{\prime\prime}$ beam. The rms
  noise is $1\sigma=17 \, \mu$Jy beam$^{-1}$. The central radio galaxy S1
  and the radio lobes filling the X-ray cavities (white circles in panel
  {\em b}) are labeled. ({\em b}) {\em VLA} 1.4 GHz B--configuration
  contours overlaid on the {\em Chandra} image in the 0.5--4 keV band.  The
  radio image has a restoring beam of
  $5^{\prime\prime}\times4^{\prime\prime}$.  The rms noise is $1\sigma=27 \,
  \mu$Jy beam$^{-1}$. Contours start at $70 \, \mu$Jy beam$^{-1}$ and then
  scale by a factor of 2. No levels at $-3\sigma$ are present in the portion
  of the image shown. Circles mark the X-ray cavities filled by the
  radio lobes ({\em a}).}
\label{fig:rbs797}
\end{figure*}

\smallskip\noindent{\bf RBS\,797.} A central minihalo was reported by Gitti
et al.\ (2006) and Doria et al.\ (2012). Our images are presented in Figure
~\ref{fig:rbs797}. Using the A--configuration image ({\em a}), we measure a
radio flux density of $S_{\rm \,BCG, \, 1.4 \, GHz}=17.9\pm0.9$ mJy 
($P_{\rm \,BCG, \,1.4 \, GHz} = (7.6\pm0.4)\times10^{24}$ W Hz$^{-1}$) for the BCG (S1), 
which includes the central compact component and the two radio lobes associated 
with a pair of X-ray cavities (circles in panel {\em b}; see  
Gitti et al.\ 2006, Doria et al.\ 2012). The flux density of whole radio emission 
detected by the B--configuration image ({\em b}) is $S_{\rm \,tot, \, 1.4 \, GHz}=23.1\pm1.1$ mJy, 
in agreement within the errors with the $24\pm0.3$ mJy reported by Doria et al.\ (2012). 
Once removed the BCG flux from the total emission, we find a residual flux of
$S_{\rm \,MH, \, 1.4 \, GHz}=5.2\pm0.6$ mJy for the minihalo (a flux density of $6.1\pm0.4$ mJy 
was estimated by Doria et al.), corresponding to 
$P_{\rm \,MH, \,1.4 \, GHz}= (2.20\pm0.24)\times10^{24}$ W Hz$^{-1}$.

%
%
%
%

\begin{figure*}
\centering
\includegraphics[width=8cm]{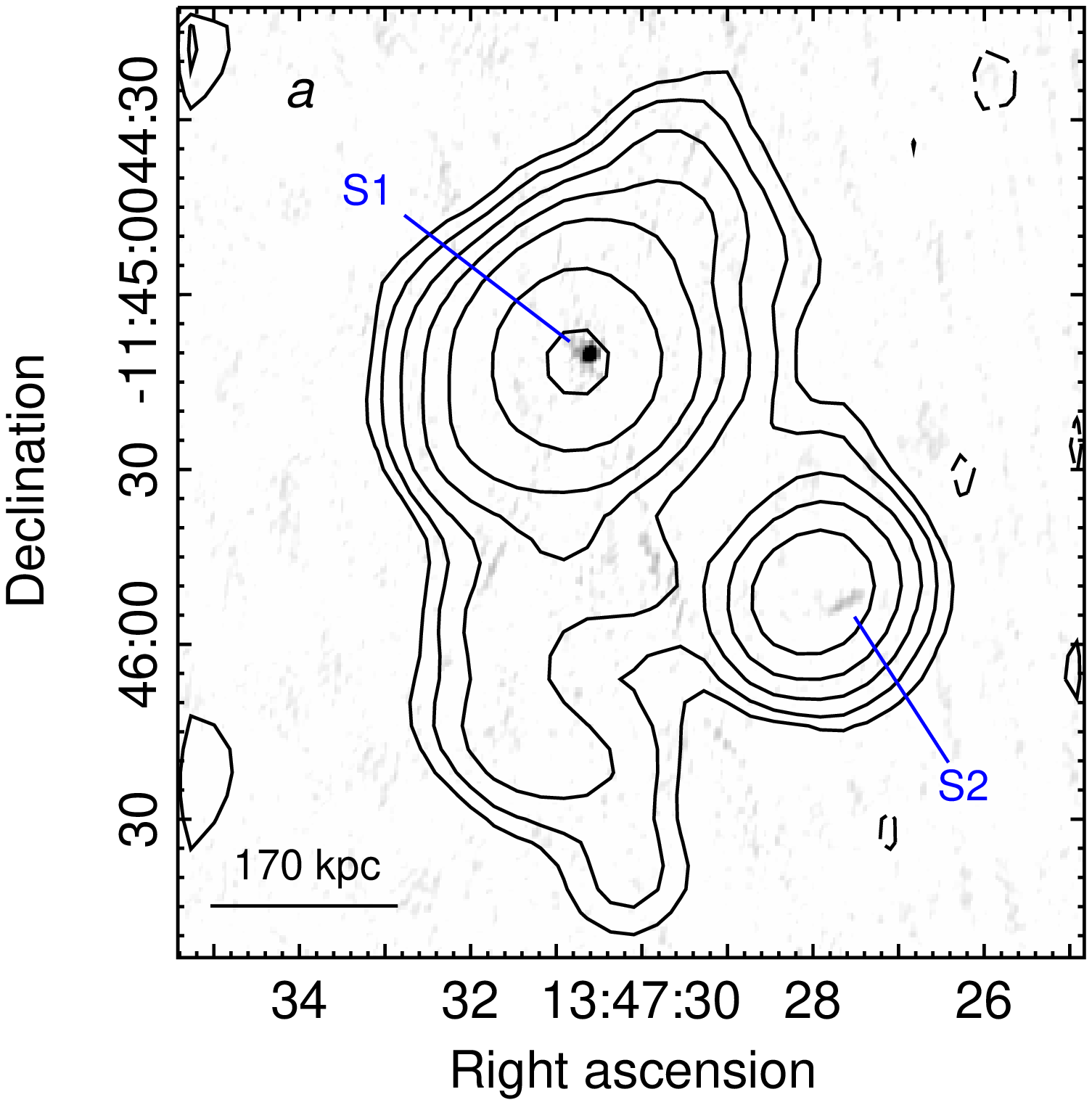}
\includegraphics[width=8cm]{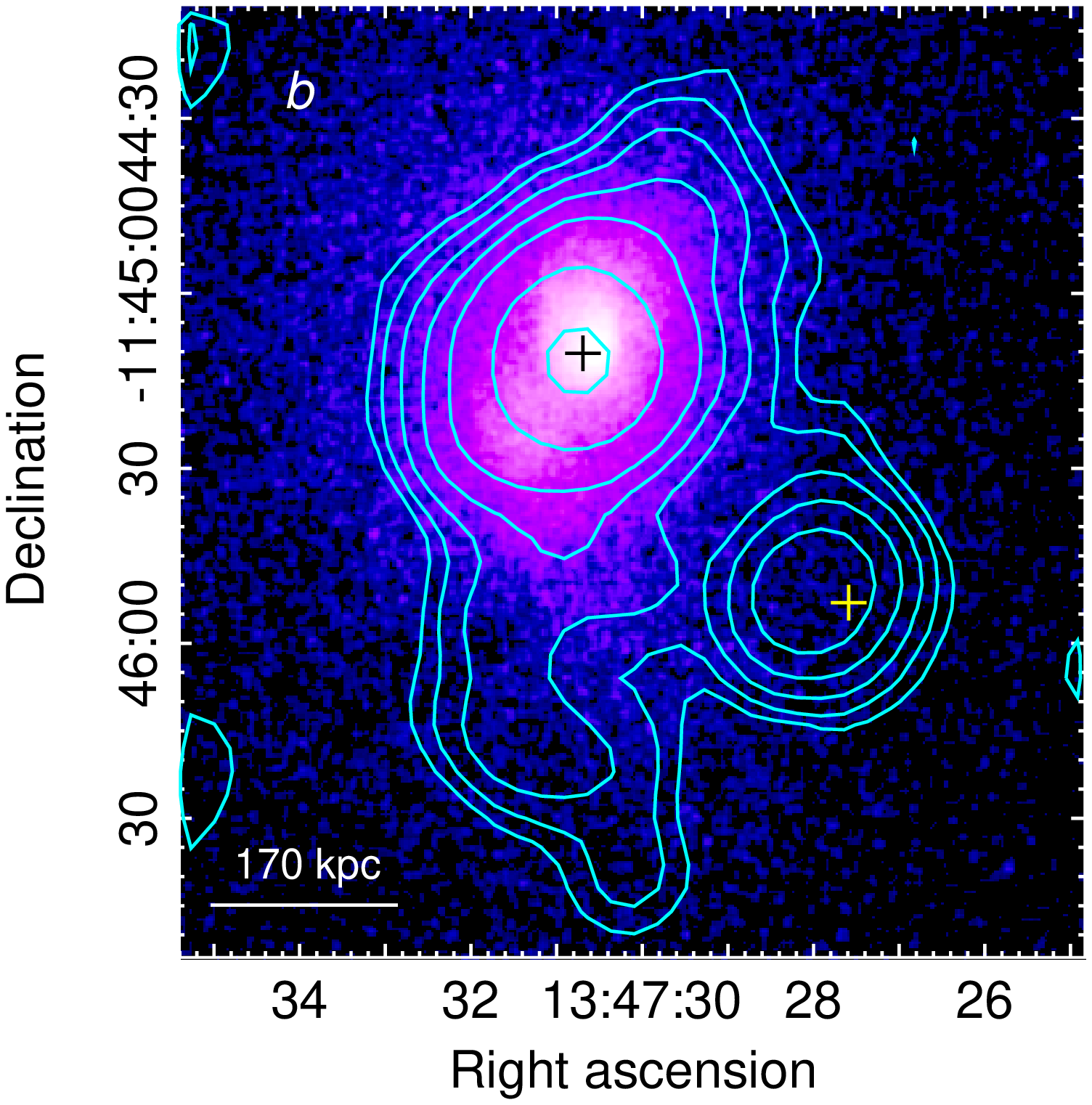}
\smallskip
\caption{RX\,J1347.5--1145. ({\em a}) {\em VLA} 1.4 GHz C--configuration 
  contours, overlaid on the 1.4 GHz A--configuration image. The restoring
  beam and rms noise are $17^{\prime\prime}$ and $1\sigma=40$ $\mu$Jy
  beam$^{-1}$ and $1^{\prime\prime}.2$ and $1\sigma=35 \, \mu$Jy
  beam$^{-1}$, respectively.  Contours are $-0.12$ (dashed), 0.12, 0.24,
  0.48, 0.96, 2, 8, 32, 128 mJy beam$^{-1}$. Individual radio galaxies are
  labeled.  ({\em b}) {\em VLA} 1.4 GHz C--configuration contours of the
  minihalo and radio galaxies (black and yellow crosses; see panel {\em a}),
  overlaid on the {\em Chandra} image in the 0.5-4 keV band.}
\label{fig:rxj1347}
\end{figure*}

\smallskip\noindent{\bf RX\,J1347.5--1145}. A central minihalo was reported
by Gitti et al.\ (2007).  Our images are presented in Figure
~\ref{fig:rxj1347}. The central radio galaxy is a point source in the 1.4
GHz image from the A configuration ({\em a}). Its flux density is
$S_{\rm \,BCG, \, 1.4 \, GHz}=30.3\pm1.5$ mJy and its luminosity is
$P_{\rm \,BCG, \,1.4 \, GHz} = (22.7\pm1.1)\times10^{24}$ W Hz$^{-1}$. 
A much fainter and extended source (S2) is also detected, with a flux of 
$S_{\rm \,S2, \,1.4 \, GHz}=2.0\pm0.1$ 
mJy within the $3\sigma$ isocontour. Its flux density increases to 
$4.9\pm0.2$ mJy 
in the lower-resolution image from the C configuration, where the source becomes 
unresolved. This flux difference
is likely due to missing extended emission in the A--configuration image.
Similar flux density values for S1 and S2 were measured by Gitti et al.\ 
(2007) using the same radio data analyzed here.  The total flux in the
C--configuration image is $S_{\rm \,tot, \, 1.4 \, GHz}=69.3\pm3.5$ mJy,
measured out to a radius of $r=80^{\prime\prime}$, where the integrated 
flux reaches saturation (\S\ref{sec:obs}). The total flux
reported by Gitti et al.\ (2007) is $\sim 25\%$ lower than our measured
flux. However, if we integrate the flux density within the $3\sigma$
isocontour of our C--configuration image, we obtain $60\pm3$ mJy, which is
consistent within $1.5\sigma$ with Gitti et al.\ (2007). After removal of S1 and S2
from the total flux, we estimate $S_{\rm \,MH, \, 1.4 \, GHz}=35.6\pm2.9$ mJy 
for the minihalo. The
error includes the uncertainty in the subtraction of S2 using its flux from
the A-- or C--configuration images. The minihalo luminosity at 1.4 GHz is
$P_{\rm \,MH, \,1.4 \, GHz}= (26.7\pm2.1)\times10^{24}$ W Hz$^{-1}$.

%
%
%
%

\begin{figure*}
\centering
\includegraphics[width=8cm]{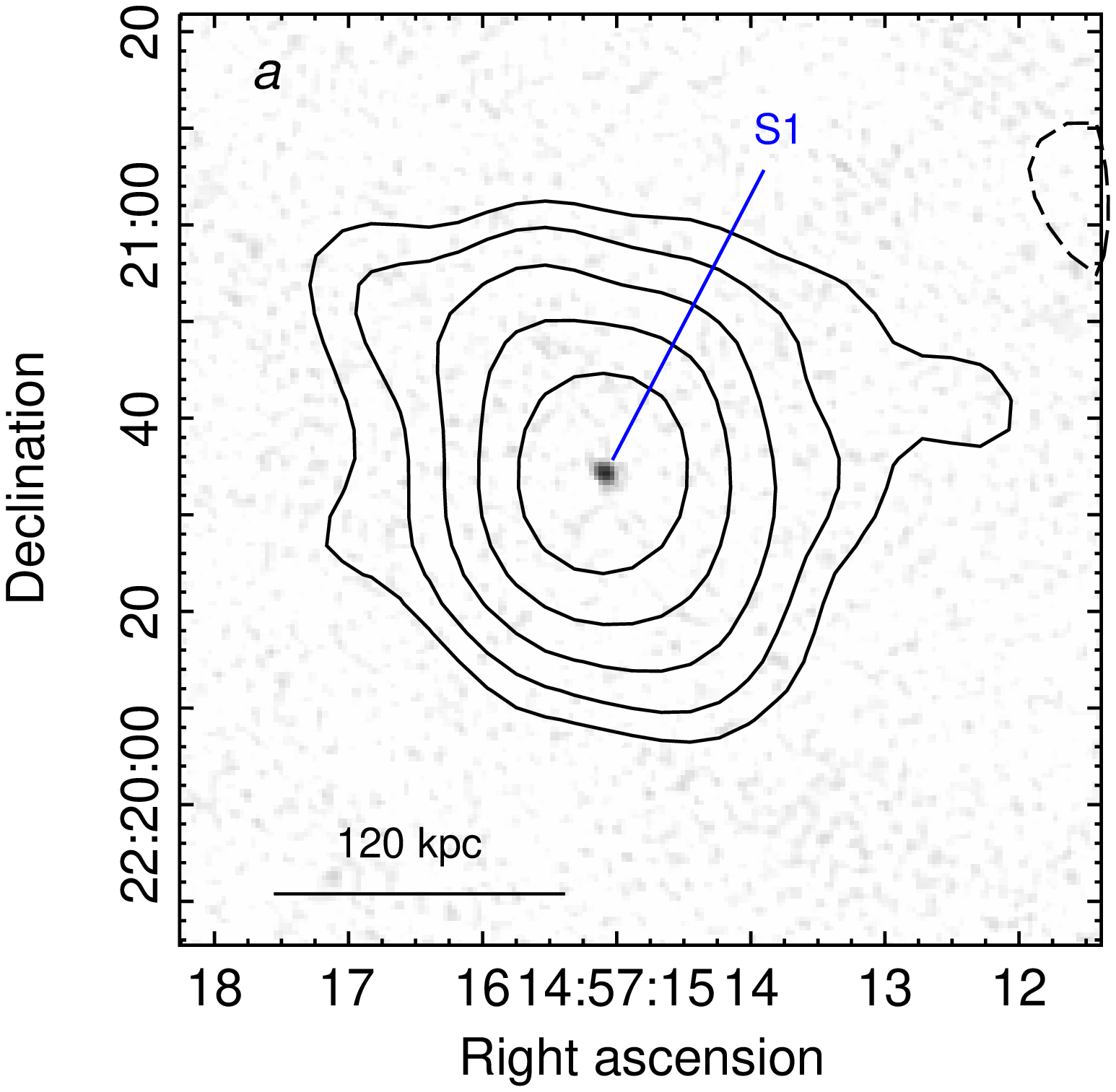}
\includegraphics[width=8cm]{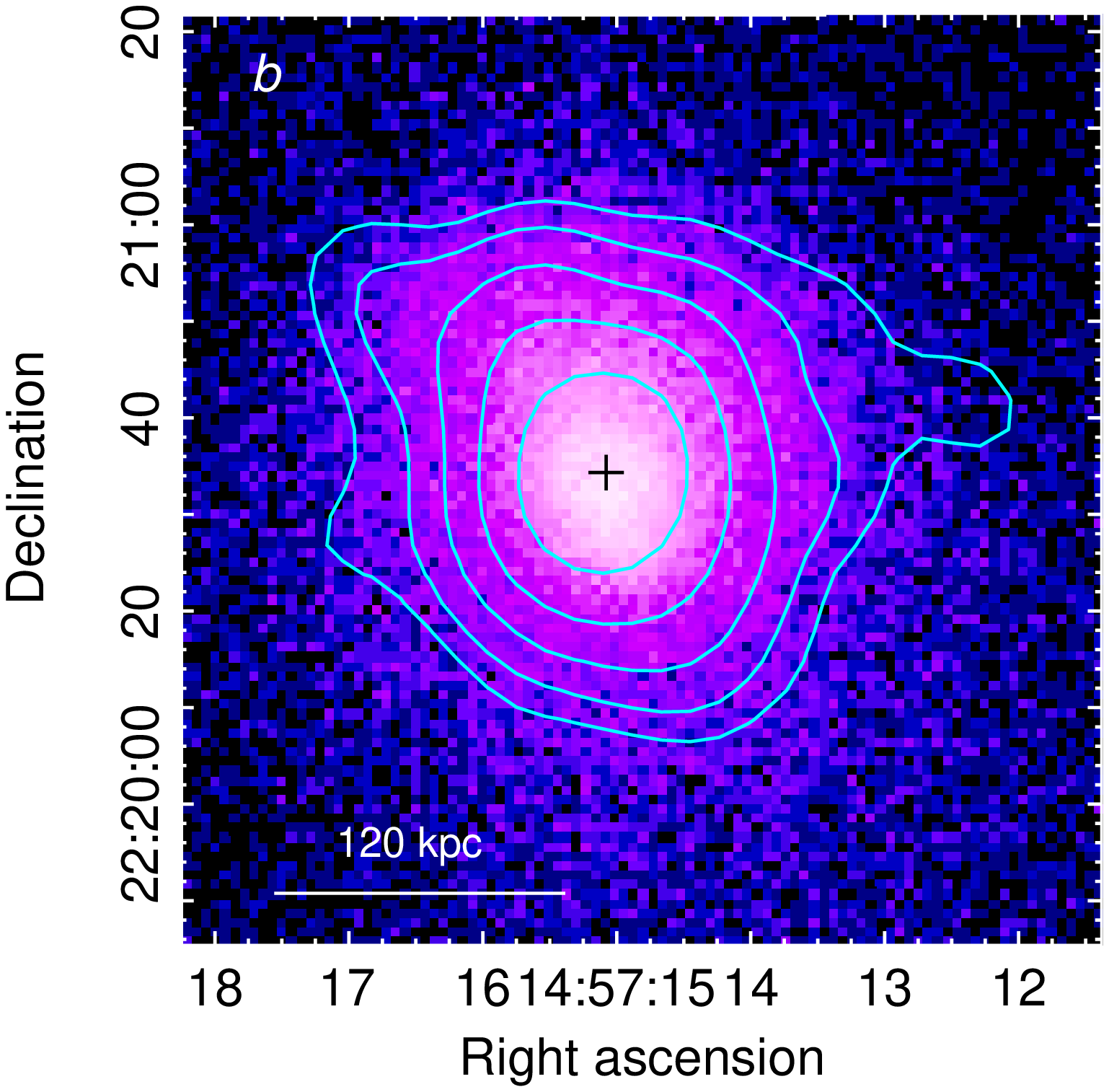}
\smallskip
\caption{MS\,1455.0+2232. ({\em a}) {\em VLA} 1.4 GHz C--configuration 
contours, overlaid on the 1.4 GHz A--configuration image. 
The restoring beam and rms noise are $16^{\prime\prime}\times15^{\prime\prime}$ 
and $1\sigma=70$ $\mu$Jy beam$^{-1}$ and $1^{\prime\prime}.5\times1^{\prime\prime}.2$ 
and $1\sigma=55 \, \mu$Jy beam$^{-1}$, respectively. Contours are $-1$ (dashed), 
1, 2, 4, 8, 16, ... $\times 3\sigma$. S1 is the central radio galaxy.
({\em b}) {\em VLA} 1.4 GHz C--configuration contours of the minihalo and radio galaxy 
(black cross), overlaid on the {\em Chandra} image in the 0.5-4 keV band.}
\label{fig:ms1455}
\end{figure*}

\smallskip\noindent{\bf MS\,1455.0+2232 (Z7160)}. The minihalo, 
first detected by Venturi et al.\ (2008) with the {\em GMRT} at 610 MHz, 
is only marginally extended in the {\em VLA} C--configuration image 
at 1.4 GHz (Figure ~\ref{fig:ms1455}). The central compact radio galaxy 
has $S_{\rm \,BCG, \, 1.4 \, GHz}=4.7\pm0.2$ mJy in the A--configuration image,
corresponding to $P_{\rm \,BCG, \,1.4 \, GHz} = (0.96\pm0.05)\times10^{24}$ W Hz$^{-1}$. 
The total integrated flux density in the C--configuration image is $S_{\rm \,tot, \, 1.4 \, GHz}=13.2\pm0.7$ mJy, the 
minihalo accounting for $S_{\rm \,MH, \, 1.4 \, GHz}=8.5\pm1.1$ mJy
($P_{\rm \,MH, \,1.4 \, GHz}= (1.75\pm0.23)\times10^{24}$ W Hz$^{-1}$). The BCG 
and minihalo flux densities at 610 MHz are $S_{\rm \,BCG, \, 610 \, MHz}=9.5\pm0.8$ 
mJy and $S_{\rm \,MH, \, 610 \, MHz}=28.7\pm3.7$ mJy, respectively (Mazzotta \& Giacintucci 2008). 
Their spectral indices between 610 MHz and 1.4 GHz are $\alpha=0.85\pm0.11$ and $\alpha=1.46\pm0.22$.

%
%
%
%

\begin{figure*}
\centering
\includegraphics[width=8cm]{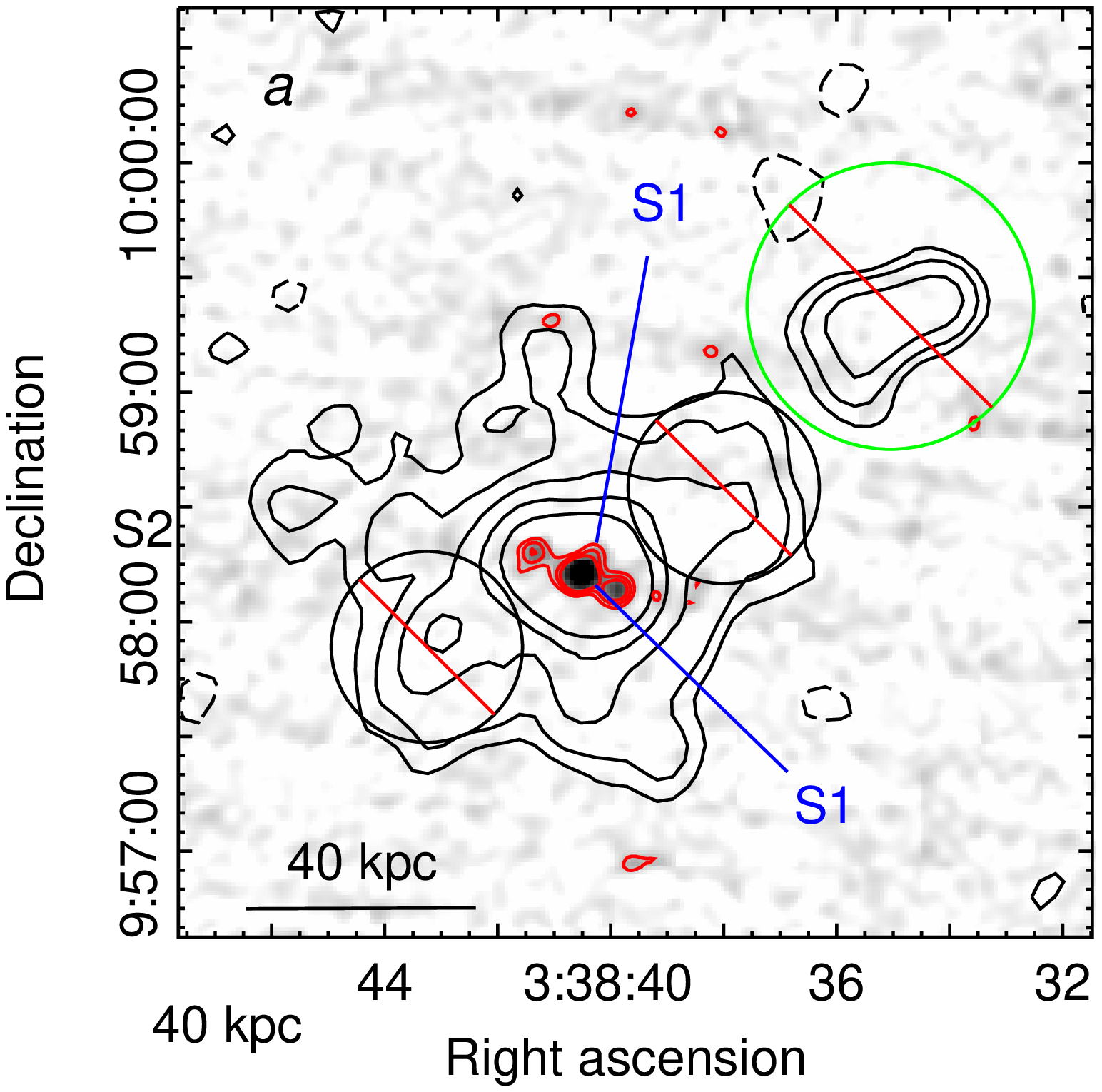}
\includegraphics[width=8cm]{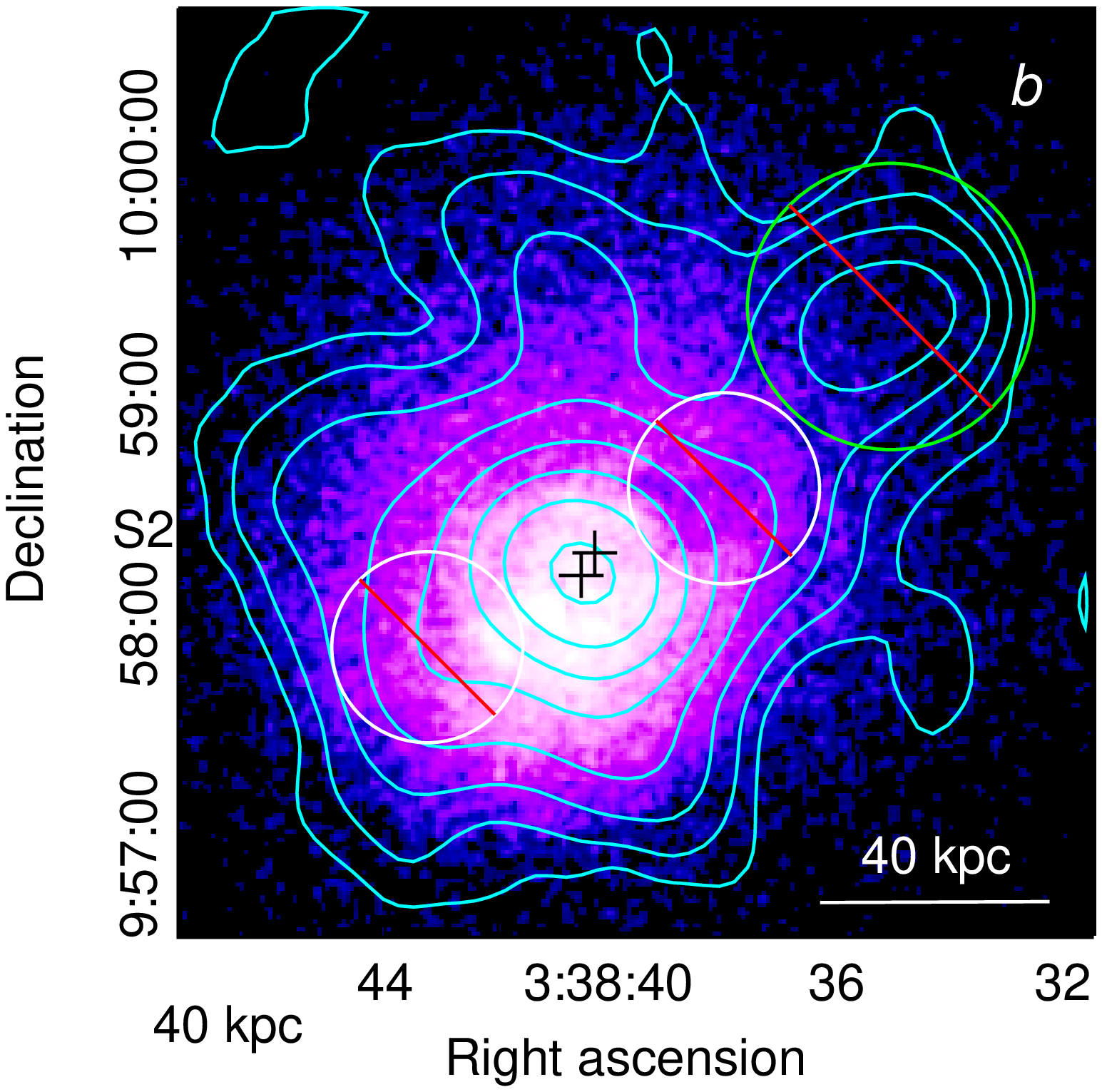}
\smallskip
\caption{2A\,0335+096. ({\em a}) {\em VLA} 1.4 GHz C--configuration 
  contours, overlaid on the 1.4 GHz B--configuration image (grey scale and
  red contours at $+3, 6$, and $10 \sigma$). The restoring beam and rms
  noise are $15^{\prime\prime}\times14^{\prime\prime}$ and $1\sigma=35$ mJy
  beam$^{-1}$ and $4^{\prime\prime}$ and $1\sigma=58 \, \mu$Jy beam$^{-1}$,
  respectively.  Black contours are $-1$ (dashed), 1, 2, 4, 8, 16, ...
  $\times 3\sigma$.  The central radio galaxy and nearby companion are
  labeled S1 and S2. Black circles mark the position of two X-ray
  cavities filled by fossil radio lobes (Giacintucci et al. in preparation).
The green circle marks the location of another possible fossil radio lobe. ({\em b}) {\em
    VLA} 1.4 GHz C+D--configuration contours of the minihalo and radio
  galaxies (black crosses; see panel {\em a}), overlaid on the {\em Chandra}
  image in the 0.5-4 keV band. The radio image has a restoring beam of
  $30^{\prime\prime}\times27^{\prime\prime}$. The rms noise is $1\sigma=40
  \, \mu$Jy beam$^{-1}$. Contours are 1, 2, 4, 8, 16, . . . $\times
  3\sigma$.  No levels at $-3\sigma$ are present in the portion of the image
  shown.  The region occupied by the X-ray cavities and fossil lobes (white
and green circles) have been excluded to measure the flux
  density of the minihalo.}
\label{fig:2a0335}
\end{figure*}

\smallskip\noindent{\bf 2A\,0335+096}. A central minihalo was reported by
Sarazin et al.\ (1995). Our 1.4 GHz images are presented in Figure ~\ref{fig:2a0335}. 
The central radio galaxy (S1) is a core-dominated, double-lobe source with a 
flux of $S_{\rm \,BCG, \, 1.4 \, GHz}=16\pm1$ mJy in the B--configuration image ({\em a}). 
Its radio power is  $P_{\rm \,BCG, \,1.4 \, GHz} = (0.058\pm0.003)\times10^{24}$ W Hz$^{-1}$.
A nearby companion (S2) has $S_{\rm \,S2, \, 1.4 \, GHz}=1.0\pm0.1$ mJy. 
Low-frequency {\em GMRT} images at 235 MHz 
and 610 MHz reveal the presence of two fossil radio lobes associated with 
a previous outburst of the central AGN, opposite with respect to S1 and aligned 
along a NorthWest-SouthEast axis (a paper is in preparation; see talk 
by Giacintucci at the Snowcluster 2018 conference\footnote{http://www.physics.utah.edu/snowcluster/archive/2018/talks/Giacintucci.pdf.}
for an image at 610 MHz). Their position is marked by white circles in 
Figure ~\ref{fig:2a0335}{\em b}. The North-West lobe fills a prominent X-ray 
cavity seen in the {\em Chandra} image (Mazzotta et al.\ 2003, Sanders et al.\ 2009). 
A distinct patch of extended emission is detected at $\sim 25^{\prime\prime}$ ($\sim 18$ kpc) 
from the cluster center (green circle). This feature becomes more prominent at lower 
frequencies due to its very steep radio spectrum and is interpreted as a fossil 
lobe from an even older AGN outburst. We measure a total flux of $S_{\rm \,tot, \, 1.4 \, GHz}=48\pm2.$ mJy in the 
combined C+D--configuration image ({\em b}). To estimate the 1.4 GHz flux density of 
the minihalo, we subtract the flux of S1 and S2 from the total emission as well 
as the flux in the fossil lobes (SE lobe: $4.1\pm0.2$ mJy, NW lobe: $2.7\pm0.1$ mJy, 
outer NW lobe: $3.1\pm0.2$ mJy). We obtain $S_{\rm \,MH, \, 1.4 \, GHz}=21.1\pm2.1$ mJy
and $P_{\rm \,MH, \,1.4 \, GHz}= (0.06\pm0.01)\times10^{24}$ W Hz$^{-1}$.

%
%
%
%

\clearpage

\startlongtable
\begin{deluxetable*}{lcc|lcc}
\tabletypesize{\scriptsize}  
\tablecaption{Flux densities of the discrete radio sources and minihalos
\label{tab:re_sources}}
 \tablehead{
    \colhead{Cluster}   & \colhead{Source} & \colhead{$S_{\rm 1.4 \, GHz}$} & \colhead{Cluster} & \colhead{Source} & \colhead{$S_{\rm 1.4 \, GHz}$}  \\
      \colhead{}        &    \colhead{}    &   \colhead{(mJy)}            &  \colhead{}      &  \colhead{}      & \colhead{(mJy)}   \\
}
\startdata
A\,1835          &   S1  & $32.4\pm1.6$      &   A\,2029       &  S1  & $513\pm26$       \\   
                 &   S2  & $1.4 \pm0.1$      &                 &  S2  & $3.3\pm0.2$       \\  
                 &   S3  & $1.7 \pm0.1$      &                 &  S3  & $1.3\pm0.1$     \\    
                 &   S4  & $0.40\pm0.02$     &                 &  S4  & $0.45\pm0.02$     \\  
                 &   S5  & $0.55\pm0.03$     &                 &  S5  & $0.31\pm0.02$     \\  
                 &   S6  & $0.56\pm0.03$     &                 &  S6  & $0.38\pm0.02$     \\  
                 &   S7  & $0.14\pm0.01$     &                 &  S7  & $4.3\pm0.2$       \\  
                 &   S8  & $0.56\pm0.03$     &                 &  S8  & $0.89\pm0.04$     \\  
                 &   S9  & $0.63\pm0.03$     &                 &  S9  & $6.6\pm0.3$       \\  
                 &  S10  & $1.6 \pm0.1$      &                 & S10  & $0.39\pm0.02$     \\  
                 &  S11  & $0.67\pm0.03$     &                 & S11  & $0.30\pm0.02$    \\   
                 &  S12  & $0.30\pm0.02$     &                 & MH   & $19.5\pm2.5$     \\   
                 &  S13  & $3.3 \pm0.2$      & && \\
                 &  S14  & $0.21\pm0.01$     & RBS\,797          &  S1  & $17.9\pm0.9$   \\   
                 &  S15  & $0.12\pm0.01$     &                   &  MH  & $5.2\pm0.6$   \\    
                 &  S16  & $0.15\pm0.01$     & && \\                                         
                 &  S17  & $0.14\pm0.01$     & RX\,J1347.5--1145 &  S1  & $30.3\pm1.5$  \\    
                 &  MH   & $6.1\pm1.3$       &                   &  S2  & $4.9\pm0.2^a$ \\    
                 &       &                   &                   &  MH  & $35.6\pm2.9$  \\    
Ophiuchus        &   S1  & $ 30.6\pm1.5$     & && \\                                                   
                 &   S2  & $  1.8\pm0.1$     & MS\,1455.0+2232   & S1 & $4.7\pm0.2$     \\       
                 &   S3  & $ 24.5\pm1.2$     &                   & MH & $8.5\pm1.1$    \\        
                 &   S4  & $ 60.6\pm3.0$     &    && \\                                                   
                 &   S5  & $  3.4\pm0.2$     & 2A\,0335+096      & S1 & $17\pm1$  \\                                            
                 &   S6  & $  5.2\pm0.3$     &                   & S2 & $1.0\pm0.1$ \\           
                 &   S7  & $  6.7\pm0.3$     &                   & MH & $21.1\pm2.1$ \\          
                 &   S8  & $  593\pm30$      &    && \\       
                 &   S9  & $  4.7\pm0.2$     &    && \\ 
                 &  S10  & $  3.3\pm0.2$     & && \\
                 &  MH   & $   62\pm9$       & && \\
\enddata
\tablecomments{Column 1: cluster name. Column 2: radio source name. Column 3: flux density at 1.4 GHz.
$^a$ Measured on the C--configuration image; its flux density on the A--configuration image is $2.0\pm0.1$.
}
\end{deluxetable*}


\end{document}